\documentclass[journal,compsoc]{IEEEtran}
\renewcommand{\normalsize}{\fontsize{9.5}{11.5}\selectfont}
\usepackage{epsfig,rotating,setspace,latexsym,amsmath,epsf,amssymb,amsfonts,bm,theorem,epstopdf,cite,mathtools,caption,subcaption,enumerate,longtable,accents,bbm,algorithm,algorithmic,graphicx,epsf,authblk,url,color,multirow,comment,tabularx,dsfont,physics,tikz}
\usepackage[inline]{enumitem}

\newtheorem{claim}{Claim}

\newtheorem{example}{Example}

\newcommand{\figref}[1]{\figurename~\ref{#1}}
\IEEEoverridecommandlockouts
\allowdisplaybreaks

\allowdisplaybreaks

\title{Incentive Attacks in BTC: Short-Term Revenue Changes and Long-Term Efficiencies}

\author{Mustafa Doger \qquad Sennur Ulukus\\
\normalsize Department of Electrical and Computer Engineering\\
\normalsize University of Maryland, College Park, MD 20742\\
\normalsize  \emph{doger@umd.edu} \qquad \emph{ulukus@umd.edu}}

\date{}

\begin{document}

\maketitle

\begin{abstract}
Bitcoin's (BTC) Difficulty Adjustment Algorithm (DAA) has been a source of vulnerability for incentive attacks such as selfish mining, block withholding and coin hopping strategies. In this paper, first, we rigorously study the short-term revenue change per hashpower of the adversarial and honest miners for these incentive attacks. To study the long-term effects, we introduce a new efficiency metric defined as the revenue/cost per hashpower per time for the attacker and the honest miners. 

Our results indicate that the short-term benefits of intermittent mining strategies are negligible compared to the original selfish mining attack, and in the long-term, selfish mining provides better efficiency. We further demonstrate that a coin hopping strategy between BTC and Bitcoin Cash (BCH) relying on BTC DAA benefits the loyal honest miners of BTC in the same way and to the same extent per unit of computational power as it does the hopper in the short-term. For the long-term, we establish a new boundary between the selfish mining and coin hopping attack, identifying the optimal efficient strategy for each parameter. 

For block withholding strategies, it turns out, the honest miners outside the pool profit from the attack, usually even more than the attacker both in the short-term and the long-term. Moreover, a Power Adjusting Withholding (PAW) attacker does not necessarily observe a profit lag in the short-term. In other words, even without a difficulty adjustment, a PAW attacker makes profits. It has been long thought that the profit lag of selfish mining is among the main reasons why such an attack has not been observed in practice. We show that such a barrier does not apply to PAW and relatively small pools are at an immediate threat. In addition to optimization of revenue ratio, we study alternative objective functions that the adversary may choose to optimize to get most out of the attack among all miners as well as possibly get immediate revenue increase even before the difficulty adjustments.
\end{abstract}

\section{Introduction}
Nakamoto consensus \cite{btc-whitepaper} relies on the Proof-of-Work (PoW) principle with the longest chain protocol to achieve consensus among decentralized nodes in a permissionless setting. The protocol prescribes the miners to mine on top of the longest chain and share the newly mined blocks immediately, which is called the honest protocol. By doing so, each miner gets a revenue proportional to the fraction of hashpower it possesses in the system. The selfish mining strategy introduced in \cite{selfish-mining} by Eyal and Sirer shows that by deviating from the honest protocol and withholding its newly mined blocks for a time, the adversary can increase its revenue ratio in the system. Due to the difficulty adjustment algorithm (DAA), this in turn implies a revenue increase for the deviant miner in Bitcoin after an initial period of loss \cite{grunspan2019-profitability-selfish-mining}. 
 
Over the course of the next decade, many researches focused on extending the work, such as optimizing the selfish mining attacks \cite{optimal-selfish, prob-selfish-mdp-method, stubborn-mining, doger2025selfishminersdoublespend}, studying the short-term revenue change of the adversary \cite{grunspan2019-profitability-selfish-mining, profit_lag}, introducing variations of intermittent/smart intermittent selfish mining to milden the short-term revenue loss of the adversary \cite{intermittent_mining, time_average_selfish_mining}. Others focus on methods to prevent or mitigate the attacks through protocol changes \cite{countering_selfish_mining, freshness_preferred, Preventing_Selfish_Creation_Time, Publish_or_Perish,decor+,sakurai2024tiebreakingrulebasedpartial,preneel_common_metrics}. A systematization of knowledge of selfish mining attacks in Bitcoin is provided in \cite{selfless_sok}. Some other studies focus on alternative strategies such as miners hopping between different but compatible cryptosystems \cite{profit_lag, coin-hopping-meshkov, coin_hop_profitability} or switching their mining rigs on/off \cite{mind_the_mining, energy_equilibria_mining} which also result in profits due to the DAA similar to selfish mining attacks. Similar to selfish mining attacks, a certain Block Withholding (BWH) attack \cite{courtois2014subversiveminerstrategiesblock,fork_after_witholding_attack,power_adjusting} shows that an adversarial miner can increase its revenue ratio by allocating some of its hashpower to pool mining while avoiding to share {the blocks it mines} with the manager whenever it encounters one.

\subsection{A Background on Incentive Attacks}
\subsubsection{Selfish Mining}
Nakamoto's seminal paper \cite{btc-whitepaper} introduced the Bitcoin protocol which prescribes participants to mine blocks on top of the longest chain of blocks by solving cryptographic puzzles and receive rewards in return. Nakamoto argued that participants are incentivized to behave honestly by showing that a double-spend attack succeeds with negligible probability if the attacker holds less than half the computational power in the network and the parties involved wait sufficiently long before confirming a block. However, Eyal and Sirer \cite{selfish-mining} showed that a Bitcoin miner can increase its share of the rewards by mining selfishly instead of honestly, i.e., withholding the new block it mines and mining on its new block privately for a time, even if it has less than half the computational power in the network. The analysis relies on a parameter called adversarial network influence, $\gamma$, which is essentially the fraction of the honest miners in the network that receive the adversarial block first when the adversary releases its withheld block right after an honest miner mines a block and shares it with peers. Modeling the system as a Markov chain, the authors prove that even when the adversary has no network influence, it is incentivized to mine selfishly if the fraction of the hashpower it has in the network, $\alpha$, is at least one third. 

Later, \cite{optimal-selfish} introduced a Markov Decision Process (MDP) framework to obtain the $\epsilon$-optimal selfish mining strategies that maximize the adversarial revenue ratio $\rho$. The use of MDP framework for Nakamoto consensus sparked a wide range of attack considerations. \cite{sompolinsky2016bitcoin} models double-spending attacks and block rewards in the same MDP framework to study the profitability of double-spending. \cite{on_the_sec_pow_gervais} creates and solves an extended quantitative MDP framework with confirmation rule, network propagation dynamics, mining costs and eclipse attack to argue about the performance and security of PoW blockchains under different parameter regimes. Since the relative value iteration of \cite{optimal-selfish} converges slow, \cite{prob-selfish-mdp-method} introduces a policy iteration method with probabilistic termination for practical implementation purposes. 

Others consider selfish miners under pure transaction fee regimes \cite{instability-no-block-reward} or multi-agent selfish miners \cite{multi-selfish,multi-selfish-simulation,multi-selfish-prop-delay}. The authors in \cite{instability-no-block-reward} observe an interesting mining behavior, called petty compliant miner, who are honest but have rational preference about competing blocks based on the transaction fee rewards they offer which can be seen as bribes. This also sparked a further investigation of selfish mining strategies under bribery \cite{sarenche2025miningpowerdestructionattacks,Deep_Bribe}. As the modeling of the agents and environment gets complex, MDP models become infeasible to study the optimal attacks due to large state/action spaces. Such considerations sparked selfish mining studies via deep neural network-based reinforcement learning where the modeling includes aspects such as transaction fees, multi-agent behavior in addition to block rewards \cite{squirrl,werlman,Deep_Bribe}. 

Concurrent with but independent from the $\epsilon$-optimal selfish mining strategies of \cite{optimal-selfish}, the authors of \cite{stubborn-mining} introduced stubborn mining variations of the selfish-mining attack, where the adversary acts selfishly in a stubborn manner under predetermined conditions to increase its revenue ratio. Using Monte Carlo simulations, the authors show that the adversary can obtain further gains on its revenue ratio compared to the selfish mining of Eyal and Sirer. A direct rigorous analysis of the revenues in stubborn mining strategies resorts to the use of Catalan numbers and is done in \cite{catalan-stubborn-grunspan}. $L$-stubborn mining strategies introduced and studied rigorously in \cite{doger2025selfishminersdoublespend}, generalize the selfish and stubborn mining strategies under a restricted state-space with a parameter $L$ that can be optimized efficiently using Catalan numbers and have direct implications on double-spending. Further, the resulting optimal revenue ratios of $L$-stubborn mining are within $10^{-2}$ of the revenue ratios obtained from the $\epsilon$-optimal selfish mining strategies of \cite{optimal-selfish}. 

The initial focus on selfish mining analysis has been the revenue ratio of the adversary \cite{selfish-mining}. This is due to the fact that the DAA of Bitcoin makes sure that the amount of blocks issued in the official chain in two weeks (corresponds to an epoch) stays approximately constant. Investigating this phenomenon, Grunspan and Perez-Marco \cite{grunspan2019-profitability-selfish-mining} show that selfish mining is not profitable before a difficulty adjustment (DA) by conducting a temporal analysis of the selfish mining attack. A similar result is obtained in \cite{Grunspan_witholding_resilience}, where it is shown that without a DA, the honest strategy is the optimal strategy. The same authors also study the profitability of stubborn mining variations of \cite{stubborn-mining} in \cite{grunspan2018profitabilitystubbornmining, grunspan2018profitabilitytrailingmining}.

The study of \cite{grunspan2019-profitability-selfish-mining} showed an interesting result, namely the selfish mining attacker suffers an initial loss period called profit lag, defined as the time until the adversarial revenue change becomes strictly positive. The profit lag of selfish mining spans at least several weeks in BTC, which is considered one of the main reasons why selfish mining is rarely, if ever, observed in the wild. Trying to fix this shortcoming of the selfish mining attack, \cite{intermittent_mining} introduces intermittent selfish mining, where the attacker mounts selfish mining attack in one epoch and mines honestly in the next. Using Monte Carlo simulations, the authors show that an attacker observes a positive revenue change at the end of every two epochs. However, it is shown in \cite{profit_lag} that the intermittent selfish mining attacker's profit continues to dip below zero after reaching positive values at the end of every two epochs. Recently, \cite{time_average_selfish_mining} proposed a smart version of the intermittent selfish mining attack, where the attacker switches between selfish and honest mining during an epoch rather than between epochs, which essentially aims to milden the profit lag issue of intermittent selfish mining attack.

\subsubsection{Coin Hopping and Active Hashpower Adjustments}
Observing how mining behavior effects the cryptographic difficulty of block creation process which in turn affects the number of blocks (hence, rewards) issued, Meshkov et al. \cite{coin-hopping-meshkov} suggests that an attacker, called coin hopper, can switch between two coins at the end of each epoch to increase its profits. Observing the shortcomings of the model used in \cite{coin-hopping-meshkov}, \cite{coin_hop_profitability} simulates the coin hopper in a more realistic setting to argue about the profitability of the attack as well as a comparison between loyal miners of each coin.

Coin hopping phenomenon is frequently observed between BTC and BCH since the two have compatible PoW mining algorithms, making the switch seamless \cite{when-difficulty-algo-breaks}. The profitability of coin hopper between BTC and BCH is rigorously studied in \cite{profit_lag}, which shows that the hopper observes an increase in revenue in BTC network without any profit lag. Others use game-theoretic approach to study coin hopping strategies between BTC and BCH \cite{btc-bch-game-coexist}. In \cite{btc-bch-game-coexist}, the authors show that even though the DAA of BCH was changed to make it more adaptive and resilient to hashpower changes, a coin hopper still poses a threat to the loyal miners of the weaker coin, i.e., BCH. In \cite{game_of_coins}, authors analyze a game scenario with multiple coins and strategic miners, who mine honestly but pick the coin they mine according to their best interest. Based on the analysis therein, a strategic miner can manipulate the rewards (e.g., transaction fees) to move the system between different equilibrium points, i.e., different hash ratios $\alpha$ for a specific coin.

Similar to the idea of coin hopping where the hopper is absent from mining in one network, smart mining \cite{mind_the_mining} proposes to increase the profits of the attacker by prescribing the attacker to turn its mining rigs on/off in even/odd epochs. The authors also point out that such an attack benefits even the honest miners. Another single-coin game theoretic framework \cite{energy_equilibria_mining} studies how miners adjust their active hashpower between epochs depending on different energy efficiencies (cost/time) they have. The authors show that when miners can adjust their hash rates over time, the only possible equilibria are those where miners turn their rigs on/off periodically.

A related study to coin hopping and adjusting active hashpower as well as the intermittent selfish mining strategies is how attackers can accumulate interest gains on loans and deposits in decentralized finance platforms \cite{stretch_squeeze_mining}. The idea is that rules on some platforms work according to the real time past whereas others rely on the number of blocks mined. Since all the mentioned studies manipulate block inter-arrival times, the attackers can increase their gains further by taking loans in one platform and depositing the borrowed funds in another.

\subsubsection{Block Withholding in Pool Mining}

Another line of work on deviant mining strategies focuses on malicious behaviors in pool mining. In Nakamoto consensus, the average time needed for a miner to receive a mining reward is inversely proportional to the fraction of the miner's hashpower with a high variance. Hence, miners join their forces together in pools to avoid fluctuating mining income. In pool mining, the miners receive PoW tasks from the pool manager. While mining, pool participants encounter and share partial PoW solutions (pPOW) that enable the manager to determine the fair share of the rewards each participant gets when a full PoW solution (fPoW) is found eventually by a miner in the pool. 

In his work, Rosenfeld investigates multiple reward systems for pool mining as well as attack vectors \cite{rosenfeld2011analysisbitcoinpooledmining}. Rosenfeld considers miners hopping between coins to take advantage of different payment schemes as well as an attack where a pool miner mounts a block withholding attack (BWH), i.e., avoids to share the fPoW with the manager whenever it encounters one, to sabotage the pool. In 2014, Eligius mining pool announced that it was subject to a BWH and lost 300 BTC at the time \cite{EligiusBlockWithholding2014}.

By extending the sabotage attack of Rosenfeld, Courtois and Bahack \cite{courtois2014subversiveminerstrategiesblock} show that an adversarial miner can increase its revenue ratio by allocating some of its hashpower to BWH  against pools \cite{courtois2014subversiveminerstrategiesblock}. Luu et al. noted some overestimations in the analysis of \cite{courtois2014subversiveminerstrategiesblock} and studied BWH rigorously in a more general framework \cite{power_splitting_pools}. Similarly, Qin et al. investigated the optimal strategies and the necessary conditions in a two-pool-system \cite{optimal-bwh}. An interesting study by Eyal showed that when pools infiltrate each other and mount withholding attacks, in the end they both suffer in terms of the revenue ratio, known as miner's dilemma \cite{miners_dilemma}. A game theoretical analysis of pools attacking each other is provided in \cite{pool-games}.

As a variation of BWH, Fork After Withholding (FAW) attack are introduced in \cite{fork_after_witholding_attack} which shows that the BWH attack can be improved by creating forks in the system between the withheld fPoW and a block of a miner outside the pool. They also show that when pools attack each other, in some cases, the larger pool can avoid the dilemma. Later, \cite{power_adjusting} introduces Power Adjusting Withholding attacks (PAW) which further increases the adversarial gains by dynamically adjusting the adversarial hashpower allocation within the pool after the adversary encounters and hides the fPoW. A limitation of the analysis both in FAW and PAW papers is that they use a simplifying assumption to resolve forks. Another interesting result is presented in \cite{selfholding}, where the authors combine the selfish mining attack and BWH which outperforms the classical selfish mining strategy.

\subsubsection{Network Influence and Fork Choice}
The parameter $\gamma$ which is determined by the fork choice rule and adversarial network influence, is a deciding factor in the amount of gains selfish mining brings compared to the honest mining strategy, i.e., the adversarial revenue ratio in selfish mining increases as $\gamma$ increases. Not only that, it also determines the profitability of selfish mining. The authors in \cite{selfish-mining} note that the current fork choice rule in BTC where honest miners are prescribed to mine on the first block they have seen in case of a fork, is more vulnerable than choices such as randomly picking a block in case of ties since investing in nodes distributed over the network is cheaper than investing in computational hardware. As a result, studies such as \cite{countering_selfish_mining, freshness_preferred, Preventing_Selfish_Creation_Time, Publish_or_Perish,decor+,sakurai2024tiebreakingrulebasedpartial} propose changes in the fork choice rules to prevent/decrease the adversarial gains from the attack. \cite{preneel_common_metrics} investigates how effective these changes are using multi-metric evaluation framework. In this paper, we simply consider the BTC fork choice rule to resolve forks and assume a fixed adversarial network influence parameter $\gamma$. We refer the reader to studies such as \cite{Impact_of_Temporary_Fork,Gobel_selfish_mine_prop_delay} which contain network simulations as well as formulas to estimate $\gamma$ using propagation delays between mining entities. 

\subsection{Our Contributions}

In this paper, we focus on the existing work on deviant mining strategies of \cite{selfish-mining,intermittent_mining, time_average_selfish_mining,stubborn-mining,doger2025selfishminersdoublespend,profit_lag, courtois2014subversiveminerstrategiesblock,fork_after_witholding_attack, power_adjusting} that undermine the incentive compatibility of Nakamoto consensus. We first study the short-term revenue change of the adversarial and honest miners for various selfish mining strategies and the alternate mining strategy of \cite{profit_lag} where the adversarial fraction of hashpower is $\alpha$ and the adversarial network influence is $\gamma$. Our results extend the analysis and the observations made in \cite{intermittent_mining,time_average_selfish_mining,profit_lag,grunspan2019-profitability-selfish-mining} regarding the short-term revenue changes.

Short-term revenue changes and profit lag of the adversaries mounting the selfish mining attack of Eyal and Sirer (and its intermittent version) \cite{selfish-mining} has been analyzed with Monte Carlo simulations in \cite{intermittent_mining} and more rigorously in \cite{grunspan2019-profitability-selfish-mining,time_average_selfish_mining,profit_lag}. However, the literature lacks a rigorous short-term analysis for general selfish mining strategies such as optimal selfish mining \cite{optimal-selfish}. Even though \cite{time_average_selfish_mining} provides a general formula for revenue changes of optimal selfish mining, one has to run the MDP model of \cite{optimal-selfish} for each parameter and perform subsequent numerical analysis such as finding average rewards of the optimal policy in the steady state or a Monte Carlo analysis. To explicitly analyze short-term revenue changes as well as the profit lags for all possible parameters without running MDPs/Monte Carlo simulations, we use $L$-stubborn mining strategies introduced in \cite{doger2025selfishminersdoublespend}. $L$-stubborn mining strategies framework are considered as a substitute for the optimal selfish mining strategies introduced in \cite{optimal-selfish} where the optimal revenue ratios are within $10^{-2}$ of the revenue ratios obtained from the $\epsilon$-optimal mining strategies of \cite{optimal-selfish} and many relevant quantities needed in the analysis can be explicitly expressed in closed from with simple formulas.

Authors of \cite{profit_lag} prove that the alternate mining strategy, which prescribes the adversary to switch between Bitcoin (BTC) and Bitcoin Cash (BCH) in regular intervals corresponding to difficulty adjustments, benefits the adversary without any revenue loss in the short-term. We contribute to the analysis by showing that, under such an attack, the loyal honest miners of BTC observe the same positive revenue change in the short-term as the adversary that hops between coins. We further note that the increase in revenue per computational power is equal for both honest and adversarial miners and proportional to the adversarial fraction of hashpower.

Although studies in \cite{courtois2014subversiveminerstrategiesblock,fork_after_witholding_attack,power_adjusting} showed that BWH attack and its variations such as FAW and PAW increase the adversarial revenue ratio, there is no study focusing on short-term revenue changes of BWH on pools. To fill the gap, we first introduce a slightly modified version of the original PAW attack that we call PAW-Type-B in order to study the attacks in a rigorous manner without the simplifying assumptions of \cite{fork_after_witholding_attack,power_adjusting} regarding the forks. After deriving the relevant quantities in PAW-Type-B, we focus on the revenue change of the adversarial and honest pool miners as well as the honest miners outside the pool. Our results indicate that the honest miners outside the pool observe a positive revenue change per computational power in the short-term, usually even more than the adversary mounting the attack. A more surprising result is, unlike selfish mining, the adversary observes no profit lag when launching PAW on pools for a wide range of parameters, especially when the pool size is small. It has been long thought that the profit lag of selfish mining is the main reason why it has not been observed in practice. Our results indicate that such a barrier does not apply to PAW and small pools are at an immediate threat.

In our analysis, to investigate the results of PAW thoroughly, we resort to numerical optimization tools and define various sets of objective functions that the adversary can optimize. For example, the previous works \cite{fork_after_witholding_attack,power_adjusting} only focus on the maximum adversarial revenue ratio. However, as we point out, such an optimization may result in higher revenue per computational power for the honest miners outside the pool than the adversary. As a result, we investigate alternative objective functions which guarantee that the adversary becomes the entity that gets most out of the attack among all miners. Optimization of another objective function, namely the revenue change just before the difficulty adjustment, makes sure that the adversary gets most out of the attack immediately for certain parameters.

Next, we turn our attention to long-term analysis of deviant mining strategies. We introduce a new efficiency metric that essentially calculates the adversarial revenue/cost per computational power per time in the long-run and compares it with the honest one. We use the new metric to compare the profitability of these strategies within a single framework. To do a valid efficiency comparison, we assume an ideal open market system that is stabilized in the long-run such that, under the attack, the honest miners that stay in the system have sustainable revenue-cost relationship. This allows us to have an overview of the existing work on the subject with a single and reasonable model compared to many different and contradicting claims about selfish mining strategies as well as the denialism in the literature. 
 
In this framework, everything can be rigorously calculated without resorting to any Monte Carlo analysis or MDP models. Hence, our results can further be utilized in analyzing other side-effects of mining strategies such as interest rate gains \cite{stretch_squeeze_mining} or considering deviant mining strategies in game theoretic frameworks such as \cite{energy_equilibria_mining,game_of_coins}, which we leave for future work. Our results also cast a doubt on the claim that orphan reporting makes sure that honest mining is the most profitable strategy in PoW systems \cite{Grunspan_witholding_resilience} in the long-run, since such a claim ignores an open market where miners are free to leave if they make losses.

The rest of the paper is organized as follows. In Section~\ref{sec::prelim}, we go over the generalized versions of selfish mining and introduce PAW-Type-B to set the stage. In Section~\ref{sec::short-term}, we analyze the short-term effects of selfish mining, intermittent selfish mining, smart intermittent mining and alternate network mining (coin hopping) attacks and compare them. In Section~\ref{sec::long_term}, we introduce a new efficiency metric to analyze and compare the same attacks in the long-term. In Section~\ref{sec::bwh}, we focus on PAW-Type-B and study its short-term and long-term effects on the honest and adversarial miners.\footnote{PAW attacks involve pools which require additional assumptions on the hashpower distribution in the network. Further, short and long-term results of PAW are closely related and make a distinction between pool and non-pool honest miners. Hence, we separate this analysis from the other incentive attacks. For coin-hopping attacks, our main reference study \cite{profit_lag} treats coin-hopping attacks in the same framework as selfish mining, hence we follow their approach.}  Finally, in Section~\ref{sec::conc_future}, we summarize our results and discuss future works.

\section{Preliminaries: Incentive Attacks}\label{sec::prelim}
\subsection{Difficulty Adjustment Algorithm}
Throughout the paper, we assume a blockchain employing the Nakamoto consensus with a difficulty adjustment algorithm (DAA) similar to that of Bitcoin (BTC). In Nakamoto consensus, every valid block $B$ has a difficulty parameter $T_h\in(0,1)$ such that $f(B)<T_h$ where $f\rightarrow(0,1)$ is a deterministic function but can be assumed to map each input to a random value. Let us call each block that becomes part of the longest chain in the long-run as a canonical block. The DAA readjusts the mining difficulty $T_h$ after every $D_0$ blocks on the longest chain such that a canonical block is generated in $\frac{\tau_0}{D_0}$ time on average, similar to the DAA employed in BTC where $D_0=2016$ blocks\footnote{A typo in the BTC code causes $D_0=2015$ instead of the originally intended $D_0=2016$, which we ignore here.} and $\frac{\tau_0}{D_0}=10$ minutes. More specifically, after every $D_0$ canonical blocks, the DAA checks the actual time passed to mine the last $D_0$ canonical blocks, call $\tau'_0$, and rescales the block difficulty $T_h$ by $\frac{\tau'_0}{\tau_0}$. Hence, the next $D_0$ canonical blocks have the difficulty $T'_h=T_h\frac{\tau'_0}{\tau_0}$.\footnote{BTC code restricts the change to be $0.25\leq\frac{T'_h}{T_h}\leq4$. However, $\alpha\leq0.5$ guarantees the change to be within $[0.5,2]$ for the system model considered here and in the related works \cite{selfish-mining,intermittent_mining,grunspan2019-profitability-selfish-mining,optimal-selfish,doger2025selfishminersdoublespend,stubborn-mining,prob-selfish-mdp-method,time_average_selfish_mining,profit_lag,courtois2014subversiveminerstrategiesblock,power_adjusting,fork_after_witholding_attack}.} We call the time between each difficulty adjustment as an epoch. Notice, the DAA makes sure that if the total mining power and behavior in the next epoch is the same as the previous one, the next epoch lasts $\tau_0$. However, if the behavior or the total power regularly changes between each epoch, no epoch would actually last $\tau_0$ time. Note that, similar to BTC, BCH also relies on the Nakamoto consensus. However, the DAA in BCH is more complex and more responsive to fluctuations in hashpower than the BTC DAA and can readjust the difficulty within hours. 

Incentive attacks on the Nakamoto consensus rely on DAA due to the following reason. Assume that an adversary has $\alpha$ fraction of the total hashpower in the system and every block pays its miner $\frac{1}{D_0}$ units of money as a coinbase reward. If the adversary follows the honest protocol, on average, it gets $\alpha$ rewards in each epoch. Now, assume that the adversary employs a withholding strategy under which the adversarial blocks are not released as soon as they are mined and makes sure that $\rho$ fraction of the canonical blocks are adversarial. If $\rho>\alpha$, after the difficulty is adjusted, the adversary will get $\rho$ rewards in each subsequent epoch. Notice, in the initial epoch where the attack starts, due to the withholding, the adversary may get less than $\alpha$ rewards, which will be compensated in the following epochs as the difficulty is readjusted.  

The initial fluctuation in the revenue is called the short-term revenue change and will be analyzed in Section~\ref{sec::short-term} for selfish mining strategies and in Section~\ref{sec::bwh} for PAW strategies. To do so, we first introduce these strategies in the remaining part of this section and provide the relevant quantities needed for the analysis. 

\subsection{Selfish Mining Attacks}
Here, we introduce a parameterized version of mining strategies studied in \cite{doger2025selfishminersdoublespend}, which encompasses honest mining, selfish mining of Eyal and Sirer \cite{selfish-mining}, stubborn mining \cite{stubborn-mining}, as well as optimal selfish mining strategies under a limited state space that perform close to the $\epsilon$-optimal mining strategies of \cite{optimal-selfish}. This parameterization gives us a framework to study the mining revenues as well as epoch lengths resulting from selfish mining strategies without any need for a complicated MDP analysis, which will be useful both for short-term and long-term analysis. We call this parameterized strategy as $L$-selfish mining (originally introduced as $L$-stubborn mining \cite{doger2025selfishminersdoublespend}). Note that, the attack description to be provided next with $L=2$ is the same as the original selfish mining attack introduced by Eyal and Sirer, hence we do not restate the original selfish mining attack elsewhere in the text.

\subsubsection{Attack Description} 
An $L$-selfish mining strategy with $L\in \mathbb{Z^+}\cup\{\infty\}$ is defined as follows: Assume a common block, which both honest and adversarial miners are trying to extend. The chain starting from genesis block until this common block is defined as the offset chain. The honest extension from (excluding) the common block onwards is called $H$-chain with length $L_H$ whereas the adversarial extension is called $A$-chain with length $L_A$. The adversary keeps its blocks on each new height $h<L$ private until the honest miners mine a block on that height, upon which the adversary matches the honest block, i.e., releases its block on that height to create a public fork with the honest block. Before the $A$-chain reaches length $L_A=L$, whenever the adversary falls one block behind the $H$-chain, i.e., $L_H=L_A+1\leq L$, the adversary gives up $A$-chain and redefines the offset chain incorporating the $H$-chain extension as the tip of the offset chain. If the $A$-chain reaches length $L_A=L$ without falling behind, the adversary keeps mining private until $L_A=L_H+1$, upon which it releases all its private blocks and redefines the offset chain incorporating the $A$-chain extension as the tip of the offset chain. An attack cycle of $L$-selfish mining starts when an offset chain is defined and ends when a new one is defined. An attack cycle is called successful if the $A$-chain reaches length $L$. Notice, under $L$-selfish mining strategy, the offset chain consists of the canonical blocks. 

\subsubsection{Relevant Quantities} 
Here, we only mention the quantities that are useful for the analysis of this paper and refer the reader to \cite{doger2025selfishminersdoublespend} for further details about $L$-selfish mining and the derivation of the related quantities. Let $\alpha$ and $\beta$ denote the fraction of adversarial and honest hashpower in the system with $\alpha+\beta=1$. Let $\gamma\in[0,1]$ be the adversarial network influence, i.e., the fraction of honest hashpower that prefers an adversarial chain over the rest when a fork happens. Let $\mathbbm{1}_{S}$ be the indicator function of the event that an attack cycle is successful and $\overline{\mathbbm{1}_{S}}$ be its complementary, i.e., $\overline{\mathbbm{1}_{S}}=1-\mathbbm{1}_{S}$. Note that when $L=\infty$, $\overline{\mathbbm{1}_{S}}=1$ since $A$-chain cannot stay longer than $H$-chain forever. Let $L_A$ and $L_H$ denote the length of $A$-chain and $H$-chain at the end of an attack cycle, respectively. Let $L_{A,u}$ denote the number of adversarial blocks that make into the offset chain in an unsuccessful attack cycle. Further, let 
\begin{align}
    P_s(L,m)&=\frac{L-m}{L+m} {L+m \choose L}\alpha^{L}\beta^{m}\\
    P_u(n,i)&=\frac{1}{n+1} {2n \choose n}\alpha^{n}\beta^{n+1}(1-\gamma)^{n-i}\gamma^{\mathbbm{1}_{i}}\\
    P_u(n)&=\frac{1}{n+1} {2n \choose n}\alpha^{n}\beta^{n+1}.
\end{align}
Then, from \cite[Section~4]{doger2025selfishminersdoublespend}, it follows that
\begin{align}
    \mathbb{E}[L_A\mathbbm{1}_{S}]&=\sum_{m=0}^{L-1} P_s(L,m)\left(L+\frac{(L-m-1)\alpha}{1-2\alpha}\right)\\
    \mathbb{E}[L_H\mathbbm{1}_{S}]&=\sum_{m=0}^{L-1} P_s(L,m)\left(m+\frac{(L-m-1)(1-\alpha)}{1-2\alpha}\right)\\
    \mathbb{E}[L_A\overline{\mathbbm{1}_{S}}]&=\sum_{m=0}^{L-1} P_u(m)m\\
    \mathbb{E}[L_H\overline{\mathbbm{1}_{S}}]&=\sum_{m=0}^{L-1} P_u(m)(m+1)\\
    \mathbb{E}[L_{A,u}]&=\sum_{m=0}^{L-1}\sum_{i=0}^{m} P_u(m,i) i.
\end{align}
Thus, in an attack cycle, in expectation, the total number of blocks mined is 
\begin{align}
    \mathbb{E}[T_O]=\mathbb{E}[L_A+L_H],
\end{align}
whereas the total number of canonical blocks mined is
\begin{align}
    \mathbb{E}[T_C]=\mathbb{E}[L_A\mathbbm{1}_{S}+L_H\overline{\mathbbm{1}_{S}}].
\end{align}
We call the total number of blocks generated in the network per canonical block as the block redundancy ratio. In $L$-selfish mining, the block redundancy ratio is
\begin{align}
    \delta_L=\frac{\mathbb{E}[T_O]}{\mathbb{E}[T_C]}.
\end{align}
The revenue ratio of an agent is defined as the fraction of the canonical blocks generated by the agent in the long-run. The revenue ratio of the adversary under $L$-selfish mining is 
\begin{align}
    \rho_L=\frac{\mathbb{E}[L_A\mathbbm{1}_{S}+L_{A,u}]}{\mathbb{E}[T_C]}.
\end{align}
Note that, $L^*=\arg\max_{L} \rho_L$ can be found efficiently using \cite[Algorithm~1]{doger2025selfishminersdoublespend}. Note also that $L=1$-selfish mining is nothing but the honest mining strategy whereas $L=2$ is the selfish mining strategy of Eyal and Sirer \cite{selfish-mining}. When $L=\infty$, the strategy is equal to the equal fork stubborn mining of \cite{stubborn-mining,catalan-stubborn-grunspan}. In this paper, we treat $L^*$-selfish mining as a replacement for the $\epsilon$-optimal mining strategies of \cite{optimal-selfish}, since the relevant quantities for the analysis performed in this paper can be obtained explicitly  and the resulting revenue ratio of the adversary can be optimized with respect to integer values of $L$. Note that, the resulting revenue ratios $\rho_{L^*}$ are within $10^{-2}$ of the $\rho^*$ obtained from the $\epsilon$-optimal mining strategies of \cite{optimal-selfish} as presented in \cite[Table~2]{doger2025selfishminersdoublespend}. On the other hand, to obtain quantities such as block redundancy ratio from $\epsilon$-optimal mining strategies one has to find the optimal strategy for each parameter using the MDP model of \cite{optimal-selfish} as well as numerically finding the stationary distribution and subsequently the average total rewards issued or perform a Monte Carlo analysis. In Appendix~\ref{sec::app::justify_substitue}, we conduct Monte Carlo analysis on the strategies obtained from the MDP of the $\epsilon$-optimal mining strategies of \cite{optimal-selfish} for extensive set of parameters which confirms $L^*$-selfish mining is a good substitute for temporal analysis with closed-form expressions instead of conducting MDP and/or Monte Carlo analysis.

\subsection{A Block Withholding Attack: PAW-Type-B}\label{sec::bwh_intro}
Next, we go over the BWH attacks against pools \cite{courtois2014subversiveminerstrategiesblock, fork_after_witholding_attack, power_adjusting}. As the attack involves pool mining, we explain how pool mining works: When a miner joins a pool, it receives a PoW task from the manager and tries to solve it. This task can be described as finding a valid nonce below a threshold to create a block, which is called full PoW (fPoW). As pools share the rewards of mined blocks between pool members, the fair share of each member has to be determined. To do so, pool manager asks members to submit partial PoWs (pPoW) regularly. pPoWs are essentially easier versions of fPoW puzzles, i.e., nonces that are not below the threshold to create a block but help the pool determine the hashpower of the member. A BWH attacker essentially is a pool miner who submits pPoW but discards fPoW, which makes sure that the member receives its fair revenue when other members create a block, but the revenue of the blocks created by him as a pool member is lost and nobody receives a share. If this attacker also mines individually with the rest of its hashpower, it can end up increasing its revenue ratio by cleverly adjusting its hashpower between the individual and pool mining.

Here, we consider PAW model of \cite{power_adjusting}, which is a generalization of the previous BWH attacks \cite{courtois2014subversiveminerstrategiesblock, fork_after_witholding_attack}. However, the model of \cite{power_adjusting} as well as the preceding model of \cite{fork_after_witholding_attack} assume some simplification in the analysis whenever there is a fork race. Hence, in this paper we consider a slightly modified PAW attack that we call PAW-Type-B which we explain and analyze without any simplifications.

\subsubsection{PAW-Type-B Attack Description} \label{sec::bwh_intro_attack}
 The attack involves an adversary who has $\alpha$ fraction and an honest pool who has $\beta$ fraction of hashpower in the system, where $\alpha+\beta<0.5$. The remaining miners, who make up $1-\alpha-\beta$ fraction of the hashpower, are assumed to be independent honest miners. 
 
 At each attack cycle, the PAW adversary mines individually and honestly using its $(1-p_1)$ fraction of hashpower, i.e., it mines on top of the longest chain and releases its valid blocks immediately. With the remaining $p_1$ fraction, it mines as part of a pool. As a member of the pool, the adversary regularly submits pPoW but does not submit fPoW. When it encounters an fPoW, instead of completely discarding it, it keeps the fPoW hidden and readjusts its hashpower distribution such that its fraction of hashpower working for the pool is now $p_2$. From this moment on:
\begin{enumerate}
    \item If the adversary finds a second fPoW using its $p_2$ fraction, discards one of the fPoW and keeps the other private (does not matter which) and continues in the same manner. 
    \item \label{step::bwh_2}However: 
\begin{enumerate}
        \item If the adversary finds a block using its $(1-p_2)$ fraction (individual honest mining power), it releases the new block and discards the previous fPoW.
        \item If other pool members find a fPoW, the adversary accepts the new fPoW and discards the previous fPoW.
        \item \label{step::bwh_2c}If someone outside the pool creates a block, the adversary releases the privately kept fPoW. The fPoW and the honest block outside the pool enter a fork race\footnote{At this point, the authors in \cite{power_adjusting} make a mild simplifying assumption that each attack cycle is independent of each other and the resulting fork race ends in favor of the adversarial fPoW with probability $c$. Here, we explicitly consider the resulting fork race for a more realistic analysis and use the same $\gamma$ assumption as in the selfish mining model. All the results and observations in Section~\ref{sec::bwh} still hold even under the simplifying assumption with some minor changes in the final values.}. At this point, the adversary changes its hashpower distribution for the pool as zero, i.e., it mines entirely on its own. The fork race resolves in one of the following ways:
\begin{enumerate}
    \item The next block is mined by the adversary (it favors fPoW as the previous block), which it releases immediately.
    \item \label{step::rational_pool_manager} The next block is mined by the pool miners (they favor fPoW as the previous block if $\mathbbm{1}_{P_R}=1$), which they release immediately.
    \item The next block is mined by an honest miner outside the pool. In this case, 
    \begin{enumerate}
        \item with probability (w.p.) $\gamma$, the block is mined on top of the fPoW (since $\gamma$ fraction of honest miners receive first/prefer the fPoW as the previous block).
        \item w.p. $1-\gamma$, the block is not mined on top of the fPoW.
    \end{enumerate}   
\end{enumerate}
\end{enumerate}   
\end{enumerate}

After Case~\ref{step::bwh_2}, the adversary reverts its hashpower distribution for the pool back to $p_1$ and a new attack cycle starts.

\subsubsection{Relevant Quantities}
As the attack cycles are independent, to find the adversarial revenue ratio $\rho_{p_1,p_2}$ (also referred simply as $\rho$), we simply divide the expected adversarial contribution to the canonical blocks created in each attack cycle, $\mathbb{E}[B_A]$, to the expected total number of canonical blocks created in each attack cycle, $\mathbb{E}[B_C]$. In other words, due to the law of large numbers,
\begin{align}
    \rho_{p_1,p_2}=&\frac{\mathbb{E}[B_A]}{\mathbb{E}[B_C]}\label{eq::adv_rho_bwh_type_b}
\end{align}
where
\begin{align}
\mathbb{E}[B_A] = & \alpha(1-p_1) + \beta \frac{\alpha p_1}{\beta + \alpha p_1} + \alpha p_1\Big( \frac{\alpha(1 - p_2)}{1 - \alpha p_2} \nonumber \\
&+  \frac{\beta}{1 - \alpha p_2}\frac{\alpha p }{\beta + \alpha p}  \nonumber \\
&+\frac{1-\alpha-\beta}{1 - \alpha p_2}\big(\frac{\alpha p}{\beta + \alpha p}(\gamma(1-\alpha-\beta)+\alpha+\beta\mathbbm{1}_{P_R})+\alpha\big)\Big),\label{eq::adv_rho_bwh_type_b_Ba}\\
\mathbb{E}[B_C]=&1+\alpha p_1\frac{1-\alpha-\beta}{1 - \alpha p_2}, \label{eq::canonical_bwh_type_b}\\
p=&\frac{p_1+p_2-\alpha p_1 p_2}{2-\alpha p_2}.\label{eq::avg_power_bwh}
\end{align}
Here, $\mathbbm{1}_{P_R}=1$ if the pool manager is rational and favors fPoW of the adversary in a fork race and $0$ otherwise (see Case~\ref{step::rational_pool_manager}). Further, $p$ is the average fraction of the power that the adversary distributes to pool mining conditioned on the event that it encounters an fPoW and keeps it private. Similarly, it can be shown that the revenue ratio of the rest of the pool members is
\begin{align}
    \rho_{pool}=\frac{\mathbb{E}[B_P]}{\mathbb{E}[B_C]}
\end{align}
where the expected honest pool members' total contribution to the canonical blocks created in each attack cycle, $\mathbb{E}[B_P]$, is
\begin{align}
\mathbb{E}[B_P] = &  \beta \frac{\beta}{\beta + \alpha p_1} + \alpha p_1 \Big( \frac{\beta}{1 - \alpha p_2}\frac{\beta}{\beta + \alpha p}+\nonumber \\ 
&\frac{1-\alpha-\beta}{1 - \alpha p_2}\big(\frac{\beta}{\beta + \alpha p}(\gamma(1-\alpha-\beta)+\alpha+\beta\mathbbm{1}_{P_R})+\beta\big)\Big).\label{eq::adv_rho_bwh_type_b_Bp}
\end{align}
For the honest miners outside the pool, the revenue ratio is
\begin{align}
    \rho_{rest}&=1-\rho-\rho_{pool}=\frac{\mathbb{E}[B_R]}{\mathbb{E}[B_C]}\label{eq::rho_rest}
\end{align}
where
\begin{align}
\mathbb{E}[B_R] = (1-\alpha-\beta)\bigg(1+\frac{\alpha p_1}{1-\alpha p_2}\big((1-\alpha-\beta)(2-\gamma)+\beta\overline{\mathbbm{1}_{P_R}})\bigg)
\end{align}
and $\overline{\mathbbm{1}_{P_R}}=1-\mathbbm{1}_{P_R}$. For the remainder of this paper, we simply assume a rational pool manager, i.e., $\mathbbm{1}_{P_R}=1$, which strictly increases the reward of the pool.\footnote{Our numerical analysis shows that the simplification of the revenue ratio calculations with parameter $c$ is approximately valid for most parameters with $c=\gamma(1-\alpha-\beta)+\alpha+\beta$ under a rational pool manager verifying the claim in \cite[Section~9]{fork_after_witholding_attack}.}
The goal of the adversary is to pick $p_1$ and $p_2$ values that maximize an objective function such as $\rho$. Note that, in general, $\rho_{p_1,p_2}$ is neither convex nor concave in $p_1$. Here, the block redundancy ratio is
\begin{align}
    \delta_{p_1,p_2}&=\frac{\mathbb{E}[B_O]}{\mathbb{E}[B_C]}\label{eq::withhold_cycle_dur_type_b}
\end{align}
where $\mathbb{E}[B_O]$ is the expected total number of blocks created in each attack cycle and is found as follows
\begin{align}
\mathbb{E}[B_O]&=\mathbb{E}[B_C]+\frac{\alpha p_1}{1-\alpha p_2}.
\end{align}
We provide the proofs for the relevant quantities of PAW-Type-B in Appendix~\ref{sec::app::bwh}.

\section{Short-term Analysis}\label{sec::short-term}
In this section, we analyze the short-term average revenue change of the adversary and honest miners under selfish mining and coin hopping strategies. For the sake of simplicity, we make the following assumptions: Each canonical block rewards $\frac{1}{D_0}$ unit of coinbase reward, i.e., in total, 1 unit of coinbase rewards are distributed per epoch. We do not consider the transaction fee rewards as a separate entity from the coinbase rewards. Before the attack starts, every miner, including the adversary, was following the honest protocol. The adversary starts its attack right after a difficulty adjustment (DA) and we call the epoch where the adversary starts to attack as the first epoch. Since we are focusing on short-term revenue change, we simply compare the revenue of the adversary under the attack with the revenue of honest strategy using the coinbase rewards as a unit. More specifically, assuming the attack starts at time $t=0$, the short-term revenue change at time $t$ is defined as the total rewards the adversary gets under the attack at time $t$ minus the total rewards the adversary would get at time $t$ if it followed the honest protocol instead.  Further, whenever we use the term revenue change in this paper, we actually refer to its expected value.

\subsection{$L$-Selfish Mining\protect\footnote{
To distinguish notions such as revenue changes and profit lags of different strategies, we use the shorthand notation $(L)$ when the adversary continually applies the $L$-selfish mining strategy.}, $(L)$}

\subsubsection{Epoch Durations}
Assuming the adversary mounts an $L$-selfish mining attack right at the start of the first epoch, we first calculate the epoch durations. Under $L$-selfish mining strategy, at each attack cycle, on average, $\mathbb{E}[T_C]$ canonical blocks are created, hence the expected number of $L$-selfish attack cycles in an epoch is simply $\frac{D_0}{\mathbb{E}[T_C]}$. On the other hand, at each $L$-selfish attack cycle, on average, $\mathbb{E}[T_O]$ blocks are mined in total by everyone. In the first epoch, a block is created in $\frac{\tau_0}{D_0}$ time units on average, since we assume everyone was following the honest protocol previously. Thus, an attack cycle in the first epoch lasts $\mathbb{E}[T_O]\frac{\tau_0}{D_0}$ time units on average. Multiplying the number of $L$-selfish attack cycles with the duration of the cycles, the first epoch lasts
\begin{align}
    t_1=\frac{\mathbb{E}[T_O]}{\mathbb{E}[T_C]}\tau_0=\delta_{L}\tau_0,
\end{align}
on average and the adversary creates $\rho_L$ fraction of $D_0$ canonical blocks.
It is straightforward to conclude that the DAA reduces the difficulty by $\delta_{L}$ (the block redundancy ratio) at the end of the first epoch. After the first difficulty adjustment, each subsequent epoch lasts $\tau_0$ on average since the adversary continues its attack without changing anything. 
\subsubsection{Revenue Changes}
We start by analyzing the revenue change of the adversary when it continuously performs $L$-selfish mining compared to the revenue it would get if it simply deployed the honest protocol, defined as $\Delta^{(L)}_{A}(t)$. Note that, the adversary would get $\delta_{L}\alpha$ rewards in $t_1=\delta_{L}\tau_0$ time if it were to follow the honest protocol. With the attack, at $t_1$, $D_0$ canonical blocks are generated in total and the adversary gets $\rho_L$ rewards. Thus,
\begin{align}
    \Delta^{(L)}_{A}(t_1)=\rho_L-\alpha\delta_{L}\leq 0 \label{eq::l-stub-delta-t1}.
\end{align}
Notice that, at the end of the first epoch, the adversary makes losses \cite{Grunspan_witholding_resilience}. After $t_1$, at each epoch, $D_0$ blocks are added to the offset chain in $\tau_0$ time and the fraction of the blocks the adversary gets increases by $(\rho_L-\alpha)$ compared to honest mining strategy, hence, we have
\begin{align}
    \Delta^{(L)}_{A}(t_1+x)=\rho_L-\alpha\delta_{L}+(\rho_L-\alpha)\frac{x}{\tau_0}. \label{eq::l-stub-delta-t-x}
\end{align}

So far, the analysis we provided is a generalization of the analysis done in \cite{profit_lag}, where we generalized the selfish mining to $L$-selfish mining. Next, we consider the revenue change of the honest miners, i.e., $\Delta^{(L)}_{H}(t)$, which is ignored in the literature. It is straightforward to see, at $t_1$, we have
\begin{align}
    \Delta^{(L)}_{H}(t_1)=(1-\rho_L)-(1-\alpha)\delta_{L} \leq 0, \label{eq::hon-l-stub-delta-t1}
\end{align}
and after $t_1$, we have
\begin{align}
 \Delta^{(L)}_{H}(t_1+x)=(1-\rho_L)-(1-\alpha)\delta_{L} +(\alpha-\rho_L)\frac{x}{\tau_0}. \label{eq::hon-slope-l-stub-delta-t1} 
\end{align}

\subsubsection{Normalized Revenue Changes}
We also define the normalized revenue changes as $\overline{\Delta}^{(L)}_{A}=\frac{\Delta^{(L)}_{A}}{\alpha}$ and $\overline{\Delta}^{(L)}_{H}=\frac{\Delta^{(L)}_{H}}{1-\alpha}$. A normalized revenue change is needed in order to understand how much an agent gains or loses per the computational power it possesses. For example, $\overline{\Delta}^{(L)}_{A}(t)=y$ means that the mining revenue obtained by the adversary under the attack at time $t$, can be obtained by the same adversarial hashpower without the attack, i.e., if the adversary were to mine honestly, at time $t+y\tau_0$. This definition holds independent of the specific values of the adversarial fraction of hashpower $\alpha$ and $\tau_0$, hence the name \textit{normalized}.

\begin{figure}[t!]
\captionsetup[subfigure]{aboveskip=0pt,belowskip=9pt}
     \centering
\begin{subfigure}[t]{0.46\columnwidth}
\centering
\includegraphics[width=\textwidth]{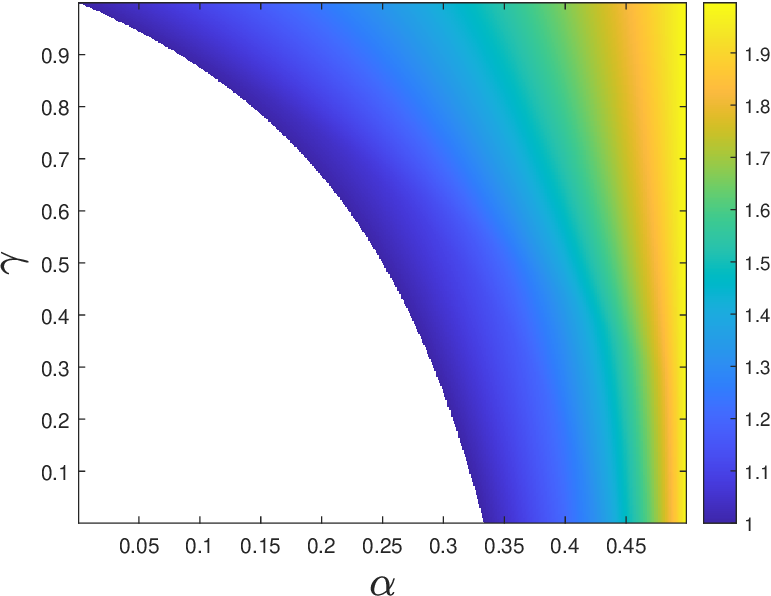}
    \caption{$\frac{\rho_{L^*}}{\alpha}$}
    \label{fig::rev_ratio_L_star}
\end{subfigure}
~
\begin{subfigure}[t]{0.46\columnwidth}
\centering
\includegraphics[width=\textwidth]{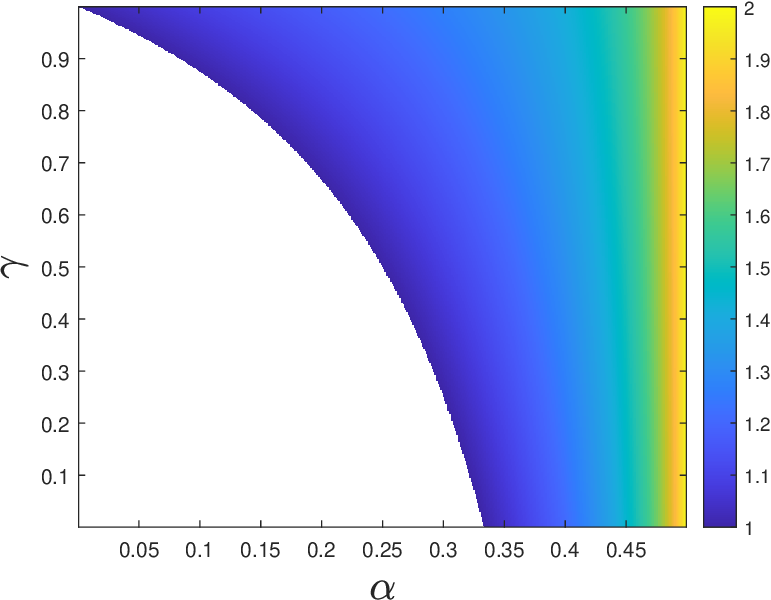}
    \caption{$\frac{\rho_{2}}{\alpha}$}
    \label{fig::rev_ratio_L_2}
\end{subfigure}
     ~
\begin{subfigure}[t]{0.485\columnwidth}
\centering
\includegraphics[width=\textwidth]{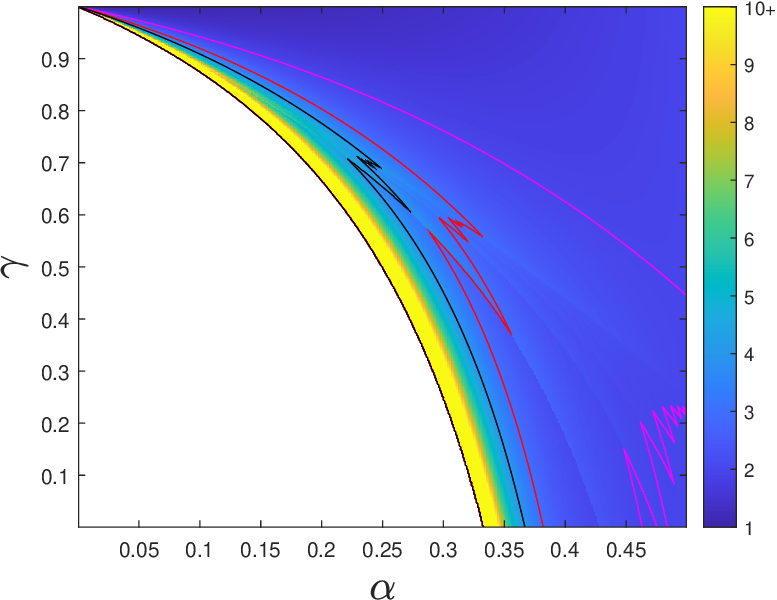}
    \caption{Profit lag, $\Delta_{1}^{L^*}$}
    \label{fig::weeks_L_star}
\end{subfigure}
\hfill
\begin{subfigure}[t]{0.485\columnwidth}
\centering
\includegraphics[width=\textwidth]{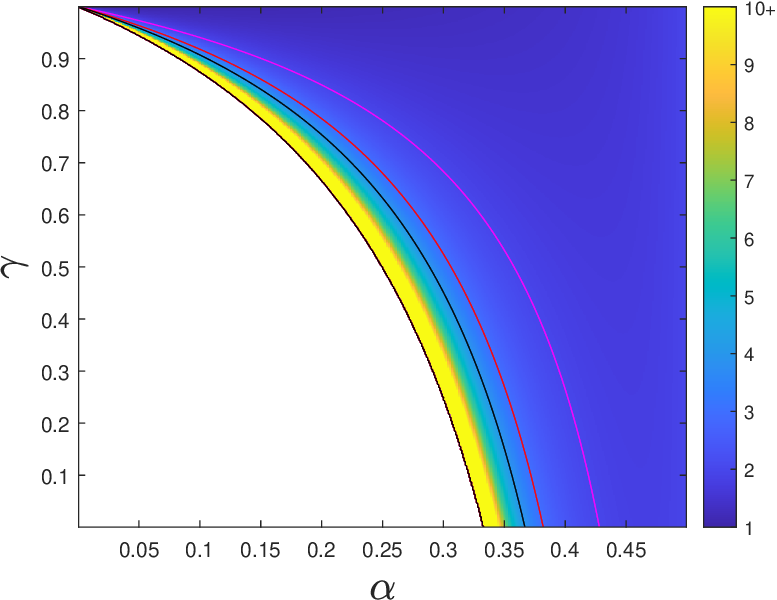}
    \caption{Profit lag, $\Delta_1^{2}$}
    \label{fig::weeks_2_star}
\end{subfigure}    
    \caption{Normalized revenue ratios and time until positive revenue in $L$- and $L^*$-selfish mining.}
	\label{fig::selfish_time}
\end{figure}
\begin{figure}[t!]
     \centering
\begin{subfigure}[t]{0.46\columnwidth}
\centering
\includegraphics[width=\textwidth]{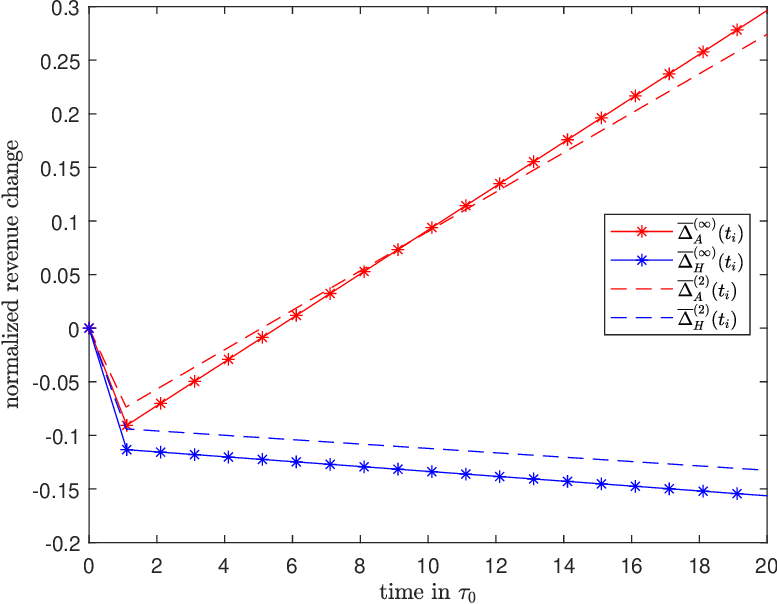}
    \caption{$(\alpha,\gamma)=(0.1,0.9)$, $L=2$, $L^*=\infty$}
    \label{fig::selfish_0.1}
\end{subfigure}
     ~
\begin{subfigure}[t]{0.46\columnwidth}
\centering
\includegraphics[width=\textwidth]{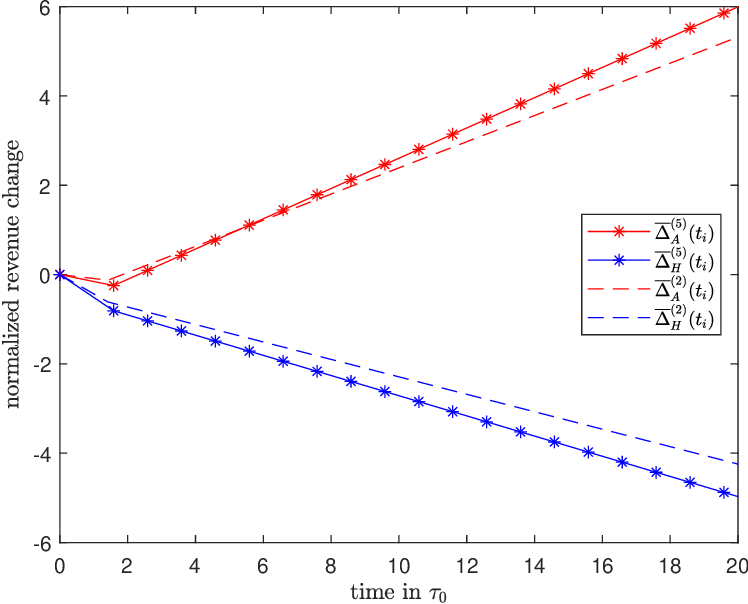}
    \caption{$(\alpha,\gamma)=(0.4,0.4)$, $L=2$, $L^*=5$}
    \label{fig::selfish_0.4}
\end{subfigure}
    \caption{Revenue changes per hashpower.}
	\label{fig::selfish_time_2}
\end{figure}

\subsubsection{Profit Lag} 
We define the time needed for the adversarial revenue difference of $L$-selfish mining to rise above zero strictly, normalized by $\tau_0$, as profit lag. In other words, the profit lag of $L$-selfish mining is 
\begin{align}
    \Delta_{1}^{L}=\frac{\inf \{\tau:\Delta_A^{(L)}(t)>0, \forall t>\tau\}}{\tau_0}.
\end{align}
Heuristically, we can claim that switching from selfish mining of Eyal and Sirer to optimized selfish mining of \cite{optimal-selfish} increases the profit lag as it increases redundancy ratio which increases the initial drop in revenue in \eqref{eq::l-stub-delta-t1}. This observation is verified by the comparing the profit lags of $2$-selfish mining and optimal $L^*$-selfish mining which is provided in \figref{fig::selfish_time}.

\subsubsection{Numerical Results} 
In \figref{fig::selfish_time}, we present the normalized revenue ratio and profit lag in terms of $\tau_0$ of $L^*$-selfish and $L$-selfish mining attacks for all $(\alpha,\gamma)$. Note that the normalized revenue ratio equals the slope of the normalized revenue change of the adversary plus one (after epoch $1$, see \eqref{eq::l-stub-delta-t-x}). The magenta and red lines in the profit lag figures are the borders of $2\tau_0$ and $3\tau_0$ whereas the region inside the black curves represent a profit lag larger than $4\tau_0$. Notice that, compared to $L=2$, $L^*$ increases the normalized revenue ratio, i.e., the slope of the revenue change in \eqref{eq::l-stub-delta-t-x} since $\rho_{L^*}>\rho_2$, however, it also increases the profit lag. The reason is, before $t_1$, $L$-selfish attack cycles create more redundancy, i.e., larger $\delta_L$ when $L$ is larger. This difference increases the initial loss of the adversary in the first epoch compared to if it followed honest mining. In subsequent epochs, i.e., after the first DA, this difference is not in effect anymore as the blockchain growth is normalized again and the revenue of $L^*$-selfish attack eventually surpasses $2$-selfish attack. For honest miners, both terms are negative in \eqref{eq::hon-l-stub-delta-t1} as well as the slope in \eqref{eq::hon-slope-l-stub-delta-t1}, hence normalized revenue is always worse for $L=L^*$ compared to $L=2$ and the difference grows further as time passes. 

Contrary to common belief, with BTC DAA, \figref{fig::weeks_2_star} shows that the selfish mining of Eyal and Sirer results in positive revenue in less than $4\tau_0$ ($8$ weeks in BTC) for a wide range of parameters, i.e., the region outside the black curves. In \figref{fig::selfish_0.1} and \figref{fig::selfish_0.4}, we display the normalized revenue change of $2$-selfish mining and $L^*$-selfish mining for two sets of parameters, i.e., $(\alpha,\gamma)=(0.1,0.9)$ and $(0.4,0.4)$. Note that, the former set falls into the region inside the black curves whereas the latter falls outside. In the former case, $L^*=\infty$, whereas in the latter case, we have $L^*=5$. Other than the general observations made earlier about the slopes and profit lags, it is immediate to see that, per hashpower, the honest miners are adversely affected by each selfish mining attack more than adversarial miners even in the first epoch. 

\subsection{Intermittent $L$-Selfish Mining\protect\footnote{
The shorthand notation $(L,1)$ is used since from one epoch to the other, the adversary switches between $L$- and $1$-selfish (honest) mining.}, $(L,1)$}\label{sec::short-inter}
An adversarial miner employing the intermittent selfish mining acts as follows: For every odd epoch, it employs $L$-selfish mining and then switches to honest mining ($1$-selfish) for an even epoch. 
\subsubsection{Epoch Durations} From the previous section, the first epoch (and all subsequent odd epochs) lasts
\begin{align}
    t_{odd}=\frac{\mathbb{E}[T_O]}{\mathbb{E}[T_C]}\tau_0=\delta_{L}\tau_0,
\end{align}
on average. On the other hand, a similar analysis shows, an even epoch lasts
\begin{align}
    t_{even}=\frac{1}{\delta_{L}}\tau_0,
\end{align}
on average with $t_{odd}\geq\tau_0\geq t_{even}$.
Here, notice that with BTC's DAA, such an attack increases the average block inter arrival time across two consecutive epochs to
\begin{align}
    \frac{\delta_L^2+1}{2\delta_L}\frac{\tau_0}{D_0}\geq \frac{\tau_0}{D_0}. \label{eq::inter-avg-block-time}
\end{align}

\subsubsection{Revenue Changes}
At the end of an even epoch $2n$, the adversarial revenue change is
\begin{align}
    \Delta^{(L,1)}_{A}(t_{2n})=\left((\rho_L+\alpha)-\alpha \frac{\delta_L^2+1}{\delta_L}\right)n,\label{eq::inter-l-delta}
\end{align}
since the adversary gets $\rho_L$ and $\alpha$ fraction of rewards for odd and even epochs, respectively, with the attack. On the other hand, with honest mining strategy, it would get $\alpha$ fraction at each epoch but there would be $n\frac{\delta_L^2+1}{\delta_L}$ epochs in total instead of $2n$. Similarly, at the end of an odd epoch $2n+1$, the adversarial revenue change is
\begin{align}
    \Delta^{(L,1)}_{A}(t_{2n+1})=\Delta^{(L)}_{A}(t_{2n})+\rho_L-\alpha\delta_{L},
\end{align}
which follows from the same idea as in \eqref{eq::l-stub-delta-t1}. For honest miners, we have
\begin{align}
    \Delta^{(L,1)}_{H}(t_{2n})&=\left((2-\rho_L-\alpha)-(1-\alpha) \frac{\delta_L^2+1}{\delta_L}\right)n,\\
    \Delta^{(L,1)}_{H}(t_{2n+1})&=\Delta^{(L)}_{H}(t_{2n})+(1-\rho_L)-(1-\alpha)\delta_{L}.
\end{align}

\begin{figure}[t!]
     \centering
\begin{subfigure}[t]{0.48\columnwidth}
\centering
\includegraphics[width=\columnwidth]{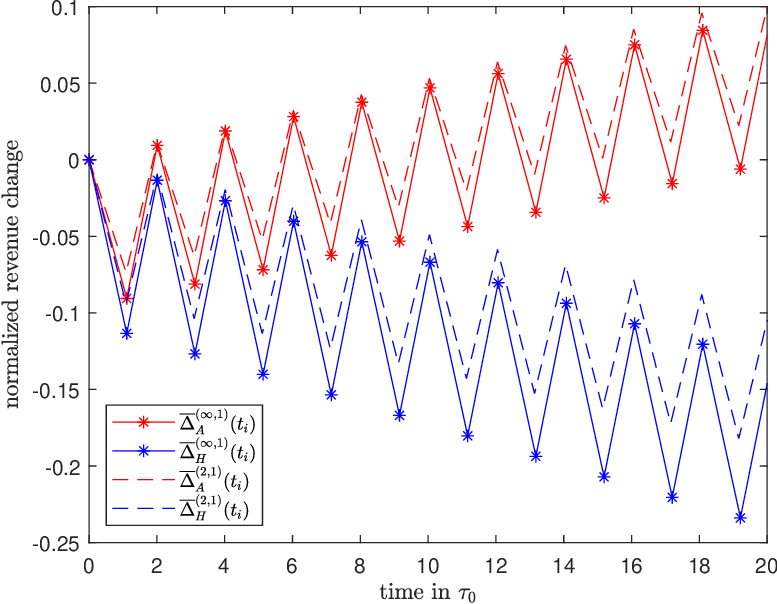}
    \caption{$(\alpha,\gamma)=(0.1,0.9)$, $L=2$, $L^*=\infty$}
    \label{fig::inter_0.1}
\end{subfigure}
     \hfill
\begin{subfigure}[t]{0.465\columnwidth}
\centering
\includegraphics[width=\columnwidth]{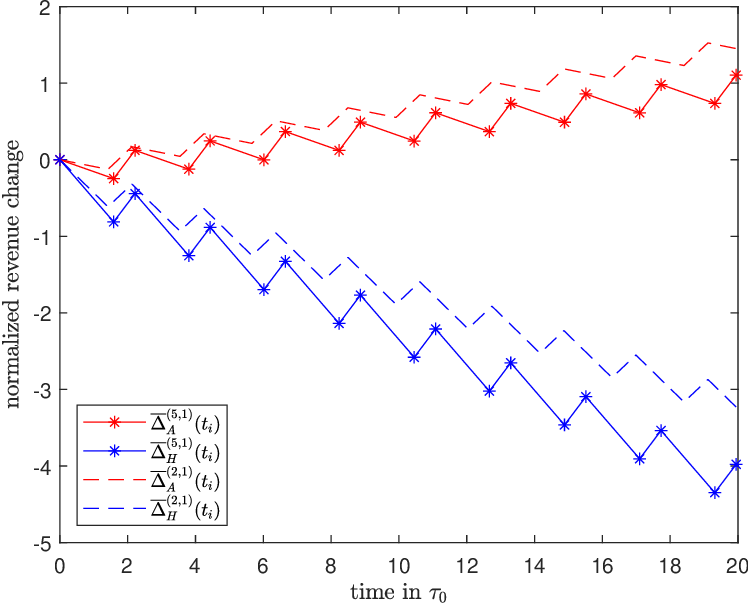}
    \caption{$(\alpha,\gamma)=(0.4,0.4)$, $L=2$, $L^*=5$}
    \label{fig::inter_0.4}
\end{subfigure}
    \caption{Revenue change per hashpower in intermittent $L$-selfish and $L^*$-selfish mining.}
	\label{fig::inter}
\end{figure}

\subsubsection{Numerical Results}
In \figref{fig::inter}, we display the normalized revenue change of intermittent $L$-selfish mining attacks for the same set of parameters discussed earlier. For the adversary, it is immediate to see, $L=2$ is strictly better than $L^*$. For $(\alpha,\gamma)=(0.1,0.9)$, $\delta_{\infty}>\delta_{2}$ despite $\rho_{\infty}>\rho_{2}$ in \eqref{eq::inter-l-delta}, which explains the difference. Although $\delta_{\infty}>\delta_{2}$ and $\rho_{\infty}>\rho_{2}$ are also true for $L$-selfish attack in \figref{fig::selfish_0.1} and \figref{fig::selfish_0.4}, this difference is only effective for the first epoch there. Here, i.e., in the intermittent $L$-selfish attack, the attack repeats itself every two epochs, hence, $L=2$ is strictly better than $L^*$. Hence, for the rest of the analysis, we only focus on $L=2$ when we consider intermittent $L$-selfish attack. Further, as correctly pointed out in \cite{profit_lag}, the profit lag is longer in intermittent $L$-selfish mining compared to $L$-selfish mining despite the revenue change being positive at the end of even epochs, since it drops back to negative values in the subsequent epochs, which we investigate in more detail next.

\subsubsection{Profit Lag}
As in $L$-selfish mining, we define $\Delta_1^{L,1}$ to be the profit lag of intermittent selfish mining, i.e.,
\begin{align}
    \Delta_1^{L,1}=\frac{\inf \{\tau:\Delta_A^{(L,1)}(t)>0, \forall t>\tau\}}{\tau_0}.
\end{align}
Further, to highlight the comparison to selfish mining, we let $\Delta_{L,1}^{L}$ denote selfish profit lag against intermittent, in other words, the time until $L$-selfish outperforms intermittent $L$-selfish mining normalized by $\tau_0$, i.e.,
\begin{align}
    \Delta_{L,1}^{L}=\frac{\inf \{\tau:\Delta_A^{(L)}(t)>\Delta_A^{(L,1)}(t), \forall t>\tau\}}{\tau_0}.
\end{align}

\begin{figure}[t!]
\captionsetup[subfigure]{aboveskip=0pt,belowskip=9pt}
     \centering
\begin{subfigure}[t]{0.46\columnwidth}
\centering
\includegraphics[width=\textwidth]{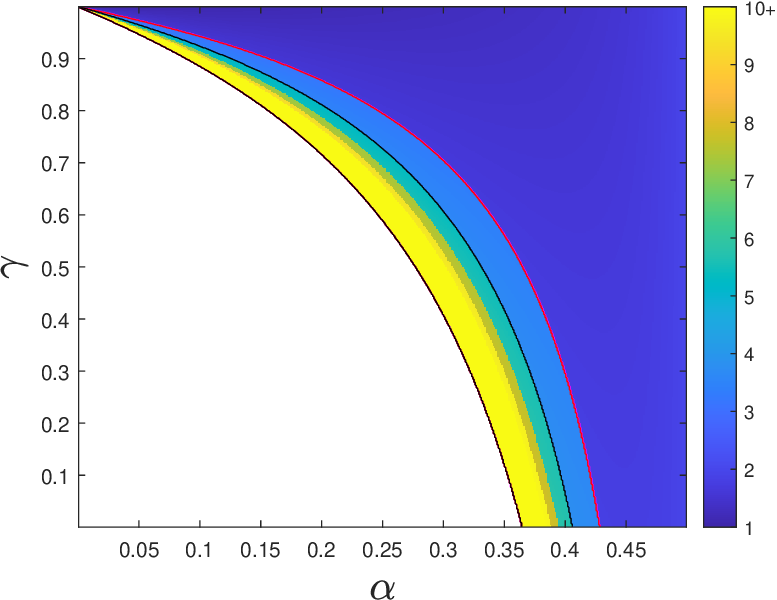}
    \caption{Intermittent profit lag against honest, $\Delta_1^{2,1}$}
    \label{fig::weeks_inter}
\end{subfigure}
~
\begin{subfigure}[t]{0.46\columnwidth}
\centering
\includegraphics[width=\textwidth]{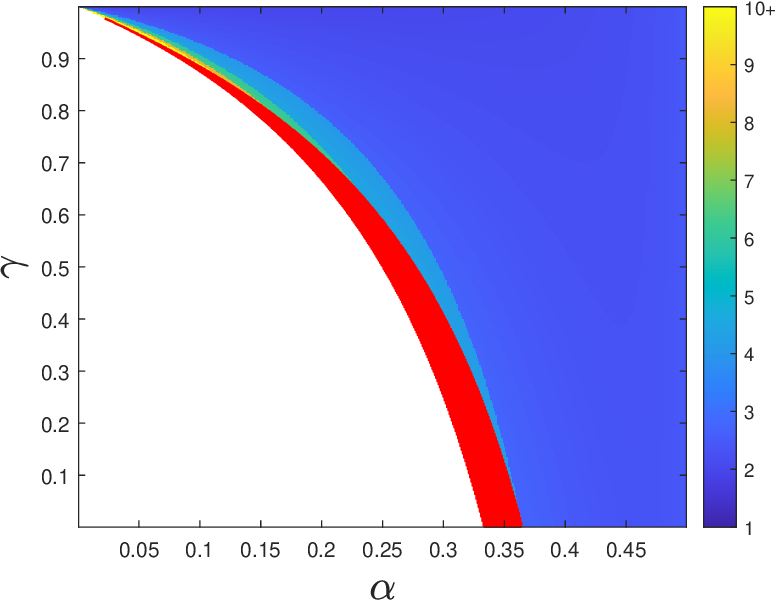}
    \caption{Selfish profit lag against intermittent, $\Delta_{2,1}^{2}$}
    \label{fig::weeks_inter_vs_self}
\end{subfigure}
~
\begin{subfigure}[t]{0.46\columnwidth}
    \centering
    \includegraphics[width=\textwidth]{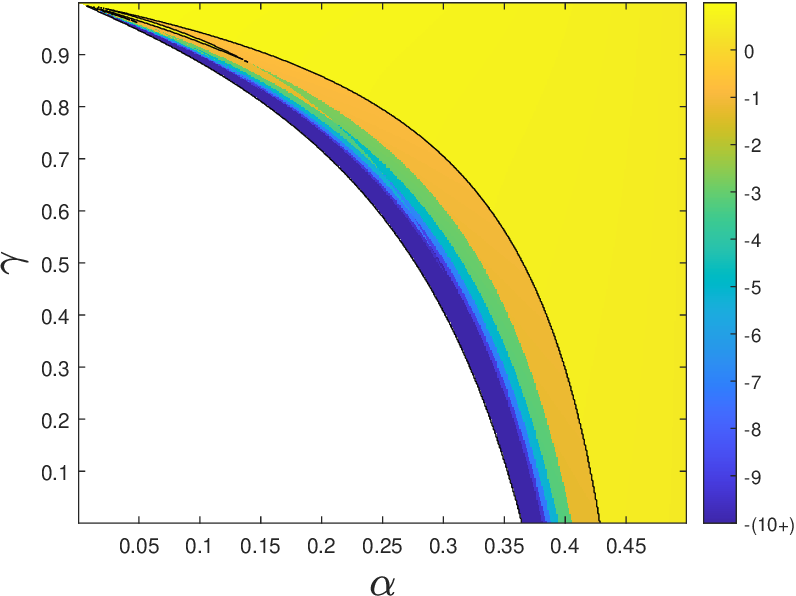}
    \caption{ $\Delta_{2,1}^{2}-\Delta_1^{2,1}$}
    \label{fig::weeks_inter_vs_self_vs_hon}
\end{subfigure}
    \caption{Intermittent selfish mining lags.}
	\label{fig::inter_time}
\end{figure}

In \figref{fig::weeks_inter}, we display the profit lag of intermittent $2$-selfish mining, where the red and black lines have the same boundary meanings as in \figref{fig::weeks_2_star}. Note that, the revenue drops between $[t_{2n}, t_{2n+1}]$ for intermittent mining, hence the profit lags have jumps in the boundaries as profit lag cannot be between $[t_{2n}, t_{2n+1}]$, i.e., the revenue change cannot rise above zero between $[t_{2n}, t_{2n+1}]$. Further, in \figref{fig::weeks_inter_vs_self}, we show the time until the revenue change of $2$-selfish mining surpasses that of intermittent $2$-selfish mining and remains above (strict), normalized by $\tau_0$. The red region corresponds to the parameters where selfish mining is more profitable than honest mining whereas intermittent is never profitable. Notice that for a wide range of parameters (all areas in the relatively darker blue shaded region of \figref{fig::weeks_inter_vs_self}), selfish mining becomes absolutely more profitable than intermittent mining in less than $2.5\tau_0$ ($5$ weeks in BTC). Although \cite{profit_lag} notes that profit lag of intermittent mining is much higher than what authors of \cite{intermittent_mining} initially thought, \figref{fig::weeks_inter_vs_self} brings another observation, that for most parameters, the initial advantage of the intermittent mining compared to selfish mining is insignificant since it does not last long. For the remaining parameters, the profit lag of intermittent selfish mining is too high, corresponding to the yellow region of \figref{fig::weeks_inter}. 

To solidify the claims above, we plot the difference $\Delta_{2}^{2,1}-\Delta_1^{2,1}$ in \figref{fig::weeks_inter_vs_self_vs_hon}, which can be interpreted as whether the intermittent mining becomes absolutely profitable before the selfish mining surpasses it. A positive value corresponds to an affirmative answer and the value itself indicates how many $\tau_0$ time after intermittent mining becomes more profitable than the honest mining, selfish mining surpasses the revenue change of intermittent mining. Here, the black line is the boundary between negative and positive values. For the parameters below the black curve, it is clear that selfish mining becomes more profitable than intermittent mining before intermittent mining becomes more profitable than honest mining. For the parameters above the black curve, even though intermittent mining becomes absolutely profitable before the selfish mining, this advantage does not last long and selfish mining surpasses intermittent within $\tau_0$ time.

\subsection{Smart Intermittent $L$-Selfish Mining\protect\footnote{
The shorthand notation $(L/1)$ is used since within each epoch, the adversary switches between $L$- and $1$-selfish mining.}, $(L/1)$}
Given $0\leq\eta\leq0.5$, an adversarial miner employing the smart intermittent selfish mining strategy acts as follows: For every odd epoch, it employs $L$-selfish mining until $\eta D_0$ canonical blocks are created and switches to the honest mining for the remaining part of the epoch. In an even epoch, it employs $L$-selfish mining until $(1-\eta)D_0$ canonical blocks are created and switches to the honest mining for the remaining part of the epoch.

\subsubsection{Epoch Durations} Since we assume that the adversary was following the honest protocol before the attack started, the first epoch lasts
\begin{align}
    t_1=\eta\tau_0+(1-\eta)\frac{\mathbb{E}[T_O]}{\mathbb{E}[T_C]}\tau_0=(\eta+(1-\eta)\delta_L)\tau_0,
\end{align}
on average. Note that in each odd epoch, $(\eta+(1-\eta)\delta_L)D_0$ blocks are created in total. On the other hand, in each even epoch, $(1-\eta+\eta\delta_L)D_0$ blocks are created in total. Thus, after the first epoch, on average, each odd epoch lasts 
\begin{align}
    t_{odd}=\frac{\eta+(1-\eta)\delta_L}{(1-\eta)+\eta\delta_L}\tau_0=\delta_{(\eta,L)}\tau_0,
\end{align}
whereas an even epoch lasts
\begin{align}
    t_{even}=\frac{1}{\delta_{(\eta,L)}}\tau_0.
\end{align}

\subsubsection{Revenue Changes} Given $L$, in two consecutive epochs, the adversary gets $\alpha+\rho_L$ revenue in total when it follows the smart intermittent selfish mining strategy, whereas it would get $\alpha(\delta_{(\eta,L)}+\frac{1}{\delta_{(\eta,L)}}))$ revenue if it were to follow honest protocol. Thus, the adversarial revenue increase is maximized when $\delta_{(\eta,L)}+\frac{1}{\delta_{(\eta,L)}}$ is minimized, i.e., $\delta_{(\eta,L)}=1$, which is achieved at $\eta=0.5$. With $\eta=0.5$, we have
\begin{align}
    \Delta^{(L/1)}_{A}(t_1)&=\frac{\rho_L-\alpha\delta_{L}}{2},\\
    \Delta^{(L/1)}_{A}(t_{n+1})&=\Delta^{(L/1)}_{A}(t_{1})+\left(\frac{\rho_L-\alpha}{2}\right)n, \quad n>0,\label{eq::smartinter-l-delta}\\
    \Delta^{(L/1)}_{H}(t_1)&=\frac{1-\rho_L-(1-\alpha)\delta_{L}}{2},\\
    \Delta^{(L/1)}_{H}(t_{n+1})&=\Delta^{(L/1)}_{H}(t_{1})+\left(\frac{\alpha-\rho_L}{2}\right)n, \quad n>0.
\end{align}

\begin{figure}[t!]
     \centering
\begin{subfigure}[t]{0.48\columnwidth}
\centering
\includegraphics[width=\textwidth]{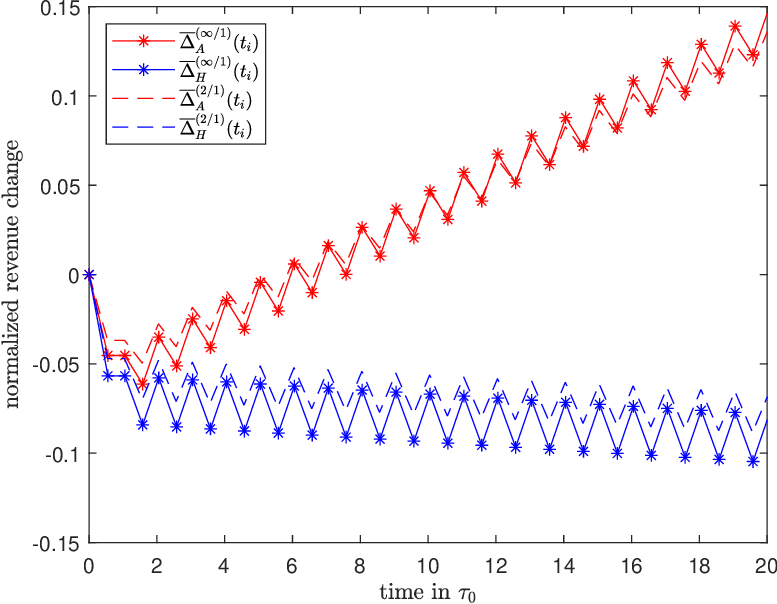}
    \caption{$(\alpha,\gamma)=(0.1,0.9)$, $L=2$, $L^*=\infty$}
    \label{fig::smartinter_0.1}
\end{subfigure}
     \hfill
\begin{subfigure}[t]{0.465\columnwidth}
\centering
\includegraphics[width=\textwidth]{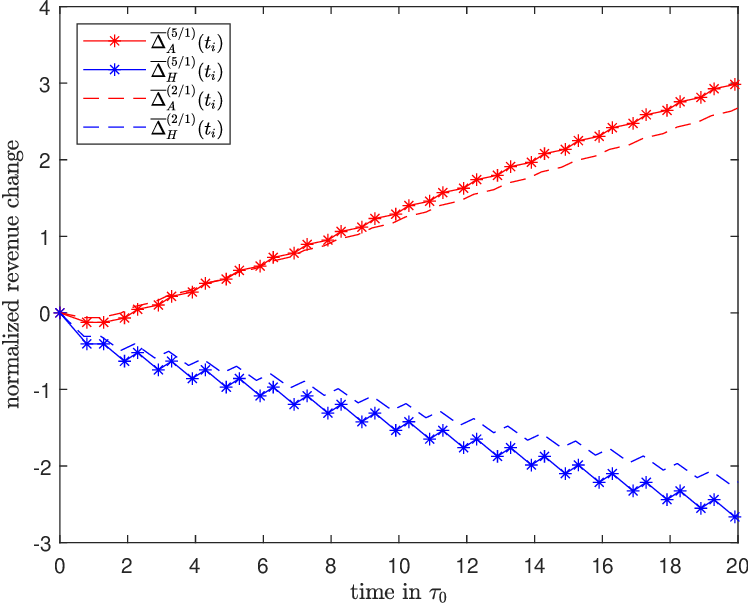}
    \caption{$(\alpha,\gamma)=(0.4,0.4)$, $L=2$, $L^*=5$}
    \label{fig::smartinter_0.4}
\end{subfigure}
    \caption{Revenue change per hashpower in smart intermittent $L$- and $L^*$-selfish mining.}
	\label{fig::smartinter}
\end{figure}

Notice that, in two consecutive epochs, smart intermittent mining achieves the same revenue as in intermittent mining. However, it makes sure that the average block inter-arrival time is $\frac{\tau_0}{D_0}$ instead of a larger value observed in \eqref{eq::inter-avg-block-time} for intermittent selfish mining which implies that the smart intermittent mining achieves the same revenue in less time compared to the intermittent selfish mining. For the sake of completeness, we display the normalized revenue change of smart intermittent selfish mining attacks in \figref{fig::smartinter} for the two sets of parameters as usual. Since the smart intermittent selfish mining has the average block inter-arrival time $\frac{\tau_0}{D_0}$, having $L^*$ instead of $L$ positively affects the average slope of the revenue change compared to the intermittent selfish mining attacks.

\begin{figure}[t!]
\captionsetup[subfigure]{aboveskip=0pt,belowskip=9pt}
     \centering
\begin{subfigure}[t]{0.46\columnwidth}
\centering
\includegraphics[width=\textwidth]{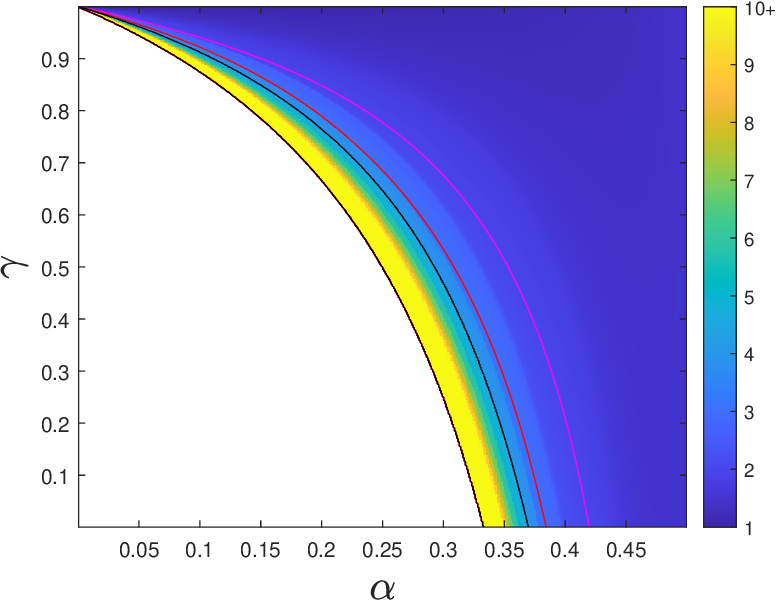}
    \caption{Smart intermittent lag against honest, $\Delta_1^{2/1}$}
    \label{fig::weeks_smartinter}
\end{subfigure}
~
\begin{subfigure}[t]{0.46\columnwidth}
\centering
\includegraphics[width=\textwidth]{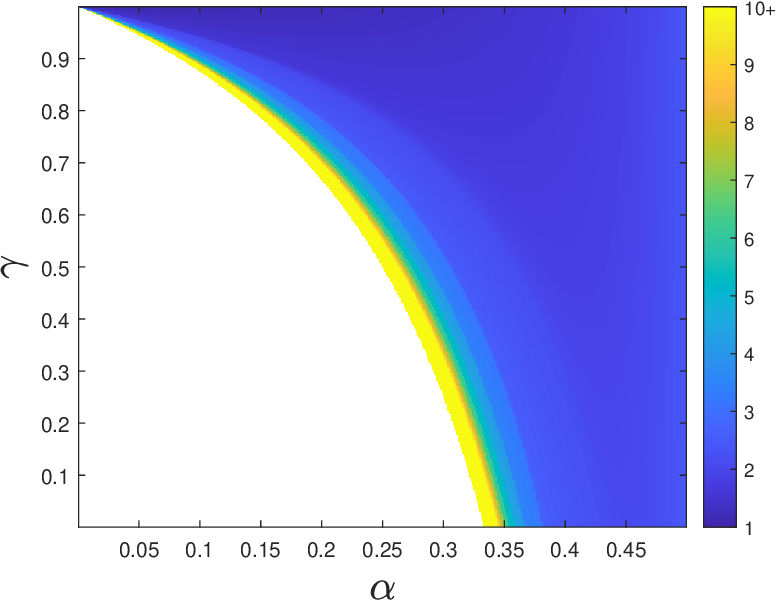}
    \caption{Selfish lag against smart intermittent, $\Delta_{2/1}^{2}$}
    \label{fig::weeks_smartinter_vs_self}
\end{subfigure}
~
\begin{subfigure}[t]{0.46\columnwidth}
    \centering
    \includegraphics[width=\textwidth]{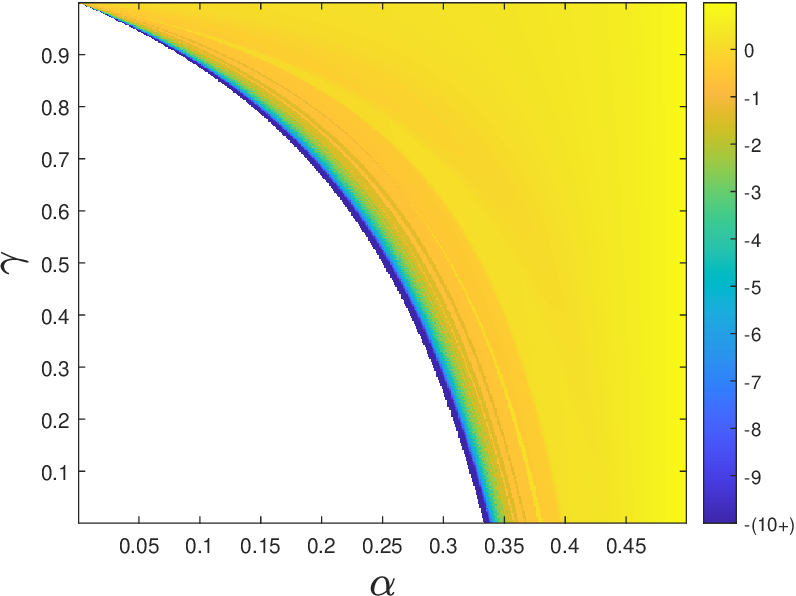}
    \caption{ $\Delta_{2/1}^{2}-\Delta_1^{2/1}$}
    \label{fig::weeks_smartinter_vs_self_vs_hon}
\end{subfigure}
    \caption{Smart intermittent selfish mining lags.}
	\label{fig::smartinter_time}
\end{figure}

\subsubsection{Profit Lag} 
We define the smart intermittent profit lag as $\Delta_1^{L/1}$ and the selfish profit lag against smart intermittent as $\Delta_{L/1}^{L}$ and display the values for each parameter in \figref{fig::smartinter_time}. Although smart intermittent mining outperforms intermittent mining, many of our observations in \figref{fig::inter_time} hold here as well, in other words, for parameters that result in negative values in \figref{fig::weeks_smartinter_vs_self_vs_hon}, selfish mining becomes more profitable than smart intermittent mining before smart intermittent mining generates positive revenue. For other parameters, the initial advantage of the smart intermittent mining compared to selfish mining is still insignificant since it does not last more than $\tau_0$.

\subsection{Alternate Network Mining\protect\footnote{
The shorthand notation $(1,0)$ is used since out of two consecutive epochs, the adversary mines one epoch in BTC.}, $(1,0)$}
Here, we analyze the alternate network mining introduced in \cite{profit_lag}. In the alternate network mining, two blockchain systems are assumed to have compatible mining algorithms so that the miners can switch freely between the two, such as Bitcoin (BTC) and Bitcoin Cash (BCH). Further, before the attack starts, the revenue per time from both networks are assumed to be the same since the opposite would imply a migration between the networks and all revenue quantities mentioned here are measured in terms of the units of BTC coinbase rewards. We focus on the first network for the short-term analysis and call it BTC for the sake of simplicity whereas the second system is called BCH. In the first epoch, the attacker leaves the BTC and mines honestly on BCH. It returns to mine honestly on BTC right after the difficulty adjustment at BTC and continues to switch between the networks in the same manner after each difficulty adjustment at BTC. On the other hand, honest miners are loyal to the network they are mining on. As in reality, we assume BCH has more adaptive difficulty adjustment algorithm, hence the changes in epoch lengths due to the migration is negligible in BCH. In other words, a coin hopping adversary cannot easily stretch or squeeze block inter-arrival times since the BCH DAA is much more responsive, which implies that the changes in revenue per time in BCH due to the coin hopping are negligible. 

\subsubsection{Epoch Durations}
Since the mining power in BTC reduces by $\alpha$ fraction in the first epoch, on average, the first epoch (and the remaining odd epochs) lasts
\begin{align}
    t_{odd}=\frac{1}{1-\alpha}\tau_0=\delta_{0}\tau_0.\label{eq::alternate_epoch_dur}
\end{align}
On the other hand, the mining power in BTC increases from $1-\alpha$ to $1$ in the second epoch compared to the first epoch, hence, on average, the second epoch (and the remaining even epochs) lasts
\begin{align}
    t_{even} = (1-\alpha)\tau_0 =\frac{1}{\delta_{0}}\tau_0.
\end{align}

\subsubsection{Revenue Changes}
We note again that the average block inter-arrival time increases with this attack. At the end of an even epoch $2n$, the adversarial revenue change is
\begin{align}
    \Delta^{(1,0)}_{A}(t_{2n})=\left((\alpha\delta_{0}+\alpha)-\alpha \left(\delta_{0}+\frac{1}{\delta_{0}}\right)\right)n=\alpha^2n.
\end{align}
The reasoning is as follows: If the adversary were to mine solely on BTC, it would get $\alpha$ fraction of blocks mined in $n(\delta_{0}+\frac{1}{\delta_{0}})\tau_0$ time, where $n(\delta_{0}+\frac{1}{\delta_{0}})D_0$ blocks would be created in total, which explains the second term. In BCH, the honest mining revenue is the same as in BTC before the attack starts and BCH is much more adaptive to hashpower changes, i.e., epoch lengths do not vary much. Hence, $\alpha$ hashpower in $\tau_0$ time still gets $\alpha$ rewards in BCH after the attack starts. Since the adversary mines honestly at BCH in odd epochs, it gets $\alpha\delta_0$ rewards in BCH every odd epoch. When it returns to BTC in even epochs, it gets $\alpha$ fraction of rewards in even epochs in BTC. 

\begin{figure}[t!]
\centerline{\includegraphics[width=0.48\columnwidth]{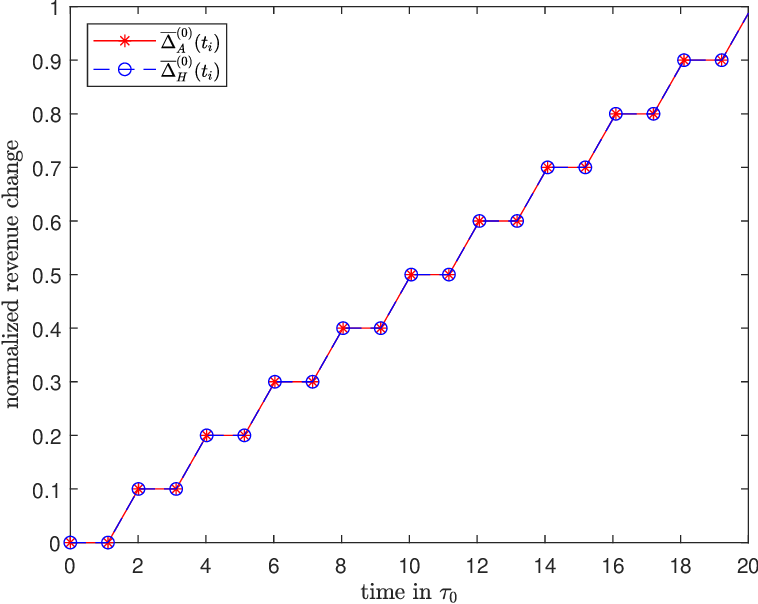}}
	\caption{Revenue change per hashpower for alternate network mining, $\alpha=0.1$.}
	\label{fig::alternate}
\end{figure}

For honest (loyal) miners of BTC, the revenue change is
\begin{align}
    \Delta^{(1,0)}_{H}(t_{2n})&=\left((1+(1-\alpha))-(1-\alpha) \left(\delta_{0}+\frac{1}{\delta_{0}}\right)\right)n\\&=((1-\alpha)-(1-\alpha)^2)n,
\end{align}
since loyal miners get all the rewards in an odd epoch in BTC. As a result, for both adversarial and honest miners, the increase normalized revenue change is $\alpha $ in even epochs, i.e.,
\begin{align}
    \overline{\Delta}^{(1,0)}_{A}(t_{2n})=\overline{\Delta}^{(1,0)}_{H}(t_{2n})=\alpha n,
\end{align}
which is plotted in \figref{fig::alternate}. This result is quite interesting as, in short-term, an increasing hash ratio of coin hopping adversary is beneficial to loyal miners of BTC even when we normalize the revenue changes by their respective hashpowers. We note that the authors of \cite{profit_lag} ignore the revenue change of honest miners in their analysis. On the other hand, authors of \cite{mind_the_mining} consider a `smart mining' algorithm where the adversary switches its mining rigs off during odd epochs, and point out that it benefits even the honest miners, which is somewhat similar to our conclusion. Nevertheless, our short-term analysis implies that there will be an influx of miners to BTC system in an open economy since BTC becomes profitable to mine for everyone when the attacker conducts alternate mining strategy. For example, loyal miners in BCH would be incentivized to leave BCH as BTC revenue is larger than $\rho$ even for loyal miners of BTC. This influx, in turn, should bring down the revenues for everyone mining in BTC. We also note that our intuition and arguments about this situation coincides with the game theoretic analysis of \cite{btc-bch-game-coexist}, where authors argue that the weaker coin could downfall due to the lack of loyal miners.

\begin{figure}[t!]
     \centering
\begin{subfigure}[t]{0.48\columnwidth}
\centering
\includegraphics[width=\textwidth]{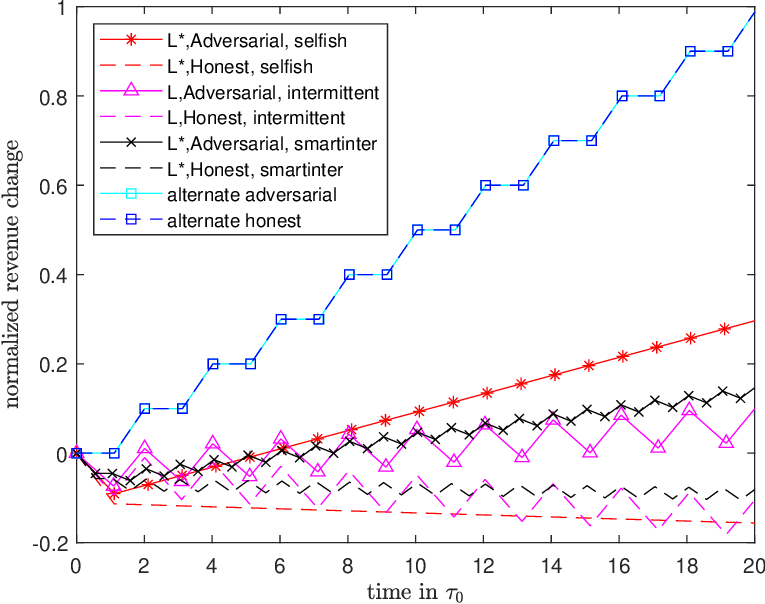}
    \caption{$(\alpha,\gamma)=(0.1,0.9)$}
    \label{fig::compare_0.1}
\end{subfigure}
     \hfill
\begin{subfigure}[t]{0.465\columnwidth}
\centering
\includegraphics[width=\textwidth]{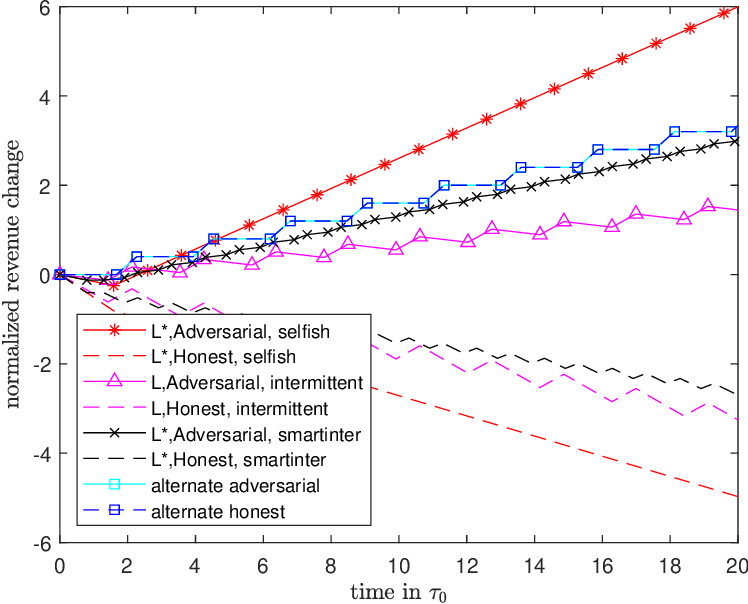}
    \caption{$(\alpha,\gamma)=(0.4,0.4)$}
    \label{fig::compare_0.4}
\end{subfigure}
    \caption{Comparison of revenue change per hashpower.}
	\label{fig::compare}
\end{figure}

\subsection{Comparison of Revenue Changes}
For the sake of completeness, in \figref{fig::compare}, we plot the normalized revenue change for both honest and adversarial miners for each strategy with the two $(\alpha,\gamma)$ parameter choices considered in this section where $x$-axis represents the time normalized by $\tau_0$. Note that, for selfish and smart intermittent selfish mining, we only plot $L=L^*$ since the revenue change eventually surpasses that of $L=2$. For $(\alpha,\gamma)=(0.1,0.9)$, it takes $5\tau_0$ for selfish mining to be absolutely more profitable than smart intermittent mining and $6\tau_0$ epochs for intermittent mining. On the other hand, it takes more than $10\tau_0$ for smart intermittent mining to become absolutely more profitable than intermittent mining. For $(\alpha,\gamma)=(0.1,0.9)$, alternate network mining is always more profitable than the remaining strategies whereas selfish mining is better when $(\alpha,\gamma)=(0.4,0.4)$. Further, selfish mining does the most damage to the honest miners in both cases and alternate mining has the same revenue change for honest and adversarial miners. When $(\alpha,\gamma)=(0.4,0.4)$, selfish mining revenue change surpasses intermittent and smart intermittent mining after $2\tau_0$ and alternate mining after $4\tau_0$ time. These values confirm our observations from Figures~\ref{fig::selfish_time}, \ref{fig::inter_time} and \ref{fig::smartinter_time} regarding the profit lags.

For selfish mining strategies above, if the coin value and the cost of mining stays the same, the honest revenue loss implies that the honest miners would leave the system and potentially migrate to another blockchain system with compatible mining algorithms to salvage their mining hardware costs. This in turn, would result even higher adversarial fraction in the system, damaging the security as well as the reputation of the chain and might even lead to the downfall of the coin. Notice that, in \cite{selfish-mining}, authors argue that selfish mining becomes more profitable with growing $\alpha$, hence, rational miners would join the attackers' pool, which would further increase the effect of selfish mining and damage the blockchain's decentralization and fairness. Here, one can argue from a different perspective, that the blockchain's decentralization and fairness is increasingly damaged since the honest miners could leave the system due to losses. From this short-term analysis, it is clear that, the system parameters cannot stay the same and the revenue analysis done here will not hold as things change. Similarly, for alternate network mining, it is clear that the system dynamics would not stay as they were before the attack since an influx of honest miners is expected as honest miners make the same increased revenue per hashrate as the adversary. Thus, a long-term analysis is needed here as well. Next, we analyze the long-term profitability of these strategies. 

\section{Long-Term Analysis} \label{sec::long_term}
\subsection{A New Metric}
\subsubsection{The Need for a New Metric}
To perform a reliable long-term analysis and comparison of selfish mining strategies, we need a compatible unified metric usable under various scenarios. Some recent papers\cite{grunspan2019-profitability-selfish-mining, Grunspan_witholding_resilience,tailstorm} use a new metric, the adversarial revenue per time (or per chain progress) to prove resilience against incentive attacks. Grunspan and Perez-Marco \cite{Grunspan_witholding_resilience} prove that if orphan blocks are accounted for in the DAA, then, following honest mining strategy results in the best revenue per time even if the relative reward of the adversary increases compared to its fair share. 

Let $\rho_{attack}$ denote the fraction of the rewards the adversary gets in the system (also called relative reward) when the adversary employs the deviant strategy and $\alpha$ denote its fair share (the fraction of hashpower it possesses). Let $\nu$ ($\mu$ resp.) denote the honest (adversarial resp.)  revenue per time when the adversary follows the honest protocol and $\nu_{attack}$ ($\mu_{attack}$ resp.) denote the honest (adversarial resp.) revenue per time when the adversary employs the deviant strategy. Essentially \cite{Grunspan_witholding_resilience, tailstorm} argue that under deviant adversarial mining strategies, the adversarial revenue per time decreases to $\mu_{attack}<\mu$, hence the system is resilient against the attacks. However, such a metric not only ignores the computational power of the adversary but also the other side of the coin, i.e., the honest revenues and its implications in the long-run. Notice that, an incentive compatible system that is not under an attack, i.e., when everyone follows the honest protocol, should satisfy
\begin{align}
    \frac{\nu}{1-\alpha}=\frac{\mu}{\alpha},
\end{align}
which is an equality between the revenue per time per computational power of the adversarial and honest miners.
When the chain is under attack such that $\mu>\mu_{attack}$, but $\rho_{attack}>\alpha$, this automatically implies that honest revenue per time per computational power decreases to
\begin{align}
    \frac{\nu_{attack}}{1-\alpha}<\frac{\nu_{attack}}{1-\rho_{attack}}= \frac{\mu_{attack}}{\rho_{attack}}<\frac{\mu_{attack}}{\alpha}<\frac{\mu}{\alpha}=\frac{\nu}{1-\alpha}, \label{eq::progress_relative}
\end{align}
where the equality $\frac{\nu_{attack}}{1-\rho_{attack}}= \frac{\mu_{attack}}{\rho_{attack}}$ follows from the definitions. As a result, since the honest revenue per time per computational power is less than the adversarial revenue per time per computational power under the attack, i.e.,
\begin{align}
    \frac{\nu_{attack}}{1-\alpha}<\frac{\mu_{attack}}{\alpha},\label{eq::progress_relative_2}
\end{align}
the honest miners are disproportionally more affected from the attack compared to the adversary per time per computational power. In other words, under the attack, the decrease in rewards of the honest miners per time per computational time is greater than that of the adversarial miner, i.e.,
\begin{align}
    \frac{\nu-\nu_{attack}}{1-\alpha}>\frac{\mu-\mu_{attack}}{\alpha}.
\end{align}

As a result, under the attack, in the long-run, when economical parameters adjust themselves in an open market, if honest mining is not profitable, i.e., mining rewards do not compensate mining costs such as hardware and electricity, honest miners will leave the system. This in turn will result in the downfall of the coin, which is even more severe than having a system that rewards miners unfairly. If the honest miners stay in the system in the long-run, this implies that economical dynamics (exchange value, mining costs, etc.) readjust themselves and honest reward per chain progress per computational power, $\frac{\nu_{attack}}{1-\alpha}$, becomes sustainable. This in turn implies that adversarial reward per chain progress per computational power is $\frac{\mu_{attack}}{\alpha}$, which is higher than its fair share, i.e.,  $\frac{\mu_{attack}}{\alpha}>\frac{\nu_{attack}}{1-\alpha}$.

Similarly, the authors in \cite{grunspan2019-profitability-selfish-mining} claim to do profit and loss analysis per unit time for the long-run, however, they use a model that assumes fixed exchange rate that is independent of mining strategies which contradicts the fact that honest mining in the long-run should be sustainable that is dependent on the adversarial strategy. Thus, to account for revenue and mining costs of the miners and to analyze and compare the incentive compatibility in the long-run under different scenarios, a new metric is needed which takes not only time (chain progress) but also the sustainability into the account. To this end, we devise a new unified metric which we call mining efficiency that accounts for the mining revenue, time and costs in the long-run. It is particularly useful when comparing strategies where difficulty adjustments result in stretching and squeezing of blockchain times as well as situations where miners migrate between coins. Next, we explain the specific details about the metric in a simple setting. Later, when we analyze the specific attack strategies, we give more details about our approach accordingly.

\subsubsection{A Unified Metric}
Let $\tau$ be some duration of time and assume that the adversary repeats the same action for each $\tau$. Assume $E$ represents the total energy cost of the system per unit time for mining during $\tau$ and $X$ represents the total rewards issued to the miners. Assume that the honest miners, who have $\beta_a$ fraction of the total active hashpower in the system during $\tau$, create $\beta_r$ fraction of the canonical blocks. Then, honest mining efficiency, is defined as
\begin{align}
    U_{H}=\frac{\beta_r}{\beta_a }\frac{X}{\tau E}.\label{eq::unified_metric}
\end{align}

Notice, the numerator $\beta_rX$ is the revenue of the agent whereas the denominator $\beta_a\tau E$ is essentially the total mining cost, both in $\tau$ time. Hence, the above quantity can be seen as the gain (revenue) of honest miners from the mining divided by the loss due to their energy costs, which is a representation of how effective honest mining is. It is easy to see that the quantity above is a representation of revenue/cost per time per computational power. In the rest of the paper, we sometimes refer to efficiency as \textit{revenue per time per computational power} by assuming that the total energy cost equals $E=1$ in order to convey an intuitive explanation of the results and arguments.

Assuming a stable mining behavior in the long-run with rational honest miners, the honest efficiency should be $U_H=1+\epsilon\approx1$. A higher quantity would result in more honest miners entering the system, changing the coin dynamics such as its exchange value, moving the efficiency value to $1+\epsilon$, whereas a lower quantity would result in some honest miners exiting the system moving the efficiency value to $1+\epsilon$. Thus, $U_H = 1 + \epsilon \approx 1$ represents a long-run competitive equilibrium induced by market forces. We abstract these forces—such as miner entry and exit, exchange-rate adjustments, and demand fluctuations—into an environmental function $F(\mathsf{Env})$, which we treat as exogenous and equilibrium-enforcing. This abstraction allows us to ignore detailed market dynamics while assuming that $F(\mathsf{Env})$ maintains $U_H \approx 1$.

Notice that if $U_H\neq1$ (happens usually when an attack starts), then, honest miners will leave or join the system, not only changing $\alpha$ and related quantities, but also the value of mining rewards with respect to the electricity costs. Finding a proper mathematical model for the interrelation of all these quantities in order to derive the final adversarial fraction of hashpower, $\alpha_{final}$, cannot be done without considering other economic factors in the real world which is too complex and out of scope of this paper. This final stabilized $\alpha_{final}$ value will most likely be different than the initial adversarial fraction of hashpower $\alpha_{start}$ when the attack started. In this section, we simply assume a stabilized final adversarial fraction of hashpower $\alpha_{final}$ (if such an equilibrium exists) and without loss of generality, we refer to it as $\alpha$. We further assume this stabilized adversarial fraction of hashpower satisfies $\alpha_{final}<0.5$. Else, such a case immediately implies downfall of the coin in question and renders further analysis unnecessary.

Since $\epsilon$ is small and can be ignored, we assume $U_H=1$ which in turn implies that the adversarial mining efficiency is
\begin{align}
    U_A=\frac{(1-\beta_r)}{(1-\beta_a)}\frac{X}{\tau E}=\frac{(1-\beta_r)\beta_a}{(1-\beta_a)\beta_r}. \label{eq::efficiency_func}
\end{align}
In the simplest case, if the adversary follows the honest mining strategy, i.e., when $\beta_r=\beta_a$, the above quantity is simply equal to $1$, i.e., $U_H=U_A$. 

Note that, the normalized revenue ratio of the adversary, $\frac{1-\beta_r}{1-\beta_a}$ measures the ratio of canonical blocks created by the adversary during the attack to the number of canonical blocks the adversary would create if no attack were launched. In selfish mining literature, normalized revenue ratio of the adversary, is always below $2$, i.e., $\frac{1-\beta_r}{1-\beta_a}=\frac{\rho}{\alpha}\leq2$ as in \figref{fig::rev_ratio_L_star} and \figref{fig::rev_ratio_L_2}, and usually treated as some form of efficiency. However, our argument here is that, in the long-run, the normalized revenue ratio of the honest miners, $\frac{1-\rho}{1-\alpha}$, should be considered first and treated as the basis for efficiency instead of treating $\frac{\rho}{\alpha}$ directly as efficiency, since honest miners would leave the system if their normalized revenue ratio is not efficient. Next, we analyze the long-term efficiency of the adversary that employs the selfish mining strategies considered in Section \ref{sec::short-term}. 

\subsection{$L$-Selfish Mining, $(L)$}\label{sec::l-selfish}
If the adversary follows $L$-selfish mining strategy continuously, then, $\rho_L$ fraction of total rewards belong to the adversary after every $\tau_0$ time (we can ignore the first epoch in the long-run as it happens only once), where $\alpha$ fraction of the energy cost is adversarial. Hence, the honest efficiency can be trivially obtained as
\begin{align}
    U^{(L)}_H=\frac{1-\rho_L}{1-\alpha}\frac{X}{\tau_0 E}.
\end{align}
 A stabilized system implies $U_{H}=1$, which in turn allows us to present the adversarial efficiency as
\begin{align}
    U_A^{(L)}=\frac{\rho_L}{\alpha}\frac{X}{\tau_0 E}=\frac{1-\alpha}{\alpha}\frac{\rho_L}{1-\rho_L}\label{eq::L_stub_efficiency_func}.
\end{align}
Notice, under selfish mining attacks, $\rho_L$ is an increasing function of $\alpha$ as explained in \cite{selfish-mining}. Thus, among the adversarial strategies with period $1$, optimal selfish mining attack, where the adversary uses its full hashpower, i.e., does not turn off any mining rig it has, maximizes the adversarial efficiency given in \eqref{eq::efficiency_func}. Using the same reasoning, simple algebra shows $L^*=\arg\max_{L} \rho_L$ maximizes the adversarial efficiency in \eqref{eq::L_stub_efficiency_func} among the $L$-stubborn mining strategies.

\subsection{Intermittent $L$-Selfish Mining, $(L,1)$}

The analysis of Section \ref{sec::short-inter}, implies that attack repeats itself every $(\delta_{L}+\frac{1}{\delta_{L}})\tau_0$ time. In the initial $\delta_{L}\tau_0$ time, $\rho_L$ fraction of the total $X$ rewards belong to the adversary whereas in the subsequent $\frac{\tau_0}{\delta_L}$ time $\alpha$ fraction of the total $X$ rewards belong to the adversary. Hence, the honest efficiency is
\begin{align}
    U^{(L,1)}_{H}=\frac{(2-\alpha-\rho_L)}{(1-\alpha) (\delta_{L}+\frac{1}{\delta_{L}})}\frac{X}{\tau_0 E},
\end{align}
which, together with $U_H=1$, implies,
\begin{align}
    U^{(L,1)}_A=\frac{(\alpha+\rho_L)}{\alpha (\delta_{L}+\frac{1}{\delta_{L}})}\frac{X}{\tau_0 E}=\frac{1-\alpha}{\alpha}\frac{\alpha+\rho_L}{2-\alpha-\rho_L}.\label{eq::intermittent_efficiency_func}
\end{align}
which is maximized for $L^*=\arg\max_{L} \rho_L$ given $\alpha$.

\subsection{Smart Intermittent $L$-Selfish Mining, $(L/1)$}
In smart intermittent $L$-selfish mining, with $\eta=0.5$, every epoch lasts $\tau_0$ time and the honest efficiency becomes
\begin{align}
    U^{(L/1)}_{H}=\frac{2-\alpha-\rho_L}{(1-\alpha)2}\frac{X}{\tau_0 E},
\end{align}
which, together with $U_H=1$, implies
\begin{align}
    U_A^{(L/1)}=\frac{\alpha+\rho_L}{\alpha2}\frac{X}{\tau_0 E}=\frac{1-\alpha}{\alpha}\frac{\alpha+\rho_L}{2-\alpha-\rho_L}, \label{eq::smart_intermittent_efficiency_func}
\end{align}
which is the same as \eqref{eq::intermittent_efficiency_func}. Notice, in short-term analysis, the smart intermittent selfish mining strategy increased the revenue gain per time of the adversary compared to the intermittent selfish mining by shortening the average epoch length. In the long-run analysis, even though the epoch lengths are still shorter for the smart intermittent selfish mining strategy, this does not bring any improvement, as all miners are affected by the increased average epoch time and the efficiency metric assumes the honest efficiency is the same ($U_H=1$) in both cases after the long-term stabilization of the coin dynamics.

\subsection{Alternate Network Mining, $(1,0)$}
In alternate network mining, we denote the unit value of BTC as $X$ and all revenue quantities mentioned here are measured in terms of $X$. We make the assumption that honest miners' efficiency in BTC and BCH have to be the same in the long-run, else honest miners would migrate. This also implies that the revenue per time per computational power of honest miners in both coins are the same in the long-run, since both networks share the same mining algorithm. Further, since the effect of hopping on epoch lengths of BCH is negligible, the adversarial miner also gets the same revenue per computational power per time in BCH as the loyal miners of BCH. On the other hand, the BTC DAA is much less responsive which is the reason why the coin hopping strategy results in more revenue per computational power per time for the adversary.

Clearly, honest miners in BTC get $(2-\alpha)X$ rewards in $2$ epochs which last $(\delta_0+\frac{1}{\delta_0})\tau_0$ time, which implies that the honest revenue per computational power per time (both in BTC and BCH) is $\frac{2-\alpha}{1-\alpha}\frac{1}{\delta_0+\frac{1}{\delta_0}}\frac{X}{\tau_0}$. The revenue of the adversary from BCH in odd epochs can be found by multiplying the honest revenue per computational power per time in BCH with the adversarial computational power $\alpha$ and time spent $\delta_0=(1-\alpha)^{-1}$ in BCH, i.e., $\frac{2-\alpha}{1-\alpha}\frac{\alpha}{1+(1-\alpha)^2}X$. Further, with alternate mining, the adversarial miner gets $\alpha X$ revenue in BTC network in even epochs. Thus, we can obtain the adversarial efficiency by summing the two revenues and dividing it by the adversarial computational power $\alpha$ times the total time spent $(\delta_0+\frac{1}{\delta_0})\tau_0$. Since the honest efficiency is
\begin{align}
    U^{(1,0)}_{H}=\frac{2-\alpha}{(1-\alpha)(\delta_0+\frac{1}{\delta_0})}\frac{X}{\tau_0 E},
\end{align}
this, together with $U_H=1$, implies
\begin{align}
    U^{(1,0)}_A=\frac{\alpha+\frac{2-\alpha}{1+(1-\alpha)^2}\frac{\alpha}{1-\alpha}}{\alpha(\delta_0+\frac{1}{\delta_0})}\frac{X}{\tau_0 E}=\frac{1-\alpha}{2-\alpha}+\frac{1}{1+(1-\alpha)^2},\label{eq::alternate_efficiency_func}
\end{align}
which is greater than $1$ for $0<\alpha<1$. 

Notice the difference between the short-term and long-term analysis results of the alternate mining. Although we assume that the honest miners' revenue per time per computational power in BTC and BCH are the same in both short-term and long-term analysis, in short-term this implies honest and adversarial miners benefit the same revenue change per computational power, whereas in long-term it implies a better efficiency for the adversarial miners. This happens for the following reason. In the short-term analysis, we first assume that honest revenue per time per computational power in BTC and BCH are the same before the attack started and then analyze the revenue change of loyal miners of BTC. On the other hand, in the long-term analysis, we first calculate the revenue per computational power per time of the honest miners in BTC with the attack taken into account and then assume that the same revenue rate holds for BCH.

\begin{figure}[t!]
     \centering
\begin{subfigure}[t]{0.48\columnwidth}
\centering
\includegraphics[width=\textwidth]{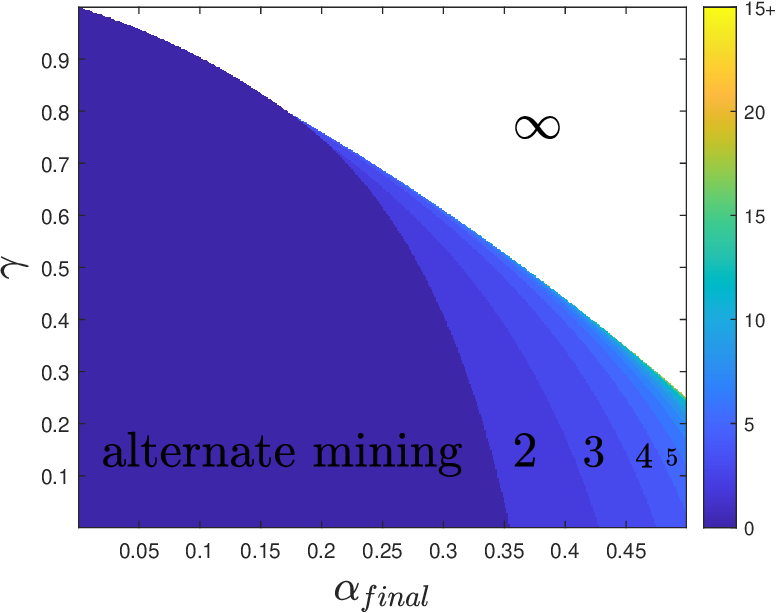}
    \caption{Best strategy}
    \label{fig::efficiency_comparison}
\end{subfigure}
     \hfill
\begin{subfigure}[t]{0.465\columnwidth}
\centering
\includegraphics[width=\textwidth]{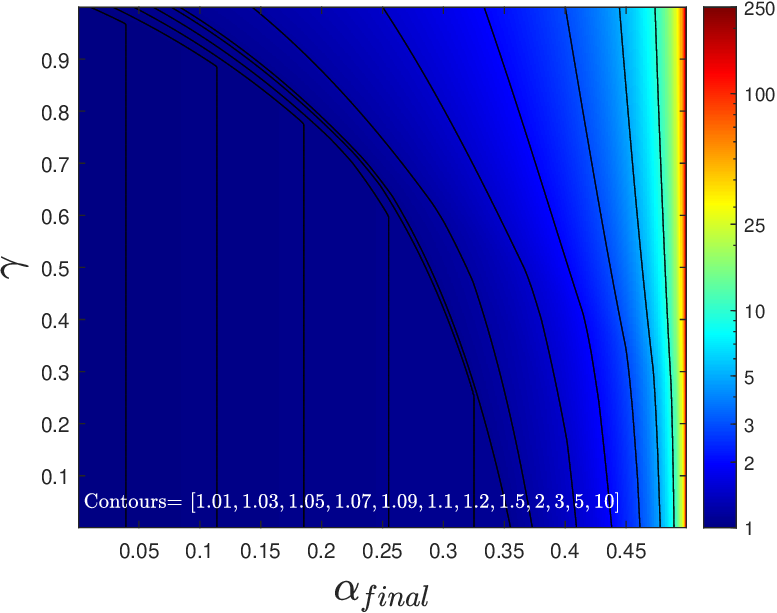}
    \caption{Efficiency of the corresponding strategy}
    \label{fig::efficiency_map}
\end{subfigure}
    \caption{Efficiencies given $\alpha_{final}$.}
	\label{fig::efficiency}
\end{figure}

\subsection{Comparison of Adversarial Efficiencies}
Starting with the same hashpower in the system, call it $\alpha_{start}$, it is not clear if any of the above strategies would result in the same $\alpha_{final}$ in the long-run when other dynamics are taken into account, such as honest miners leaving the system and the change in mining costs and revenues. Hence, it is not possible to claim that one strategy is better than the other in the long-run even if the initial hash ratio $\alpha_{start}$ is the same. However, in the long-run, if the final hash ratios (i.e., $\alpha_{final}$'s) are known for each model, using our model, their performances can be compared. 

Assuming the same final adversarial fraction of hashpower $\alpha_{final}$ in the long-run, intermittent and smart intermittent selfish strategies are outperformed by the constant effort selfish strategy, which can be seen by comparing \eqref{eq::L_stub_efficiency_func} with \eqref{eq::intermittent_efficiency_func} or \eqref{eq::smart_intermittent_efficiency_func} since $\rho_{L^*}\geq \alpha$. However, note that, with the intermittent selfish mining, the epoch lengths are stretched and squeezed alternately with the parameters $\delta_L$ and $\frac{1}{\delta_L}$. This in turn, can be utilized to calculate interest gains on loans and deposits similar to the analysis of \cite{stretch_squeeze_mining}. In fact, $\delta_{L}\geq 1$ is increasing in $L$ which in turn increases interest-rate gains from blockchain stretching \cite{stretch_squeeze_mining}. On the other hand, $\rho_L$ is quasiconcave in $L$ as proven in \cite[Theorem~3]{doger2025selfishminersdoublespend}, however, for many parameters of interest, $L^*=\infty$ or $\rho_{L^*} \approx \rho_\infty$ \cite{catalan-stubborn-grunspan,doger2025selfishminersdoublespend}. Hence, picking $L=\infty$ could be more profitable than $L^*$ when interest-rate gains are taken into account together with the block rewards. Moreover, in this case, intermittent $L$-selfish mining could be even more profitable than $L$-selfish mining. Similar arguments about stretching and squeezing are valid for alternate network mining as well.

Observe that, selfish mining has a profitability threshold and is not always profitable compared to honest mining. The same threshold is valid here and can be used to determine if the adversarial efficiency in \eqref{eq::L_stub_efficiency_func} (and \eqref{eq::intermittent_efficiency_func}, \eqref{eq::smart_intermittent_efficiency_func}) exceeds $1$. On the other hand, the adversarial efficiency for alternate network mining is greater than $1$ for all $0<\alpha<1$. Thus, unlike selfish mining, alternate network mining is profitable even for a small home miner. 

\subsubsection{Numerical Results}
In \figref{fig::efficiency_comparison}, we provide the best strategy assuming they have the same final adversarial hashrate $\alpha_{final}$. In other words, for each $(\alpha_{final},\gamma)$ parameter, the strategy that results in the highest efficiency is given in \figref{fig::efficiency_comparison} and the corresponding efficiency of that strategy is provided in \figref{fig::efficiency_map}. The regions labeled with numbers in \figref{fig::efficiency_comparison} correspond to the regions where $L$-selfish mining is the strategy that results in the highest efficiency and the number stands for the $L$ that achieves maximum efficiency. We note that the region where alternate mining is the best strategy is larger than the region where honest mining outperforms selfish mining. For example, when $\gamma=0$, selfish mining of Eyal and Sirer ($L$=2) outperforms honest mining for $\alpha>1/3$, whereas selfish mining of Eyal and Sirer does not outperform alternate network mining when $\alpha_{final}<0.355$. 

Inside \figref{fig::efficiency_map}, we also depict the contours using black lines, with the value of each contour level indicated in white at the bottom left of the figure. First, notice the straight vertical parts of the contours. This happens, because those regions correspond to alternate mining strategy in \figref{fig::efficiency_comparison} and the efficiency of alternate mining strategy does not depend on the $\gamma$ value. Further, in the regions of $(\alpha_{final},\gamma)$, where the alternate mining is the best strategy, the efficiency is always below $1.1$, which means that the adversary is at most $10\%$ more efficient than the honest miners under the attack. 

On the other hand, for larger $(\alpha_{final},\gamma)$ values, $L^*$-selfish mining is optimal and the efficiency values scale much larger than normalized revenue ratio originally used in the selfish mining literature as a form of efficiency. As an example, for $(\alpha_{final},\gamma)=(0.4,0.5)$, the normalized revenue ratio is $\frac{\rho_L}{\alpha_{final}}=1.41$ whereas we get $U_A=1.95$, which means that the adversary is $95\%$ more efficient than the honest miners under the attack. Our calculations are based on $L^{*}$-selfish mining and $\epsilon$ optimal strategy of \cite{optimal-selfish} can slightly increase the efficiencies and push the boundary between alternate mining and selfish mining in \figref{fig::efficiency_comparison}  to the left.

\section{Block Withholding Attacks} \label{sec::bwh}
Here, we first analyze the short-term revenue change associated with PAW-Type-B using the same assumptions as in Section~\ref{sec::short-term}. Then, we will use the newly introduced metric in Section~\ref{sec::long_term} to analyze the long-term efficiency of the attack.

\subsection{Short-Term Analysis}
We first note that, on average, the first epoch lasts
\begin{align}
    t_1=\delta_{p_1,p_2}\tau_0.
\end{align}
As a result, the revenue change of the adversary at the end of the first epoch is 
\begin{align}
    \Delta_A(t_1)&=\rho-\alpha\delta_{p_1,p_2}.\label{eq::bwh_init_loss_adv}
\end{align}
After the first epoch, DAA makes sure that subsequent epochs last $\tau_0$, hence, the revenue change after $t_1$ is
\begin{align}
    \Delta_A(t_1+x)&=\rho-\alpha\delta_{p_1,p_2}+(\rho-\alpha)\frac{x}{\tau_0}.
\end{align}

The revenue change analysis can be done for the pool miners as 
\begin{align}
    \Delta_P(t_1)&=\rho_{pool}-\beta\delta_{p_1,p_2},\\
    \Delta_P(t_1+x)&=\rho_{pool}-\beta\delta_{p_1,p_2}+(\rho_{pool}-\beta)\frac{x}{\tau_0},
\end{align}
and similarly, for the honest miners outside the pool as
\begin{align}
    \Delta_R(t_1)=&\rho_{rest}-(1-\alpha-\beta)\delta_{p_1,p_2},\\
    \Delta_R(t_1+x)=&\rho_{rest}-(1-\alpha-\beta)\delta_{p_1,p_2}\nonumber\\&+(\rho_{rest}+\alpha+\beta-1)\frac{x}{\tau_0}.
\end{align}

\subsubsection{PAW Attack is Profitable even without DAA}
Using martingale theory, \cite{grunspan2019-profitability-selfish-mining} shows that if the adversary follows selfish mining, it always makes losses at the end of the first epoch. The result is generalized for all adversarial strategies in \cite[Theorem~4.4]{grunspan2019-profitability-selfish-mining} and \cite[Corollary~3.4]{Grunspan_witholding_resilience} which is restated next.

\begin{claim}\label{claim::grunspan}
\textbf{(Grunspand \& Perez-Marco\cite{Grunspan_witholding_resilience,grunspan2019-profitability-selfish-mining})} Without difficulty adjustments, the optimal strategy is the honest one.
\end{claim}

However, in PAW attacks, unlike selfish mining, for some $\alpha,\beta$ and $\gamma$ values with certain choices of $p_1$ and $p_2$, the revenue change of the adversary can be always positive, i.e., \eqref{eq::bwh_init_loss_adv} can be greater than zero even in the first epoch. In Appendix~\ref{sec::app::daa}, we explain why this claim fails under PAW attacks. The next example shows such a case.

\begin{example}\label{ex::bwh-type_b}
    Let the total fraction of adversarial hashpower be $\alpha=0.2$ and the fraction of hashpower of the honest pool miners be $\beta=0.04$, where the adversary launches PAW-Type-B attack against the pool with $\gamma=0.75$, $p_1=0.05$ and $p_2=0.99$. Using \eqref{eq::adv_rho_bwh_type_b} and \eqref{eq::withhold_cycle_dur_type_b} we get $\rho=0.2040$ and $\delta_{p_1,p_2}=1.0124$, respectively. As a result, from \eqref{eq::bwh_init_loss_adv}, the adversarial revenue change at $t_1$ is $\Delta_A(t_1)=1.6\cdot 10^{-3}$. In other words, under PAW-Type-B attack, the revenue of the adversary can increase even in the first epoch before the difficulty is readjusted, which disproves Claim~\ref{claim::grunspan}.
\end{example}

\subsubsection{Objective Function Variations}
First, note that $\frac{\rho}{\alpha}$ ($\frac{\rho_{pool}}{\beta}$, $\frac{\rho_{rest}}{1-\alpha-\beta}$ resp.) minus $1$ is the slope of the adversarial revenue change $\Delta_A(t_1+x)$ ($\Delta_P(t_1+x)$, $\Delta_R(t_1+x)$ resp.) after the first epoch. Fractions such as $\frac{\rho}{\alpha}$, $\frac{\rho_{pool}}{\beta}$ and $\frac{\rho_{rest}}{1-\alpha-\beta}$ are called normalized revenue ratio since they display the revenue ratio per the computational power. Since $1\geq\gamma\geq 0$, we have $\mathbb{E}[B_R]\geq(1-\alpha-\beta)\mathbb{E}[B_C]$, which together with \eqref{eq::rho_rest} implies that
\begin{align}
\rho_{rest}\geq 1-\alpha-\beta,
\end{align} 
independent of $p_1$ and $p_2$.\footnote{The same argument also holds for the original PAW attack under the simplified modeling of \cite{power_adjusting}.} Hence, the revenue ratio of the honest miners outside the pool can only increase with this attack. Further, since $\rho_{rest}\geq 1-\alpha-\beta$, the slope of $\Delta_R$ is always positive after $t_1$, i.e., the honest miners outside the pool observe a revenue increase after the first epoch.

With $p_1$ and $p_2$ chosen optimally to maximize $\rho$, it is trivial to show that the adversary can pick $p_1=0$ and $p_2=0$, hence, 
\begin{align}
\rho^*=\max_{p_1,p_2}\rho_{p_1.p_2}\geq\alpha.
\end{align} 
It turns out, the values $p_1$ and $p_2$ maximizing $\rho$ may result in higher revenue ratio per computational power for the honest miners outside the pool compared to that of the adversary. In other words, honest miners outside the pool can benefit from the attack even more than the adversary itself. Thus, here, the adversary might be willing to optimize $\frac{\rho}{\rho_{rest}}$ instead, i.e.,
\begin{align}
    \frac{\rho}{\rho_{rest}}^{\dagger}=\max_{p_1,p_2}\frac{\rho}{\rho_{rest}}.
\end{align} 
This is due to the fact that the choices of $p_1$ and $p_2$ maximizing $\rho$ may result in 
\begin{align}
    \frac{\rho_{rest}}{1-\alpha-\beta}\geq \frac{\rho}{\alpha} \geq1\geq  \frac{\rho_{pool}}{\beta},
\end{align} 
whereas maximizing $\frac{\rho}{\rho_{rest}}$ always results in 
\begin{align}
    \frac{\rho}{\alpha} \geq \frac{\rho_{rest}}{1-\alpha-\beta}\geq1\geq \frac{\rho_{pool}}{\beta}.
\end{align}

It is also worth noting that the adversary can try to maximize $\rho$ under the constraint $\frac{\rho}{\alpha} \geq \frac{\rho_{rest}}{1-\alpha-\beta}$. This is not necessarily the same as maximization of $\rho$. In fact, some cases where the objective function is quasiconcave in $p_1$ and $p_2$ implies that maximizing $\rho$ under the constraint $\frac{\rho}{\alpha} \geq \frac{\rho_{rest}}{1-\alpha-\beta}$ results in $\frac{\rho}{\alpha}= \frac{\rho_{rest}}{1-\alpha-\beta}\geq 1$. Here, we only focus on maximum revenue ratio $\rho$ and maximum relative revenue ratio $\frac{\rho}{\rho_{rest}}$ and leave the study of constrained optimization to the interested reader.

As we discussed earlier, under PAW-Type-B, for some parameters, the adversary can make profit even in the first epoch. Hence, for those $\alpha,\beta$ and $\gamma$ parameters, the adversary may prefer maximizing the revenue change in the first epoch, $\Delta_A(t_1)$, with respect to $p_1,p_2$, i.e.,
\begin{align}
\Delta_A(t_1)^{\star}=\max_{p_1,p_2}\Delta_A(t_1)=\max_{p_1,p_2}(\rho_{p_1,p_2}-\alpha\delta_{p_1,p_2}).    \label{eq::bwh_init_loss_adv_max}
\end{align}

Consider the average adversarial fraction of hashpower wasted at the pool obtained in Appendix~\ref{sec::app::bwh} as 
\begin{align}
    p_w=\frac{\delta_{p_1,p_2}-1}{\alpha\delta_{p_1,p_2}}.
\end{align}
Let $p_w^*$, $p_w^{\dagger}$, $p_w^{\star}$ be the average hashpower wasted when the adversary maximizes $\rho$, $\frac{\rho}{\rho_{rest}}$, $\Delta_A(t_1)$, respectively. Our numerical optimization analysis shows that $p_w^{\star}\leq p_w^{\dagger}\leq p_w^{*}$. In other words, consider the hashpower wasted at the pool when the adversary is tying to maximize its revenue ratio $\rho$. By wasting less hashpower at the pool, the adversary achieves its maximal relative revenue ratio $\frac{\rho}{\rho_{rest}}^{\dagger}$, whereas by wasting even less hashpower at the pool, the adversary achieves maximum revenue change at the end of the first epoch, i.e., $\Delta_A(t_1)^{\star}$.

\begin{figure}[t!]
\captionsetup[subfigure]{aboveskip=0pt,belowskip=9pt}
     \centering
\begin{subfigure}[t]{0.46\columnwidth}
\centering
\includegraphics[width=\textwidth]{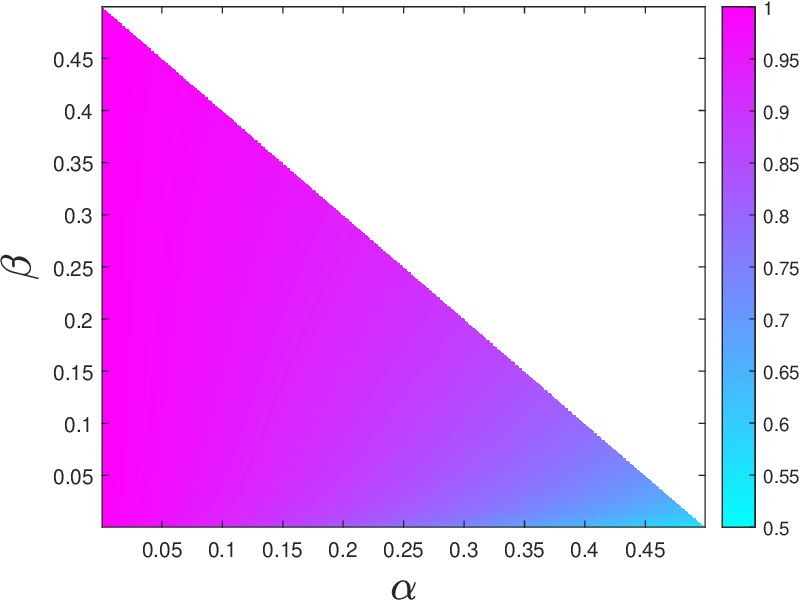}
    \caption{Pool miners, $\gamma=0$}
    \label{fig::rev_ratio_bwh_gamma_0_pool}
\end{subfigure}
~
\begin{subfigure}[t]{0.46\columnwidth}
\centering
\includegraphics[width=\textwidth]{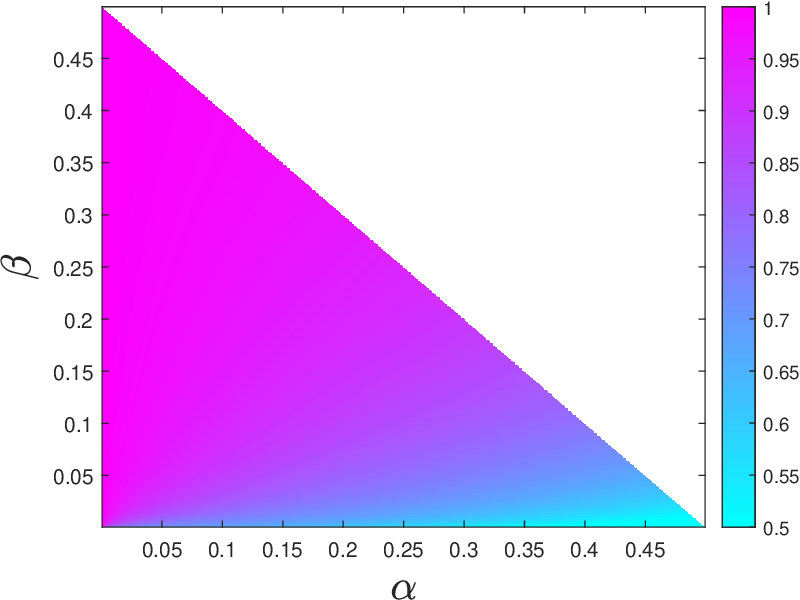}
    \caption{Pool miners, $\gamma=0.5$}
    \label{fig::rev_ratio_bwh_gamma_05_pool}
\end{subfigure}
~
\begin{subfigure}[t]{0.46\columnwidth}
\centering
\includegraphics[width=\textwidth]{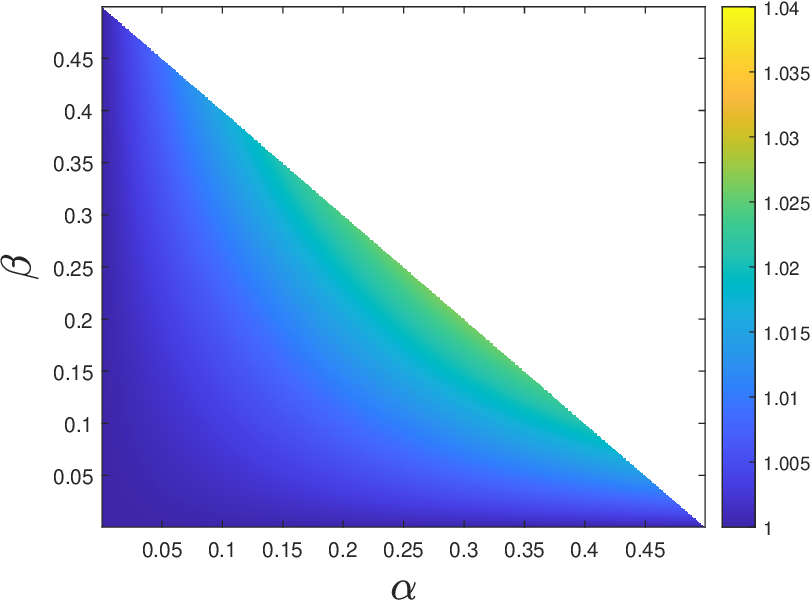}
    \caption{The adversary, $\gamma=0$}
    \label{fig::rev_ratio_bwh_gamma_0_adv}
\end{subfigure}
~
\begin{subfigure}[t]{0.46\columnwidth}
\centering
\includegraphics[width=\textwidth]{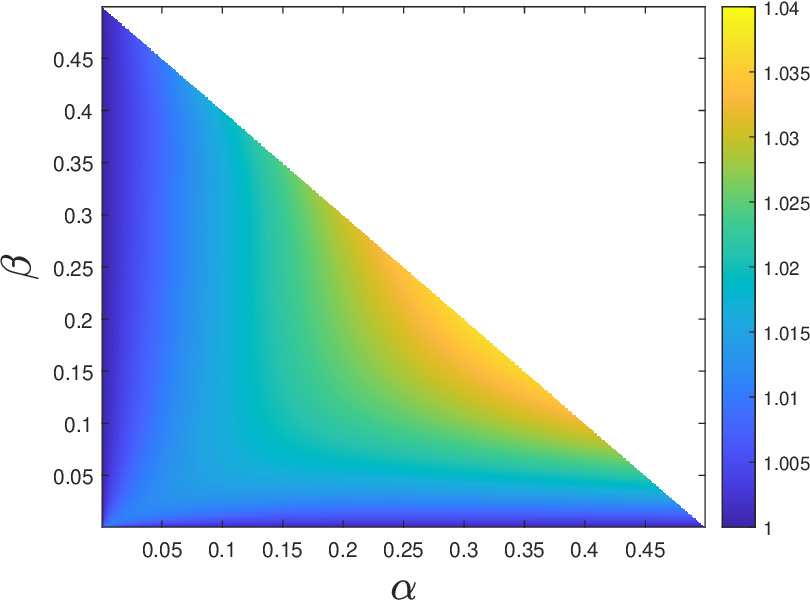}
    \caption{The adversary, $\gamma=0.5$}
    \label{fig::rev_ratio_bwh_gamma_05_adv}
\end{subfigure}
~
\begin{subfigure}[t]{0.46\columnwidth}
\centering
\includegraphics[width=\textwidth]{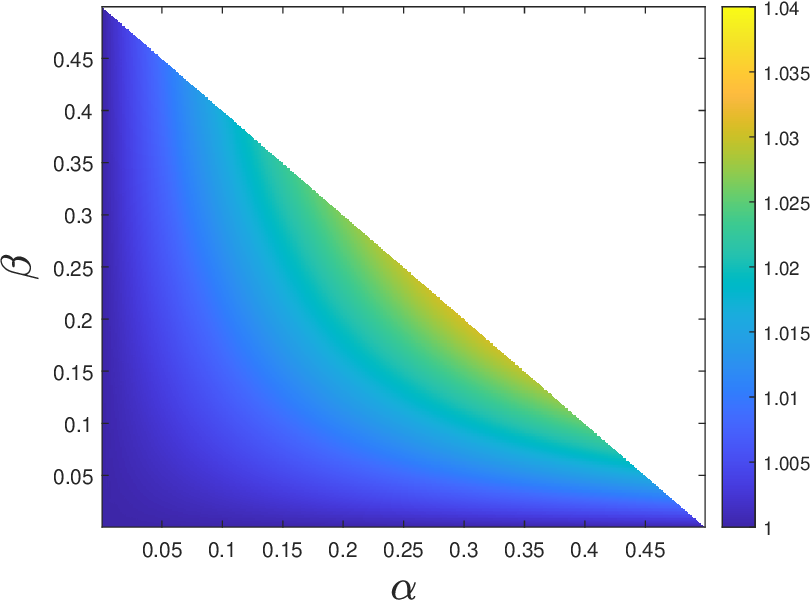}
    \caption{Other miners, $\gamma=0$}
    \label{fig::rev_ratio_bwh_gamma_0_rest}
\end{subfigure}
~
\begin{subfigure}[t]{0.46\columnwidth}
\centering
\includegraphics[width=\textwidth]{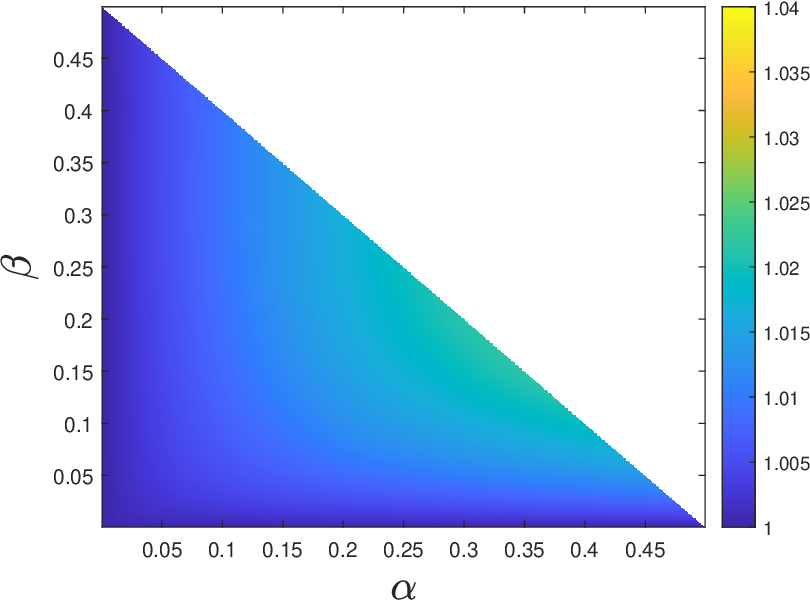}
    \caption{Other miners, $\gamma=0.5$}
    \label{fig::rev_ratio_bwh_gamma_05_rest}
\end{subfigure}

    \caption{Normalized revenue ratios in PAW-Type-B.}
	\label{fig::withhold_rev_ratios}
\end{figure}

\begin{figure}[t!]
     \centering
    \includegraphics[width=0.8\columnwidth]{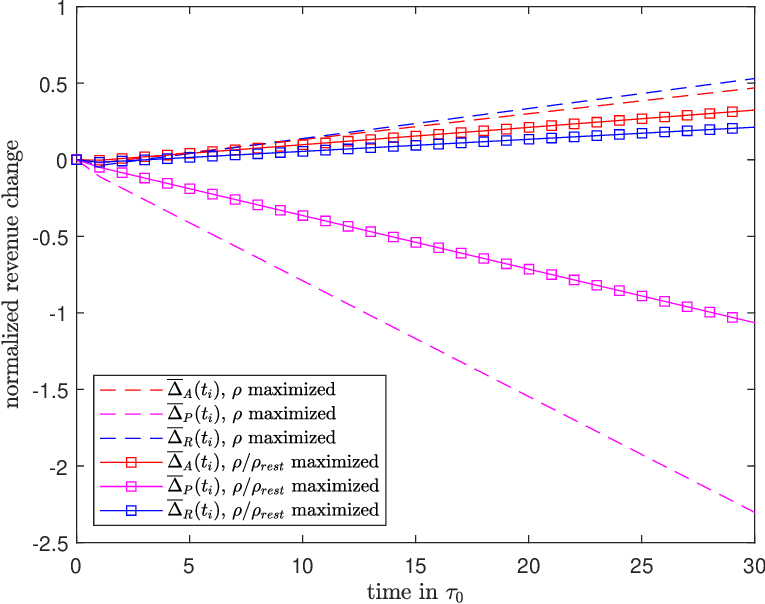}
    \caption{Normalized revenue change, $\alpha=0.2$, $\beta=0.2$, $\gamma=0.1$}
    \label{fig::block_prof_diff}
\end{figure}

\begin{figure}[t]
\captionsetup[subfigure]{aboveskip=0pt,belowskip=9pt}
    \centering
\begin{subfigure}[b]{0.46\columnwidth}
\centering
\includegraphics[width=\textwidth]{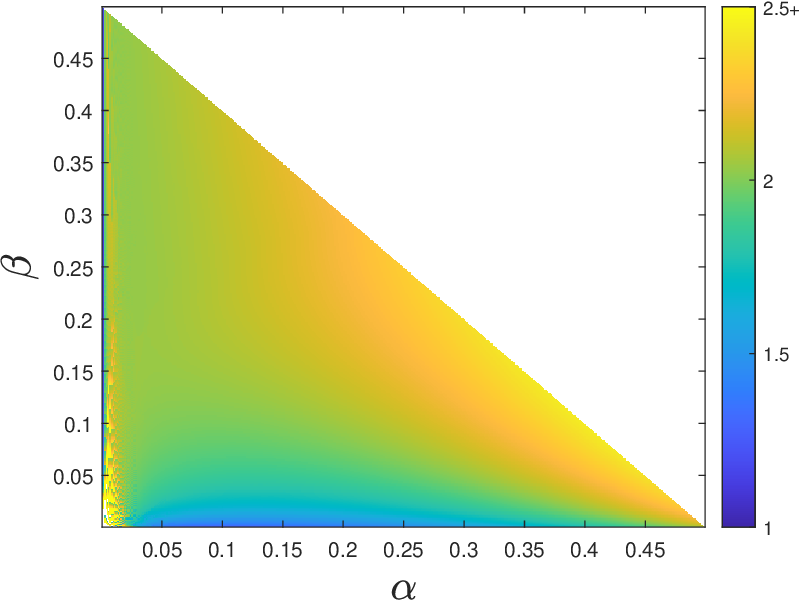}
    \caption{The adversary, $\gamma=0$}
    \label{fig::weeks_withhold_bwh_gamma_0_adv}
\end{subfigure}
~
\begin{subfigure}[b]{0.46\columnwidth}
    \centering{
    \includegraphics[width=\textwidth]{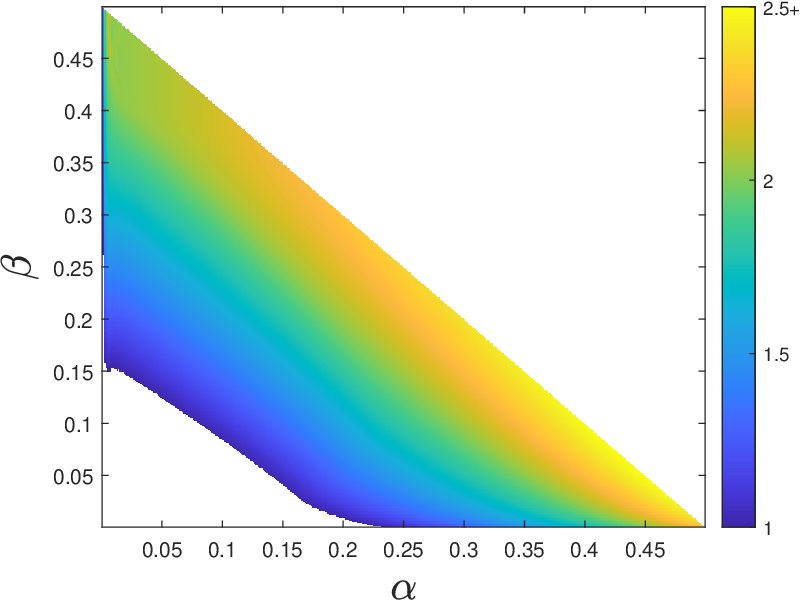}}
    \caption{The adversary, $\gamma=0.5$}
    \label{fig::weeks_withhold_bwh_gamma_05_adv}
\end{subfigure}
~
\begin{subfigure}[b]{0.46\columnwidth}
    \centering{
    \includegraphics[width=\textwidth]{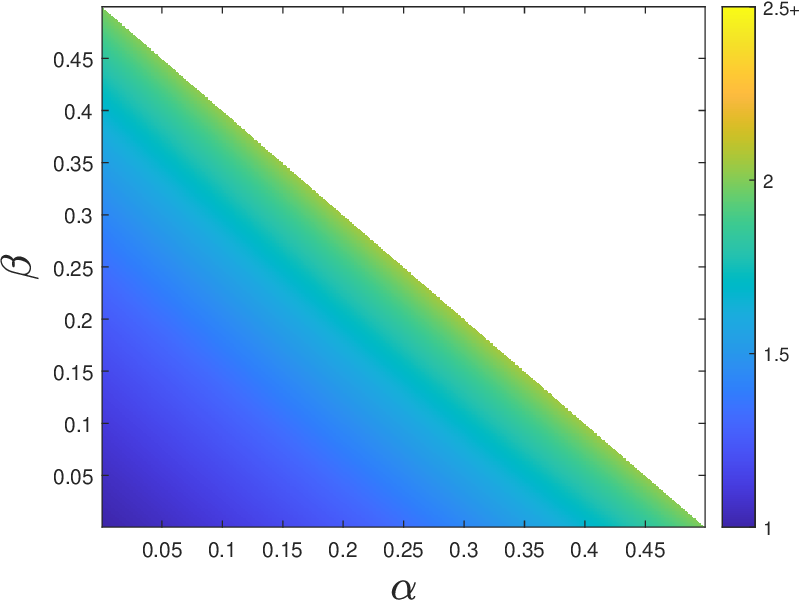}}
    \caption{Other miners, $\gamma=0$}
    \label{fig::weeks_withhold_bwh_gamma_0_rest}
\end{subfigure}
~
\begin{subfigure}[b]{0.46\columnwidth}
    \centering{
    \includegraphics[width=\textwidth]{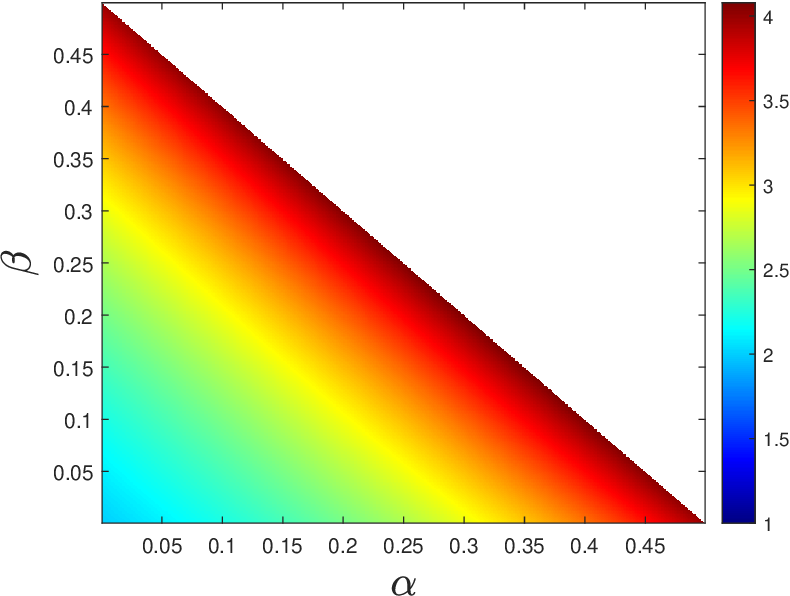}}
    \caption{Other miners, $\gamma=0.5$}
    \label{fig::weeks_withhold_bwh_gamma_05_rest}
\end{subfigure}
    \caption{Profit lags in terms of $\tau_0$ in PAW-Type-B.}
	\label{fig::withhold_short}
\end{figure}

\subsubsection{Numerical Results} 
In \figref{fig::withhold_rev_ratios}, we present the normalized revenue ratio of the adversary ($\frac{\rho}{\alpha}$), the honest pool miners ($\frac{\rho_{pool}}{\beta}$) and the remaining honest miners ($\frac{\rho_{rest}}{1-\alpha-\beta}$) under PAW-Type-B attack for all $\alpha,\beta$ with $\gamma=\{0,0.5\}$ parameters where the adversary picks $p_1$ and $p_2$ to maximize $\rho$, i.e., 
\begin{align}
\rho^*=\max_{p_1,p_2}\rho_{p_1,p_2}.
\end{align} 
As we assume $\alpha+\beta<0.5$, we only plot the regions satisfying the condition. It is clear that an increasing $\alpha$ and $\gamma$ also increases the normalized revenue ratio for the adversary whereas the relationship of the normalized adversarial revenue ratio with respect to the pool size $\beta$ is not that straightforward when $\gamma=0.5$ and $\alpha$ small. As we proved earlier, the revenue ratio of the honest miners outside the pool as well as that of the adversary increases under this attack, hence their normalized revenue is always above $1$. Thus, the honest pool miners burden all the losses in terms of the revenue ratio.

In \figref{fig::block_prof_diff}, we pick $\alpha=0.2$, $\beta=0.2$ and $\gamma=0.1$ and present the normalized revenue changes of each miner type under PAW-Type-B attack with dashed lines when the adversary picks $p_1$ and $p_2$ values that maximize its revenue ratio $\rho$. It is clear that the honest pool miners make losses for all epochs whereas the adversary and the honest miners outside the pool make losses only in the first epoch. Notice that the normalized revenue change of the honest miners outside the pool has a higher slope than the normalized revenue change of the adversary, i.e., their normalized revenue ratio is the highest. In other words, after the first epoch, the honest miners outside the pool make more profit per computational power than the adversary. Hence, eventually, the revenue per computational power of the honest miners outside the pool surpasses that of the adversary. Note that, instead of picking $p_1$ and $p_2$ maximizing $\rho$, if the adversary maximizes $\frac{\rho}{\rho_{rest}}$, this makes sure that, it has more profit per computational power than any other miner but less than what it gets if it were to maximize $\rho$. We display the normalized revenue changes with square markers when the adversary picks $p_1$ and $p_2$ values that maximize $\rho/\rho_{rest}$ and makes sure that the adversary has the highest profit per computational power among all miners.

\begin{figure}[t!]
\captionsetup[subfigure]{aboveskip=0pt,belowskip=9pt}
     \centering
\begin{subfigure}[t]{0.46\columnwidth}
    \centering
    \includegraphics[width=\textwidth]{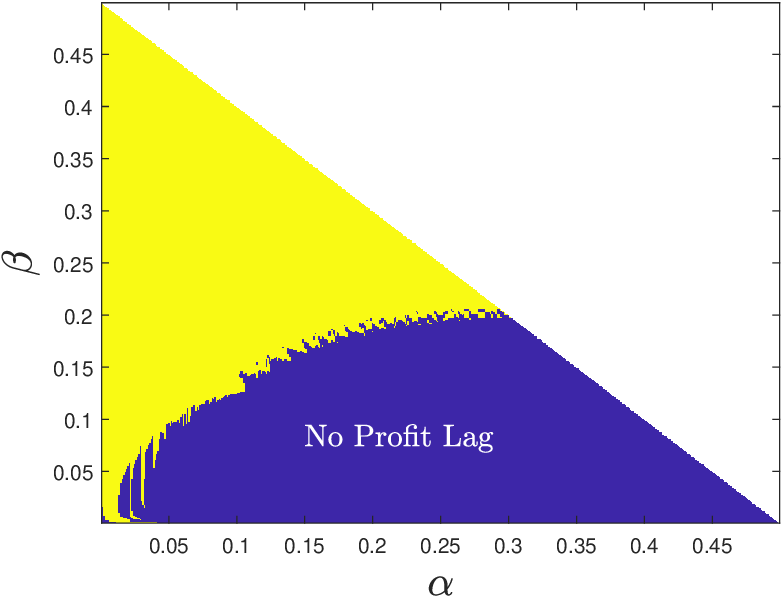}
    \caption{$\gamma=0$}
    \label{fig::no_prof_lag_gamma_0}
\end{subfigure}
~
\begin{subfigure}[t]{0.46\columnwidth}
    \centering
    \includegraphics[width=\textwidth]{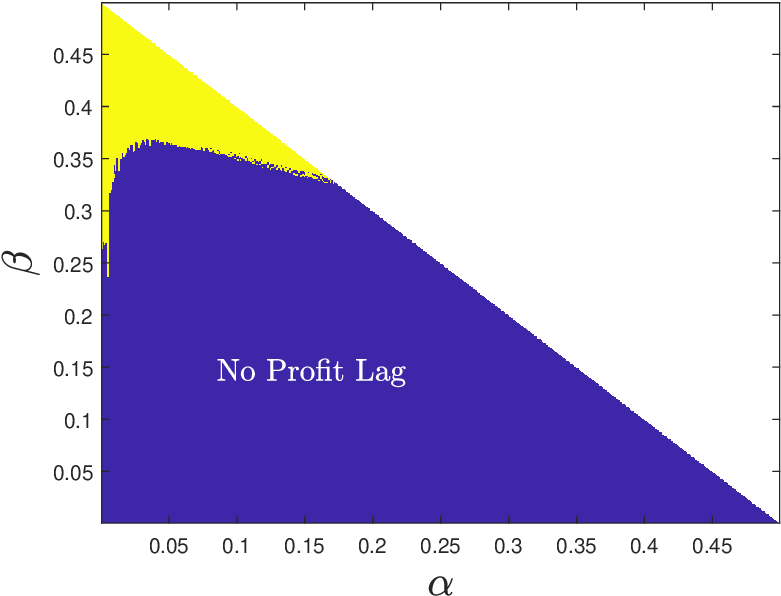}
    \caption{$\gamma=0.5$}
    \label{fig::no_prof_lag_gamma_05}
\end{subfigure}
~
\begin{subfigure}[t]{0.46\columnwidth}
\centering
\includegraphics[width=\textwidth]{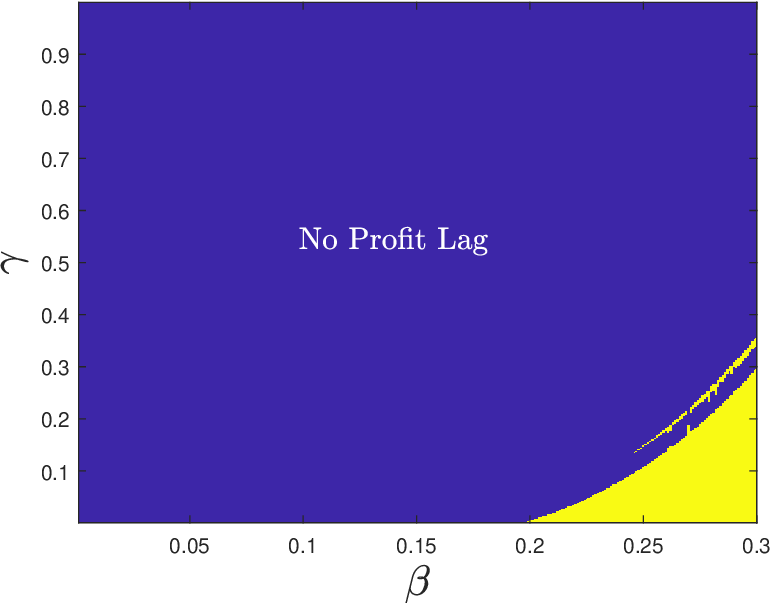}
    \caption{$\alpha=0.2$}
    \label{fig::no_prof_lag_alpha}
\end{subfigure}
~
\begin{subfigure}[t]{0.46\columnwidth}
    \centering
    \includegraphics[width=\textwidth]{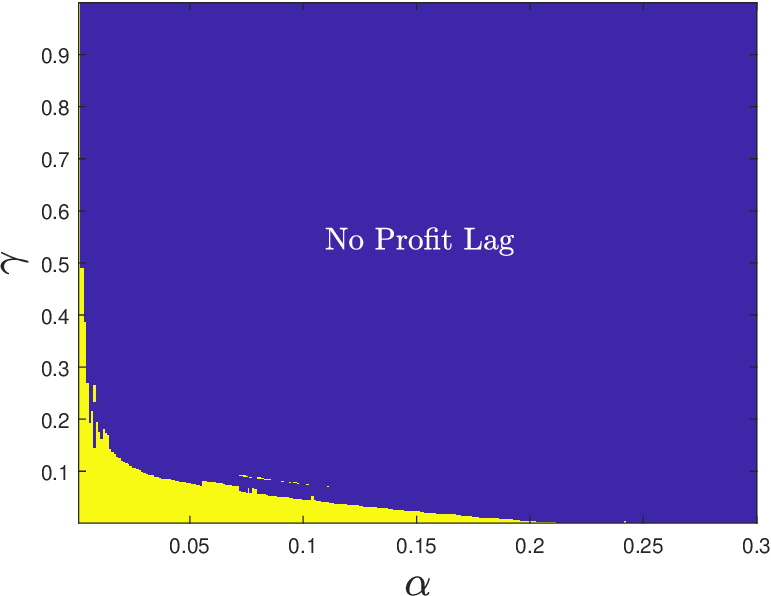}
    \caption{$\beta=0.2$}
    \label{fig::no_prof_lag_beta2}
\end{subfigure}
    \caption{Regions where the revenue of PAW-Type-B attacker increases in the first epoch under maximizing \eqref{eq::bwh_init_loss_adv_max}.}
	\label{fig::no_prof_lag_regions}
\end{figure}

\subsubsection{Profit Lag}
In \figref{fig::weeks_withhold_bwh_gamma_0_adv} (\figref{fig::weeks_withhold_bwh_gamma_05_adv} resp.), we display the profit lag of the adversary in terms of $\tau_0$ under PAW-Type-B where we fix $\gamma=0$ ($\gamma=0.5$ resp.) for all $\alpha$, $\beta$ values and with $p_1$ and $p_2$ that maximize $\rho$. It is clear from \figref{fig::withhold_short} that the adversarial profit lag is always below $2.5\tau_0$ (approximately $5$ weeks in BTC). Further, decreasing $\gamma$ increases the profit lag of the adversary as expected. On the other hand, increasing pool size increases the profit lag of the adversary. An interesting observation from \figref{fig::withhold_short} is that, as the adversarial fraction of hashpower $\alpha$ increases, the profit lag increases as well. For the sake of completeness, \figref{fig::withhold_short} also shows the profit lag of the honest miners outside the pool as they profit from the attack just like the adversary and their profit lag seems to depend directly on $\gamma$ and $\alpha+\beta$.

The white regions in lower left corner of \figref{fig::weeks_withhold_bwh_gamma_05_adv} correspond to the parameters where maximizing $\rho$ results in always positive revenue change, i.e., no adversarial revenue loss in \eqref{eq::bwh_init_loss_adv}. In other words, in these regions, the honest strategy is not optimal even without a DAA. Note that, for other non-white regions, there may be choices of $p_1$ and $p_2$ values that result in $\rho>\alpha$ without any revenue loss in \eqref{eq::bwh_init_loss_adv}. In other words, there may be other regions where the honest strategy is not optimal even without a DAA. To identify all such regions, we simply maximize \eqref{eq::bwh_init_loss_adv_max} with respect to $p_1,p_2$ using numerical tools. The identified regions are colored blue in \figref{fig::no_prof_lag_regions} for various set of $\alpha$, $\beta$ and $\gamma$ parameters. A common trait in \figref{fig::no_prof_lag_regions} is, i.e., when the goal of the adversary is to maximize \eqref{eq::bwh_init_loss_adv_max}, as the pool size decreases, the PAW attacker does not observe a profit loss in the first epoch. Here, even when $\gamma=0$, attacking a small pool can give an adversary immediate gains in the short-term without any lag.

In any case, the short-term analysis suggests that honest pool miners are at a high disadvantage and make major losses. Hence, we turn our attention to long-term analysis as we did for the selfish mining strategies.

\subsection{Long-Term Analysis}
Notice that, in this attack, there are two groups of honest miners. One is the honest miners in the pool and the other is the honest miners outside the pool. To properly use the efficiency metric we introduced previously, we need to pick the correct set of miners whose efficiency is stabilized to $1$. As we assume a long-term stable system, the honest miners whose efficiency is the lowest has to be set to $1$ since they do not leave the system. It can be trivially shown that the efficiency of the honest pool members is the lowest when the adversary picks $p_1$ and $p_2$ values in its own favor. Hence, we assume $U_P=1$ throughout this section, where
\begin{align}
    U_P=\frac{\rho_{pool}}{\beta}\frac{X}{\tau_0E}.
\end{align}
As a result, for the adversarial miner, we have
\begin{align}
    U_A=\frac{\rho}{\alpha}\frac{X}{\tau_0E}=\frac{\rho}{\alpha}\frac{\beta}{\rho_{pool}},\label{eq::block_withhold_adv_utility}
\end{align}
and for the honest miners outside the pool, we have
\begin{align}
    U_R=\frac{1-\rho-\rho_{pool}}{1-\alpha-\beta}\frac{X}{\tau_0E}=\frac{1-\rho-\rho_{pool}}{1-\alpha-\beta}\frac{\beta}{\rho_{pool}}.\label{eq::block_withhold_oth_utility}
\end{align}

\subsubsection{Objective Function Variations}
With $p_1$ and $p_2$ chosen optimally to maximize $\rho$, it is trivial to show that $U_A\geq1$ and $U_R\geq 1$. $U_R\geq 1$ is due to the fact that $\rho_{rest}\geq 1-\alpha-\beta$ and $U_A\geq1$ is due to the fact that the adversary can pick $p_1=0$ and $p_2=0$, hence $\rho\geq\alpha$. However, it turns out, the values $p_1$ and $p_2$ maximizing $\rho$ are not necessarily the same as the values $p_1$ and $p_2$ maximizing $U_A$. This is due to the fact that in \eqref{eq::block_withhold_adv_utility} both $\rho$ and $\rho_{pool}$ depend on $p_1$ and $p_2$ but the values maximizing the adversarial revenue ratio $\rho$ appearing in the numerator do not necessarily minimize $\rho_{pool}$ in the denominator. In comparison, in constant effort selfish mining strategies such as $L$-selfish mining considered in Section~\ref{sec::l-selfish}, maximizing the adversarial revenue ratio also maximizes the adversarial efficiency, e.g., in \eqref{eq::L_stub_efficiency_func}, a strategy maximizing the numerator $\rho_L$ also minimizes the denominator ($1-\rho_L$). 

Thus, here, in the long-run, the adversarial strategy that maximizes the revenue per time per computational power normalized by that of the pool miners, is not the same as the adversarial strategy that maximizes the adversarial revenue ratio. Further, since we assume $U_P=1$, optimizing $U_A$ is the same thing as optimizing $U_A/U_P$. On the other hand, the adversary might be willing to optimize $U_A/U_R$ instead. This is due to the fact that the choices of $p_1$ and $p_2$ maximizing $U_A/U_P$ may result in $U_R\geq U_A \geq U_P$ whereas maximizing $U_A/U_R$ always results in $U_A \geq U_R \geq U_P$. Note that, \eqref{eq::block_withhold_adv_utility} and \eqref{eq::block_withhold_oth_utility} imply that optimizing $U_A/U_R$ is the same as optimizing $\frac{\rho}{\rho_{rest}}$. In other words, this is analogous to what we discussed in short-term, e.g., in \figref{fig::block_prof_diff} the adversary might prefer the lines with square markers instead of the dashed lines as it gets the most relative revenue per computational power among all miners.

Let $p_w^{\ddagger}$, $p_w^{\dagger}$, $p_w^{*}$ be the average hashpower wasted when the adversary maximizes $U_A/U_P$, $U_A/U_R$, $\rho$, respectively. Our numerical analysis shows that $p_w^{\dagger}\leq p_w^{*}\leq p_w^{\ddagger}$. In other words, consider the hashpower wasted at the pool when the adversary is tying to maximize its revenue ratio $\rho$. By wasting even more hashpower at the pool, the adversary achieves its maximal efficiency $U_A$, whereas by wasting less hashpower at the pool, the adversary achieves maximum relative efficiency $U_A/U_R$ with $U_A\geq U_R\geq U_P$. Note that, the adversary can try to maximize $U_A$ under the constraint $U_A\geq U_R$ which is not necessarily the same as maximization of $U_A/U_R$. Some cases where the objective function is quasiconcave in $p_1$ and $p_2$ implies that maximizing $U_A$ under the constraint $U_A\geq U_R$ results in $U_A/U_R=1$. Again, we only focus on maximum efficiency $U_A$ and maximum relative efficiency $U_A/U_R$ and leave the study of constrained optimization to the interested reader.

\begin{figure}[t!]
\captionsetup[subfigure]{aboveskip=0pt,belowskip=9pt}
     \centering
\begin{subfigure}[t]{0.46\columnwidth}
\centering
\includegraphics[width=\textwidth]{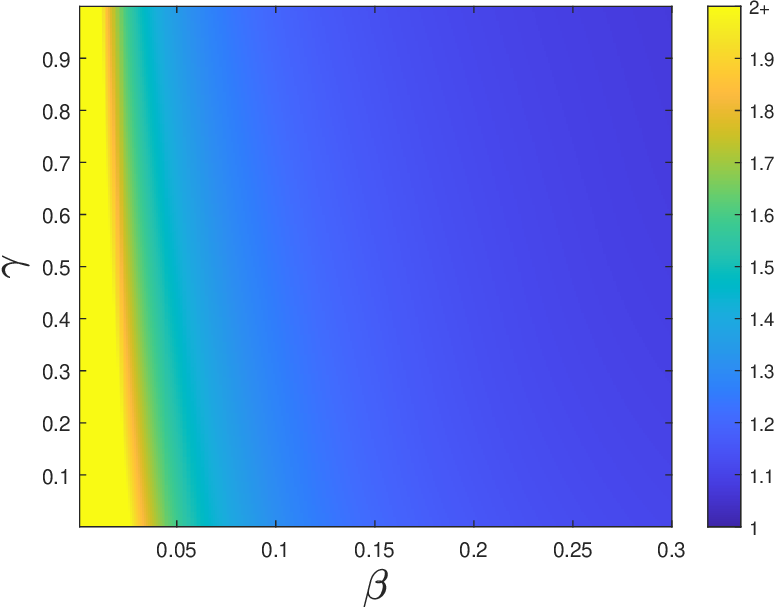}
    \caption{$U_A^{\ddagger}$, adversarial efficiency}
    \label{fig::util_adv_vs_pool2}
\end{subfigure}
~
\begin{subfigure}[t]{0.46\columnwidth}
\centering
\includegraphics[width=\textwidth]{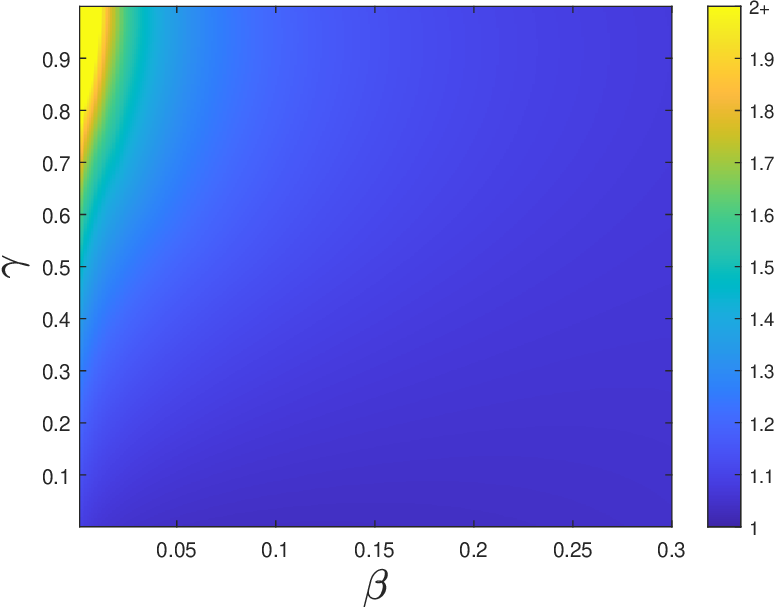}
    \caption{$U_A^{\dagger}$, adversarial efficiency}
    \label{fig::util_adv_vs_pool}
\end{subfigure}
\hfill
\begin{subfigure}[t]{0.46\columnwidth}
\centering
\includegraphics[width=\textwidth]{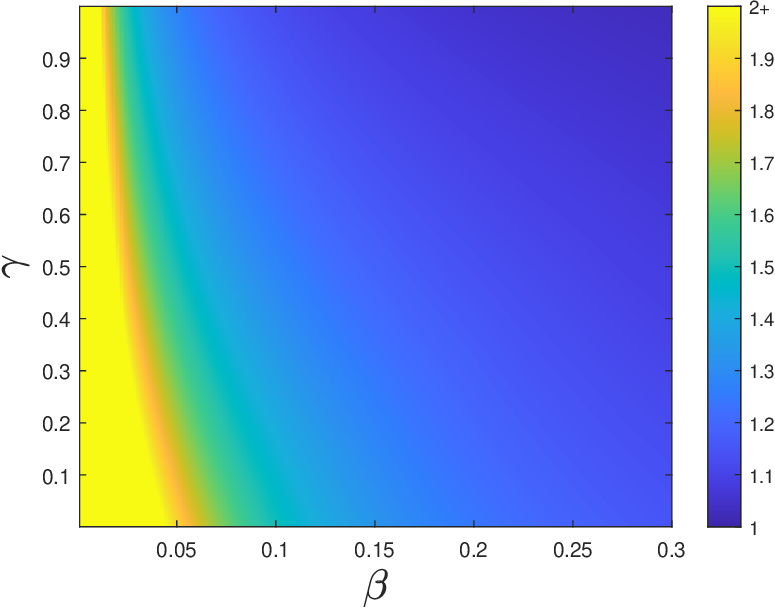}
    \caption{$U_R^\ddagger$, efficiency of miners outside the pool}
    \label{fig::util_hon_vs_pool2}
\end{subfigure}
~
\begin{subfigure}[t]{0.46\columnwidth}
\centering
\includegraphics[width=\textwidth]{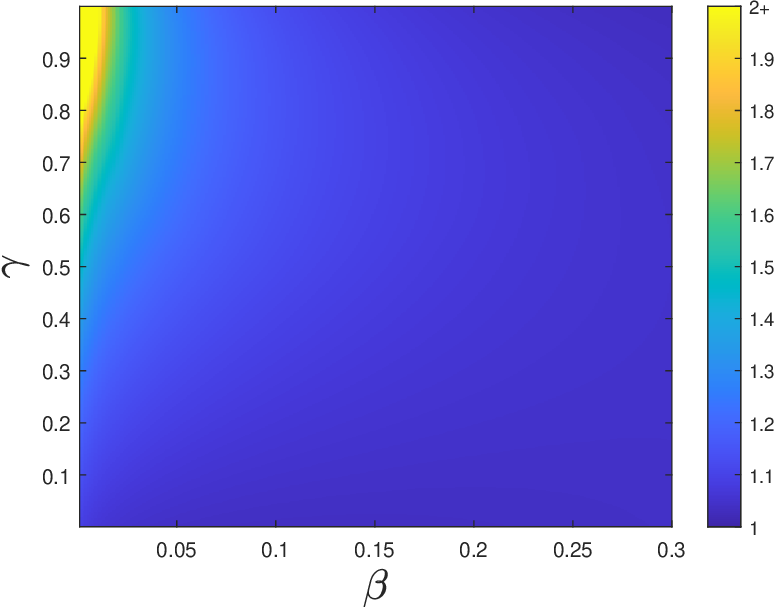}
    \caption{$U_R^{\dagger}$, efficiency of miners outside the pool}
    \label{fig::util_hon_vs_pool}
\end{subfigure}
\hfill
\begin{subfigure}[t]{0.46\columnwidth}
\centering
\includegraphics[width=\textwidth]{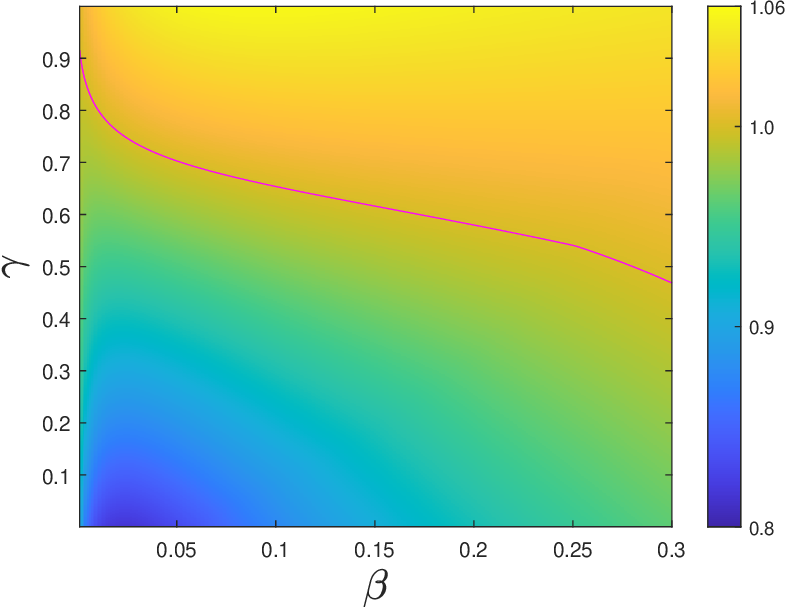}
    \caption{$U_A^{\ddagger}/U_R^{\ddagger}$, adversarial efficiency scaled by the efficiency of miners outside the pool}
    \label{fig::util_hon_vs_adv2}
\end{subfigure}
~
\begin{subfigure}[t]{0.46\columnwidth}
\centering
\includegraphics[width=\textwidth]{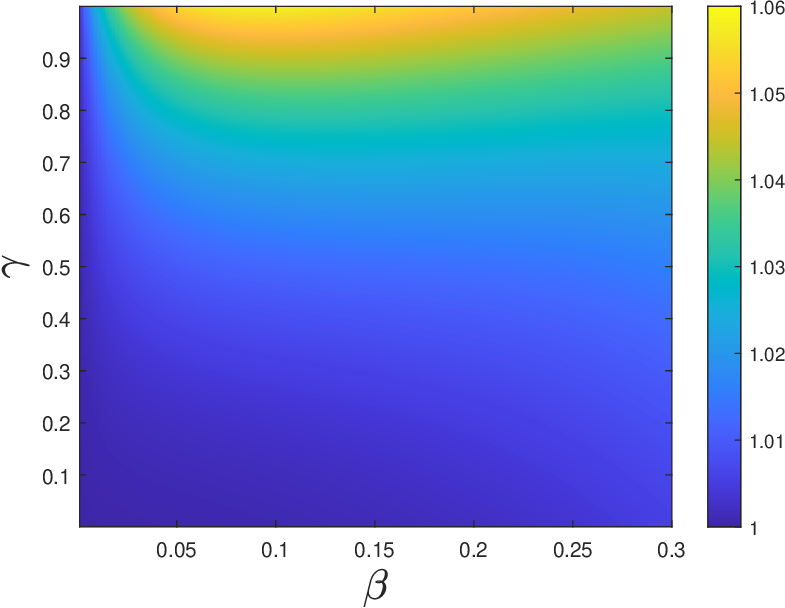}
    \caption{$U_A^{\dagger}/U_R^{\dagger}$, adversarial efficiency scaled by the efficiency of miners outside the pool}
    \label{fig::util_hon_vs_adv}
\end{subfigure}

    \caption{Efficiencies for $\alpha=0.2$, varying $\beta$ and $\gamma$.}
\label{fig::efficiency_block_withhold}
\end{figure}

\subsubsection{Numerical Results}
We denote the adversarial efficiency obtained by maximizing $U_A$ as $U_A^{\ddagger}$ and the corresponding efficiency of honest miners outside the pool as $U_R^{\ddagger}$. Similarly, $U_A^{\dagger}$ denotes the adversarial efficiency obtained by maximizing $U_A/U_R$, and $U_R^{\dagger}$ denotes the corresponding efficiency of honest miners outside the pool. Note that, in both cases, we assume that the pool miners' efficiency is $U_P=1$. In \figref{fig::efficiency_block_withhold}, we present the efficiency of adversarial miner, $U_A^{\ddagger}$ and $U_A^{\dagger}$, as well as the honest miners outside the pool, $U_R^{\ddagger}$ and $U_R^{\dagger}$. We assume adversarial fraction of hashpower to be $\alpha=0.2$ and varying pool hashpower as well as varying $\gamma$ values. 

When the adversary aims to maximize $U_A$, we see that the adversarial efficiency $U_A^{\ddagger}$ drops with both increasing hashpower of the pool $\beta$, as well as increasing network power $\gamma$, which is counterintuitive. On the other hand, the efficiency of miners outside the pool $U_R^{\ddagger}$ drops with increasing $\gamma$ and $\beta$ as expected. These two results imply that increasing $\gamma$ and $\beta$ works in the favor of the honest pool miners when $U_A$ is maximized. Further, the honest miners outside the pool gain the most from the attack even more than the adversary as long as $\gamma$ is not too high, as it can be observed in \figref{fig::util_hon_vs_adv2}, where the magenta line is the border between values below and above $\frac{U_A^{\ddagger}}{U_R^{\ddagger}}=1$.

When the adversary aims to maximize $\frac{U_A}{U_R}$, both the adversarial efficiency $U_A^{\dagger}$ and the efficiency of miners outside the pool, $U_R^{\dagger}$ drops with increasing $\beta$ and decreasing $\gamma$. \figref{fig::util_hon_vs_adv} shows that the adversary has the best efficiency under such a maximization. Here, by comparing \figref{fig::util_adv_vs_pool} with \figref{fig::util_adv_vs_pool2} for lower $\gamma$ values, we see that the adversary gives up potential efficiency gains by wasting less hashpower in order to make sure $U_A\geq U_R$.

\section{Conclusion and Discussion}\label{sec::conc_future}
Next, we summarize the short-term and long-term analysis and observations we made throughout the paper (also see Table~\ref{tab::summaries}). Then, we discuss possible future directions.

\begin{table*}[t]
\centering
\begin{tabular}{|
  >{\centering\arraybackslash}m{2cm}|
  >{\centering\arraybackslash}m{2.5cm}|
  m{11cm}|
}
\hline
\multirow{5}{=}{\rule{0pt}{18ex}Short-Term}
 & Selfish & Switching from selfish mining to optimal selfish mining slightly increases profit lag but compensated by increased revenue gain slope later.
 In either case, the profit lag is less than $8$ weeks for most $(\alpha,\gamma)$ in BTC. \\
\cline{2-3}
 & Intermittent & For most $(\alpha,\gamma)$, selfish mining becomes more profitable than the intermittent variations within less than $5$ week time in BTC. For remaining parameters, profit lag of the intermittent mining is too high anyway. \\ 
\cline{2-3}
 & Hopping & The revenue increase per hashpower of the honest miners loyal to BTC is exactly the same as the hopper (between the BTC and BCH), the increase is directly proportional to the adversarial fraction of hashpower $\alpha$. \\
\cline{2-3}
 & PAW ($\rho$ optimized) & Profit lag is less than $5$ weeks in BTC (for small $\beta$, no profit lag at all). Honest miners outside the pool gain more revenue change than the adversary per hashpower for most parameters. \\
 \cline{2-3}
 & PAW (alternative optimizations) & If $\frac{\rho}{\rho_{rest}}$ maximized, adversary has more profit per hashpower than the rest but less than what it gets if it were to maximize $\rho$. If immediate revenue change is optimized, no profit lag for adversary for most parameters. \\
\hline
\multirow{3}{=}{\rule{0pt}{12ex}Long-Term}
 & Selfish & Optimizing the efficiency in selfish mining is the same as optimizing the revenue ratio. The efficiency of selfish mining also outperforms intermittent variations if adversary ends up with the same hash ratio in long-run. \\
\cline{2-3}
 & Hopping & Coin hopper (between the BTC and BCH) always has a better efficiency than the loyal honest miners of BTC. New boundaries which indicate where selfish mining/alternate mining is more efficient provided in \figref{fig::efficiency_comparison}.\\
\cline{2-3}
 & PAW Attack & If the adversary maximizes its efficiency, the honest miners outside the pool have better efficiency unless $\gamma$ is high enough. Instead, maximization of the relative efficiency of the adversary makes sure the adversary has the best efficiency among all miners. Not surprisingly, the resulting efficiency is lower than if the adversary purely focused on its own efficiency.  \\
\hline
\end{tabular}
\caption{Summary of the analysis and observations.}
\label{tab::summaries}
\end{table*}

\subsection{Short-Term Results}
\subsubsection{Selfish Mining Variations}
Our results indicate that switching from the selfish mining strategy of Eyal and Sirer to the optimal selfish mining strategies \cite{optimal-selfish} increases the initial revenue loss of the adversary and honest miners but also increases the subsequent adversarial revenue gain. As a result, profit lag, defined as the time until the adversarial revenue change becomes strictly positive, is slightly increased. In either case, we show that under the Bitcoin (BTC) difficulty adjustment algorithm (DAA), the profit lag is less than $8$ weeks for a wide range of $(\alpha,\gamma)$ parameters. 

We further extend the observations made in \cite{profit_lag} regarding the profit lag in intermittent selfish mining strategies. We show that, for the majority of $(\alpha,\gamma)$ parameters, the selfish mining of Eyal and Sirer becomes absolutely more profitable than the intermittent selfish mining within less than $5$ week time in BTC. For the remaining parameters, the profit lag of the intermittent selfish mining is too high anyway. Although intermittent smart selfish mining performs slightly better than the intermittent selfish mining, similar observations hold, i.e., selfish mining becomes more profitable than smart intermittent mining before smart intermittent mining creates a positive revenue.

\subsubsection{Coin Hopping}
Our extended observations of the alternate mining strategy of \cite{profit_lag}, which prescribes the adversary to switch between Bitcoin (BTC) and Bitcoin Cash (BCH), indicate that the honest miners also profit from the attack in the short-term. Further, the revenue increase per hashpower of the honest miners who stay loyal to the coin is exactly the same as the adversarial miner hopping between the coins and the increase is directly proportional to the adversarial fraction of hashpower $\alpha$. Similar observations were made regarding the miners turning their rigs on and off in  \cite{mind_the_mining}, where even the miners who do not switch their rigs have a short-term revenue increase. Moreover, this further implies a downfall of the weaker coin (BCH) since the honest miners of BCH would migrate to the BTC as they make more revenue. This observation also coincides with the game-theoretic analysis of \cite{btc-bch-game-coexist}, where authors study game theoretic scenario with some miners staying loyal to their coin whereas others are hopping. 

\subsubsection{PAW Attack}
In PAW, the short-term analysis shows that when the adversary dynamically adjusts its power to maximize its revenue ratio, the profit lag is quite small (less than $5$ weeks in BTC). More interestingly, unlike selfish mining, for small $\beta$, there are cases where there is no profit lag at all for the adversary.  This result disproves the claim made in \cite{Grunspan_witholding_resilience}, stating that, without difficulty adjustment, the optimal strategy is the honest one. Another interesting observation is that the honest miners outside the pool not only observe a positive revenue change per computational power but also more than the adversary mounting the attack for most parameters. Hence, we also consider the relative revenue ratio between the adversary and the honest miners outside the pool that can be optimized to make sure that the adversarial revenue change per computational power is better than any other honest miner. Such an optimization results in less adversarial hashpower allocated to the pool. However, it not only reduces the revenue gain for honest miners outside the pool, but also the revenue gain for the adversary itself.

\subsection{Long-Term Results}
In the long-run, the final and stabilized fraction of the hashpower of the malicious miner, $\alpha_{final}$ depends on the attack it mounts and its subsequent impact on the honest miners who are free to leave and join the system as well as the market share of the specific cryptocurrency with respect to the others. Analyzing such a complicated system is out of scope of this paper, hence, we provide a rigorous analysis and a framework to compare the models assuming a stabilized final hash ratio for the malicious miner where the honest miners who stay in the system have a sustainable revenue/cost relationship. If mounting different attacks end up with different final hash ratio in the long-run, that are known, they can still be compared.

\subsubsection{Selfish Mining vs Coin Hopping}
Using the efficiency metric together with the assumption of sustainability of honest mining, we conclude that, optimizing the efficiency of the adversary in the selfish mining strategies is the same as optimizing the revenue ratio of the adversary. Hence, the efficiency of the original selfish mining attacks \cite{selfish-mining,optimal-selfish} outperforms the intermittent mining \cite{intermittent_mining} and smart intermittent mining \cite{time_average_selfish_mining} if the malicious miner ends up with the same hash ratio in the long-run. However, we note that, this condition is not necessarily true as we mentioned earlier that the final and stabilized fraction of the hashpower of the malicious miner $\alpha_{final}$ depends on the attack it mounts.

Turning the attention to the alternate mining strategy of \cite{profit_lag}, assuming rational and sustainable miners in both BTC and BCH network in the long-run, coin hopping adversary always has a better efficiency (revenue/cost per hashpower per time) than the loyal honest miners. This in turn implies that, for $(\alpha,\gamma)$ parameters for which selfish mining is not efficient, alternate mining is still efficient. In fact, we provide new boundaries which indicate where selfish mining/alternate mining is better. For example, when $\gamma=0$, the efficiency of selfish mining of Eyal and Sirer ($L$=2) outperforms honest mining for $\alpha_{final}>1/3$, whereas the efficiency of selfish mining of Eyal and Sirer does not outperform alternate network mining when $\alpha_{final}<0.355$. 

\subsubsection{PAW Attack}
For PAW attacks, unlike the selfish mining strategies, it turns out that the optimal hashpower allocation maximizing the revenue ratio of the adversary is not the same as the allocation maximizing the adversarial efficiency. This happens, because there are two groups of honest miners, i.e., honest miners of the pool and outside the pool. Assuming a stabilized system with sustainable pool miners, in the long-run, not only the adversarial miner mounting the attack but also the honest miners outside the pool have better efficiency than the pool miners. In fact, if the adversary purely focuses on maximizing its efficiency or revenue ratio, the honest miners have better efficiency than the adversary mounting the attack unless the network influence of the adversary in fork races, $\gamma$ is high enough. Thus, we also provide a maximization of the relative efficiency of the adversary that makes sure the adversary has the best efficiency among all miners. Not surprisingly, the resulting efficiency is lower than if the adversary purely focused on its own efficiency. 

\subsection{Discussion and Future Work}
In the long run analysis, we used an efficiency metric that abstracts away economic factors in the real world such as electricity costs, coin dynamics, exchange rates to compare different strategies. Essentially, the goal in our efficiency metric is to treat the honest efficiency as a reference point in long term and find how much more/less efficient is the adversarial attack itself. This way, we are able to abstract away all the complex dynamics and create a simple but meaningful metric. The idea of modeling more complex systems, is meaningful and an interesting future research direction. 

The smart mining strategy studied in \cite{mind_the_mining} is a strategy where the attacker turns its mining rigs on/off in even/odd epochs instead of hopping between coins in even/odd epochs. The study of \cite{mind_the_mining} considers fixed costs (hardware, etc.) in addition to the energy costs, where the attacker bears the fixed costs every epoch whereas the energy costs are only born in the epochs where the rigs are on.  Note that, all mining rewards in \cite{mind_the_mining} are based on the coin dynamics before the attack starts, hence the result claimed in \cite{mind_the_mining} is essentially a short-term result. In this paper, we do not focus on smart mining nor the fixed costs for the reasons explained next. 

We do not consider fixed costs for the following reasons: In the short-term analysis of this paper, the fixed costs cancel each other when we compare two strategies in the revenue change metric, hence fixed costs can be ignored. In terms of mining costs in the long-term, fixed costs scale with the hashpower of the mining agent just like the energy costs. Hence, even though we call $E$ as the total energy cost of the system per unit time in the introduction of the unified efficiency metric in \eqref{eq::unified_metric}, since fixed costs can be divided by the total duration to yield a per-unit-time contribution, they can be absorbed into $E$ and do not alter the formulation.

We do not focus on smart mining since alternate network mining is essentially an enhanced version of smart mining, i.e., alternate network mining outperforms smart mining. Notice, their costs and revenues in even epochs are the same. In odd epochs, both attackers bear the fixed costs whereas the coin hopper additionally bears the short term energy costs. Since we assume that BCH DAA is responsive and honest mining in BCH is sustainable, the coin hopper is compensated for its fixed and energy costs in odd epochs by mining honestly on BCH in odd epochs. On the other hand, the smart miner bears the fixed costs in odd epochs without any compensation.

Throughout the paper we purely focused on coinbase rewards. Even though the transaction fee rewards can also be absorbed into the coinbase rewards on average, attacks such as bribery to petty-compliant miners through the transaction fees are not considered here, which can change the revenues and the subsequent analysis \cite{instability-no-block-reward, Deep_Bribe, werlman}. Note that, even if we purely focus on revenue ratio and ignore revenue changes and efficiencies, such an analysis incorporating the transaction fees requires an MDP model with over $100$ million states or a state-of-the-art deep reinforcement learning framework \cite{Deep_Bribe,werlman}. Another revenue source for the adversary is the potential interest gains from blockchain stretching and squeezing in alternate mining and intermittent mining strategies as considered in \cite{stretch_squeeze_mining}, which we do not study. We leave such considerations to the future work. 

When we analyzed PAW attacker and the revenue changes, we assumed a single coordinated adversarial entity against pools. Many studies in the literature consider multiple pools attacking each other \cite{miners_dilemma, fork_after_witholding_attack, power_adjusting}. The study of \cite{power_adjusting} shows that PAW attack can avoid so-called miner's dilemma established in \cite{miners_dilemma} when pools attack each other. These mentioned studies use the revenue ratio as a metric when arguing about whether a dilemma exists. It would be interesting to study if a dilemma is avoidable in terms of the revenue changes in short-term as well as the efficiency metric established in this paper for the long-term analysis when multiple pools attack each other, which we leave to the future work. 

In \cite{game_of_coins}, authors analyze a game scenario with multiple coins and strategic miners, who mine honestly but pick the coin they mine according to their best interest. Based on the analysis therein, a strategic miner can manipulate the rewards (e.g., transaction fees) to move the system between different equilibrium points, i.e., different hash ratios $\alpha$ for a specific coin. However, \cite{game_of_coins} does not consider malicious attacks such as selfish mining in their game analysis and it is not clear if the result holds when there are not only strategic miners but also malicious attacks. Incorporating the efficiency metrics we provide in this paper together with the corresponding attacks and combining it with the game theoretical model of \cite{game_of_coins} is an interesting future research direction. Such a work, can contribute more meaning to our analysis since we provide the best adversarial mining efficiency for each equilibrium point $\alpha_{final}$ with our method, but we do not have a direct result to show that such an equilibrium point can be achieved through manipulations. 

\appendices

\section{Epoch Lengths in Optimal Selfish Mining}\label{sec::app::justify_substitue}
As we mentioned earlier, $L$-selfish mining framework can be used as a replacement of $\epsilon$-optimal MDP of \cite{optimal-selfish} (OSM) to find quantities such as epoch lengths which in turn can be used to find profit lags, revenue changes etc with simple formulas for all parameters. Here, we conduct a Monte Carlo analysis of the $\epsilon$-optimal selfish mining attacks by using the optimal strategies obtained from the implementation in \cite{mustafadgr} for extensive set of parameters and show that $L$-selfish mining framework is a good substitute for OSM. For each $(\alpha,\gamma)$ parameter, we run the OSM strategy for $10$ million block times and report the obtained block redundancy ratio, i.e., we assume $D_0=10$ million blocks and report $\delta_{OSM}$, where $\delta_{OSM}$ is simply obtained by dividing $10$ million to the number of canonical blocks out of all the $10$ million blocks. 

The results are reported in Table~\ref{tab::compare_with_mdp_vary_gamma} together with the theoretical block redundancy ratios of selfish mining of Eyal and Sirer \cite{selfish-mining} and $L^*$-selfish mining of \cite{doger2025selfishminersdoublespend}. The results clearly show that epoch lengths of $L^*$-selfish mining is a good substitute for the epoch lengths of optimal selfish mining attack. Note also that, the resulting revenue ratios $\rho_{L^*}$ are within $10^{-2}$ of the $\rho^*_{OSM}$ obtained from the $\epsilon$-optimal mining strategies of \cite{optimal-selfish} as presented in \cite[Table~2]{doger2025selfishminersdoublespend}. Moreover, revenue changes $\Delta(t)$ and profit lags are solely expressible in terms of $\alpha,\rho$ and $\delta$ for all $t$. For the sake of completeness, in Table~\ref{tab:intertwined_estimation_errors_final}, we present the relative discrepancy between the profit lags and initial revenue losses of $L^*$-selfish mining and OSM. Note that, $D_1$ is the discrepancy in profit lags, i.e.,  how much profit lag of $L^*$-selfish mining deviates from the profit lag of OSM normalized by the profit lag of OSM, whereas $D_2$ is the discrepancy in the initial revenue losses at the end of the first epoch.

\begin{table*}[h] 

\begin{center}\resizebox{\textwidth}{!}{
\begin{tabular}{|c ||c |c |c ||c |c |c ||c |c |c ||c |c |c ||c |c |c ||c |c |c |} 
 \hline
  & \multicolumn{3}{c||}{ $\gamma=0.0$}& \multicolumn{3}{c||}{ $\gamma=0.2$}& \multicolumn{3}{c||}{ $\gamma=0.4$}& \multicolumn{3}{c||}{ $\gamma=0.5$}& \multicolumn{3}{c||}{ $\gamma=0.6$}& \multicolumn{3}{c|}{ $\gamma=0.8$}   
 \\\hline
   \multicolumn{1}{|c||}{$\alpha$}& $\delta_{2}$&$\delta_{L^*}$& $\delta_{OSM}$ &$\delta_{2}$&$\delta_{L^*}$& $\delta_{OSM}$ &$\delta_{2}$&$\delta_{L^*}$& $\delta_{OSM}$ &$\delta_{2}$&$\delta_{L^*}$& $\delta_{OSM}$ &$\delta_{2}$&$\delta_{L^*}$& $\delta_{OSM}$ &$\delta_{2}$&$\delta_{L^*}$& $\delta_{OSM}$   \\ [0.5ex]
 \hline
0.10
& 1.092 & 1.000 & 1.000
& 1.092 & 1.000 & 1.000
& 1.092 & 1.000 & 1.000
& 1.092 & 1.000 & 1.000
& 1.092 & 1.000 & 1.000
& 1.092 & 1.000 & 1.000 \\

0.15
& 1.134 & 1.000 & 1.000
& 1.134 & 1.000 & 1.000
& 1.134 & 1.000 & 1.000
& 1.134 & 1.000 & 1.000
& 1.134 & 1.000 & 1.000
& 1.134 & 1.134 & 1.134 \\

0.20
& 1.176 & 1.000 & 1.000
& 1.176 & 1.000 & 1.000
& 1.176 & 1.000 & 1.000
& 1.176 & 1.000 & 1.000
& 1.176 & 1.000 & 1.000
& 1.176 & 1.250 & 1.250 \\

0.25
& 1.220 & 1.000 & 1.000
& 1.220 & 1.000 & 1.000
& 1.220 & 1.000 & 1.000
& 1.220 & 1.000 & 1.000
& 1.220 & 1.220 & 1.219
& 1.220 & 1.333 & 1.333 \\

0.30
& 1.269 & 1.000 & 1.000
& 1.269 & 1.000 & 1.000
& 1.269 & 1.269 & 1.269
& 1.269 & 1.269 & 1.269
& 1.269 & 1.376 & 1.381
& 1.269 & 1.429 & 1.428 \\

0.35
& 1.330 & 1.330 & 1.333
& 1.330 & 1.330 & 1.332
& 1.330 & 1.418 & 1.447
& 1.330 & 1.458 & 1.505
& 1.330 & 1.538 & 1.532
& 1.330 & 1.538 & 1.539 \\

0.40
& 1.419 & 1.419 & 1.418
& 1.419 & 1.516 & 1.520
& 1.419 & 1.585 & 1.631
& 1.419 & 1.667 & 1.656
& 1.419 & 1.667 & 1.664
& 1.419 & 1.667 & 1.667 \\

0.45
& 1.576 & 1.671 & 1.650
& 1.576 & 1.710 & 1.750
& 1.576 & 1.818 & 1.807
& 1.576 & 1.818 & 1.817
& 1.576 & 1.818 & 1.817
& 1.576 & 1.818 & 1.819 \\
\hline
\end{tabular}}
\end{center}
\caption{ Comparison of the first epoch lengths of selfish mining, $L^*$-selfish mining and MDP model of \cite{optimal-selfish} (assume $\tau_0=1$).}
\label{tab::compare_with_mdp_vary_gamma}
\end{table*}

\begin{table*}[h]

\begin{center}\resizebox{\textwidth}{!}
{\begin{tabular}{|c||c|c||c|c||c|c||c|c||c|c||c|c|}
\hline
 & \multicolumn{2}{c||}{$\gamma=0.0$} 
 & \multicolumn{2}{c||}{$\gamma=0.2$} 
 & \multicolumn{2}{c||}{$\gamma=0.4$} 
 & \multicolumn{2}{c||}{$\gamma=0.5$} 
 & \multicolumn{2}{c||}{$\gamma=0.6$} 
 & \multicolumn{2}{c|}{$\gamma=0.8$} \\
\hline
$\alpha$ & $D_1$ & $D_2$ & $D_1$ & $D_2$ & $D_1$ & $D_2$ & $D_1$ & $D_2$ & $D_1$ & $D_2$ & $D_1$ & $D_2$ \\
\hline
0.10 
& 0 & 0 
& 0 & 0 
& 0 & 0 
& 0 & 0 
& 0 & 0 
& 0 & 0 \\
0.15 
& 0 & 0 
& 0 & 0 
& 0 & 0 
& 0 & 0
& 0 & 0 
& -0.1421 & -0.0104 \\
0.20 
& 0 & 0 
& 0 & 0 
& 0 & 0 
& 0 & 0 
& 0 & 0 
& 0.0051 & 0.0024 \\
0.25 
& 0 & 0 
& 0 & 0 
& 0 & 0 
& 0 & 0 
& -0.0041 & -0.0009 
& -0.0039 & -0.0057 \\
0.30 
& 0 & 0 
& 0 & 0 
& -0.0199 & -0.0042 
& -0.0010 & -0.0007 
& -0.0341 & -0.0111 
& -0.0021 & -0.0063 \\
0.35 
& -0.2632 & -0.0406 
& -0.0367 & -0.0130 
& 0.0243 & 0.0752 
& 0.0392 & 0.1136 
& -0.0249 & -0.0477 
& 0.0006 & 0.0044 \\
0.40 
& -0.0521 & -0.0703 
& -0.0529 & -0.0654 
& 0.0178 & 0.0696 
& -0.0451 & -0.1224 
& -0.0138 & -0.0564 
& -0.0008 & -0.0061 \\
0.45 
& -0.0532 & -0.2160 
& 0.0178 & 0.0482 
& -0.0405 & -0.2349 
& -0.0145 & -0.1175 
& -0.0063 & -0.0783 
& -0.0022 & -0.1017 \\
\hline
\end{tabular}}
\end{center}
\caption{Relative discrepancy between profit lags and initial revenue losses of $L^*$-selfish mining and MDP model of \cite{optimal-selfish}.}
\label{tab:intertwined_estimation_errors_final}

\end{table*}

\section{Proofs of the Relevant Quantities in PAW-Type-B}\label{sec::app::bwh}

Consider how an attack cycle unfolds from the start:
\begin{enumerate}
    \item w.p. $(1-\alpha-\beta)$, the honest miners outside the pool find a block, they get the full block reward.
    \item w.p. $(1-p_1)\alpha$, the adversary creates a block on its own, receiving full block reward.\label{step::reward_2}
    \item w.p. $\beta$, an honest pool miner finds a fPoW, the block reward is shared between the adversary ($\frac{\alpha p_1}{\beta+\alpha p_1}$ fraction) and the honest pool miners ($\frac{\beta}{\beta+\alpha p_1}$ fraction).\label{step::reward_3}
    \item w.p. $\alpha p_1$, the adversary encounters fPoW and hides it, readjusts its hashpower distribution. In the subsequent part of this attack cycle (called sub-cycle),{\footnote{Here, the active total hashpower in the system is $1-\alpha p_2$ since the adversary keeps discarding fPoW with $\alpha p_2$ power until someone else finds a block. Hence, we rescale the subsequent event probabilities with $1-\alpha p_2$.}} \label{step::reward_4}
    \begin{enumerate}
        \item w.p. $\frac{(1-p_2)\alpha}{1-\alpha p_2}$, the adversary eventually creates a block on its own, receiving full block reward.\label{step::reward_4a}
        \item w.p. $\frac{\beta}{1-\alpha p_2}$, an honest pool miner eventually creates a block, the block reward is shared between the adversary ($\frac{\alpha p}{\beta+\alpha p}$ fraction) and the honest pool miners ($\frac{\beta}{\beta+\alpha p}$ fraction).\label{step::reward_4b}
        \item w.p. $\frac{1-\alpha-\beta}{1-\alpha p_2}$, the honest miners outside the pool find a block, the adversary releases fPoW, and the fork race starts: \label{step::reward_4c}
        \begin{enumerate}
            \item w.p. $\alpha$, the adversary mines a block on top of fPoW on its own. The adversary gets 1 full block reward for this block. Moreover, the reward of fPoW is distributed between the adversary ($\frac{\alpha p}{\beta+\alpha p}$ fraction) and the honest pool miners ($\frac{\beta}{\beta+\alpha p}$ fraction).\label{step::reward_4ci}
            \item w.p. $\beta$, the pool mines a block on top of fPoW. The pool (no adversarial contribution) gets 1 full block reward for this block. Moreover, the reward of fPoW is distributed between the adversary ($\frac{\alpha p}{\beta+\alpha p}$ fraction) and the honest pool miners ($\frac{\beta}{\beta+\alpha p}$ fraction).\footnote{For the sake of simplicity, assume a rational pool manager. Otherwise, the pool (no adversarial contribution) gets 1 full block reward and the rest of the honest miners also get 1 block reward.}\label{step::reward_4cii}
            \item w.p. $(1-\alpha-\beta)\gamma$, the honest miners outside the pool mine a block on top of fPoW. The reward of fPoW is distributed between the adversary ($\frac{\alpha p}{\beta+\alpha p}$ fraction) and the honest pool miners ($\frac{\beta}{\beta+\alpha p}$ fraction). \label{step::reward_4ciii}
        \end{enumerate}
        
    \end{enumerate}
        
\end{enumerate}

For now, assume $p$ is given, we will provide the formula with a proof later. For the adversary, $\mathbb{E}[B_A]$ can be obtained by multiplying the adversarial rewards with each event probability, i.e., from Case~\ref{step::reward_2} $(1-p_1)\alpha$, Case~\ref{step::reward_3} $\beta\frac{\alpha p_1}{\beta+\alpha p_1}$, Case~\ref{step::reward_4a} $\alpha p_1 \frac{(1-p_2)\alpha}{1-\alpha p_2}$,  Case~\ref{step::reward_4b} $\alpha p_1 \frac{\beta}{1-\alpha p_2}\frac{\alpha p}{\beta+\alpha p}$, Case~\ref{step::reward_4ci} $\alpha p_1 \frac{1-\alpha-\beta}{1-\alpha p_2}\alpha(1+\frac{\alpha p}{\beta+\alpha p})$, Case~\ref{step::reward_4cii} $\alpha p_1 \frac{1-\alpha-\beta}{1-\alpha p_2}\beta\frac{\alpha p}{\beta+\alpha p}$ and Case~\ref{step::reward_4ciii} $\alpha p_1 \frac{1-\alpha-\beta}{1-\alpha p_2}\gamma(1-\alpha-\beta)\frac{\alpha p}{\beta+\alpha p}$. For the honest pool miners, similar calculations trivially give $\mathbb{E}[B_P]$. For $\mathbb{E}[B_R]$, one can either use $\mathbb{E}[B_C]-\mathbb{E}[B_A]-\mathbb{E}[B_P]$ or explicitly follow the steps as we did for $\mathbb{E}[B_A]$.

In each cycle, Case~\ref{step::reward_4c} extends the canonical chain by $2$ blocks whereas the other events extend the canonical chain by $1$ block. Since the probability of Case~\ref{step::reward_4c} is $\alpha p_1 \frac{1-\alpha-\beta}{1-\alpha p_2}$, we get $\mathbb{E}[B_C]$ as in \eqref{eq::canonical_bwh_type_b}. To prove $\mathbb{E}[B_O]$, note that in addition to canonical blocks, Case~\ref{step::reward_4}, which happens w.p. $\alpha p_1$, always incurs $1$ fPoW that is lost. On top of that, it takes $\frac{1}{1-\alpha p_2}-1$ more trials (geometric distribution) as the adversary keeps discarding fPoWs with $\alpha p_2$ power until (excluding) someone else finds a block. Hence, $\mathbb{E}[B_O]=\mathbb{E}[B_C]+\frac{\alpha p_1}{1-\alpha p_2}$.

It remains to prove the formula for $p$ as given in \eqref{eq::avg_power_bwh}. If everyone including the adversary published a block and fPoW immediately as they are mined, a cycle would last $1$ unit of block interval time in expectation. Note, $p$ is the average fraction of the power that the adversary distributes to pool mining conditioned on the event that it encounters an fPoW and keeps it private, i.e., conditioned on the occurrence of Case~\ref{step::reward_4}. The adversarial power distribution for pool is $p_1$ until Case~\ref{step::reward_4} happens, which takes on average in $1$ unit of block interval time, however, the adversary keeps the fPoW private and as a result, the cycle lasts longer than usual. We call the first part of the cycle that lasts until encountering the first fPoW as main-cycle and its duration as $D_1$. Clearly $\mathbb{E}[D_1]$ is equal to $1$ unit of block interval time, i.e., $\mathbb{E}[D_1]=1$.  We denote the indicator of event that the attack enters the Case~\ref{step::reward_4} as $\mathbbm{1}_{D_2}$, where $\mathbb{E}[\mathbbm{1}_{D_2}]=\alpha p_1$. In the subsequent part of the attack cycle, the adversarial power distribution for pool is $p_2$ and it takes $\frac{1}{1-\alpha p_2}$ more trials (geometric distribution) as the adversary keeps discarding fPoWs with $\alpha p_2$ power until (including) someone else finds a block. Hence, after Case~\ref{step::reward_4}, it takes $\mathbb{E}[{D_2}|\mathbbm{1}_{D_2}=1]=\frac{1}{1-\alpha p_2}$ unit block times until one of Case~\ref{step::reward_4a}, Case~\ref{step::reward_4b} or Case~\ref{step::reward_4c} happens. Thus,
\begin{align}
    p=\frac{p_1\mathbb{E}[D_1]+p_2\mathbb{E}[{D_2}|\mathbbm{1}_{D_2}=1]}{\mathbb{E}[D_1]+\mathbb{E}[{D_2}|\mathbbm{1}_{D_2}=1]}=\frac{p_1+p_2-\alpha p_1 p_2}{2-\alpha p_2}.
\end{align}
This analysis also confirms the result of \cite[Theorem~5.1]{power_adjusting} regarding $p$ in one victim pool case.

Finally, since $\delta_{p_1,p_2}$ is the block redundancy ratio, one can obtain the average hashpower reduction $\alpha p_w$ by rearranging
\begin{align}
    \delta_{p_1,p_2}=\frac{1}{1-\alpha p_w},\label{eq::alternate_p_w}
\end{align}
using a similar reasoning as \eqref{eq::alternate_epoch_dur}.

\section{Profitability of PAW Without DAA}\label{sec::app::daa}
Here, we  explain why the claim in \cite[Corollary~3.4]{Grunspan_witholding_resilience} and \cite[Theorem~4.4]{grunspan2019-profitability-selfish-mining} fails to cover the PAW against pools. A similar claim is also present in \cite[Section~1]{sarenche2025miningpowerdestructionattacks} stating that \textit{no mining power destruction attacks (including PAW) can be more profitable than following the honest strategy during the initial difficulty epoch}. Note that, both \cite[Theorem~4.4]{grunspan2019-profitability-selfish-mining} and \cite[Corollary~3.4]{Grunspan_witholding_resilience} use the same method to argue in favor of the claim but the analysis in \cite{Grunspan_witholding_resilience} is more organized for a first-time reader. The claim is essentially proven by bounding the number of canonical adversarial blocks by the number of blocks mined by the adversary in total using martingale theory. As no-DAA is assumed, the adversarial revenue in turn is bounded by the total number of blocks mined by the adversary. 

In PAW, when the adversary encounters a fPoW in the main-cycle and hides it, the sub-cycle starts. During the sub-cycle, if the adversary increases its fraction of hashpower working for the pool, i.e., if $p_1<p_2$ and the honest miners outside the pool find a block, w.p. $1-(1-\gamma)(1-\alpha-\beta)$, not only the fPoW-block becomes part of the longest chain in the long-run, but also the partial reward of the adversary increases for that particular fPoW. On the other hand, the proofs in \cite{Grunspan_witholding_resilience, grunspan2019-profitability-selfish-mining} do not consider pool mining, where the share of rewards the adversary gets can increase based on the hashpower allocation during the withholding. Note that, without the dynamic power adjustment, the claim still holds for BWH on pools and FAW attacks, however, dynamic power adjustment of PAW invalidates the claim for certain parameters since the share of the rewards received by pool miners change during the adversarial withholding at the sub-cycle.

For the sake of completeness, we also provide an example set of parameters for the original PAW attack under the model of \cite{power_adjusting} with the simplified assumption of $c$ where there is no profit lag for the adversary.

\begin{example}\label{ex::bwh}
    Let the total fraction of adversarial hashpower be $\alpha=0.2$ and the fraction of hashpower of the honest pool miners be $\beta=0.04$, where the adversary launches PAW attack against the pool with $c=0.75$, $p_1=0.04$ and $p_2=0.93$. The attack results in $\rho=0.2032$ and $\delta_{p_1,p_2}=1.0098$, hence, from \eqref{eq::bwh_init_loss_adv}, the adversarial revenue change at $t_1$ is $\Delta_A(t_1)=1.2\cdot 10^{-3}$. In other words, under the PAW modeling of \cite{power_adjusting}, the revenue of the adversary can increase even in the first epoch before the difficulty is readjusted, which disproves Claim~\ref{claim::grunspan}.
\end{example}

We also verify both Example~\ref{ex::bwh-type_b} and Example~\ref{ex::bwh} using a Monte Carlo simulation under 500 million unit block times without any difficulty adjustment, i.e., we run each simulation 500 million time steps where a block (not necessarily a canonical one) is created at each time step. A comparison between the simulated values and the theoretical ones verify each other and show that the adversary can indeed increase its revenue under PAW attacks without any difficulty adjustments.

\bibliographystyle{IEEEtran}
\bibliography{blockchain}
\end{document}